\documentclass{jfm}

\begin{document}

\newtheorem{lemma}{Lemma}
\newtheorem{corollary}{Corollary}

\shorttitle{
Universal features of turbulence and its non-equilibrium TDs} 
\shortauthor{N. Reinke et al.} 

\title{On universal features of the turbulent cascade in terms of non-equilibrium thermodynamics}

\author
 {
 Nico Reinke\aff{1,2} \corresp{\email{peinke@uni-oldenburg.de}},
 Andre Fuchs \aff{1,2} 
 Daniel Nickelsen\aff{1,3,4},
  \and 
  Joachim Peinke\aff{1,2}
  }
\affiliation
{
\aff{1} Institute of Physics, University of Oldenburg
\aff{2} ForWind, University of Oldenburg, Carl-von-Ossietzky-Str. 9-11, 26129 Oldenburg, Germany
\aff{3} National Institute for Theoretical Physics (NITheP), Stellenbosch 7600, South Africa
\aff{4} Institute of Theoretical Physics, University of Stellenbosch, South Africa
}

\maketitle


\begin{abstract}

Features of the turbulent cascade are investigated for various datasets from three different turbulent flows, namely free jets as well as wake flows of a regular grid and a cylinder.  
The analysis is focused on the question as to whether fully developed turbulent flows show universal small scale features.
Two approaches are used to answer this question. Firstly, 2-point statistics, namely structure functions of longitudinal velocity increments and secondly, joint multi-scale statistics of these velocity increments are analysed. 
The joint multi-scale characterisation encompasses the whole cascade in one joint probability density function.
On the basis of the datasets, evidence of the Markov property for the turbulent cascade is shown, which corresponds to a three point closure that reduces the joint multi-scale statistics to simple conditional probability density functions (cPDF).
The cPDF are described by the Fokker-Planck equation in scale and its Kramers-Moyal coefficients (KMCs).
KMCs are obtained by a self-consistent optimisation procedure from the measured data and result in a Fokker-Planck equation for each dataset. 
The knowledge of these stochastic cascade equations enables to make use of the concepts of non-equilibrium thermodynamics and thus to determine the entropy production along individual cascade trajectories. In addition to this new concept, it is shown that the local entropy production is nearly perfectly balanced for all datasets by the integral fluctuation theorem (IFT).  Thus the validity of the IFT can  be taken as a new law of the turbulent cascade and at the same time independently confirms that the physics of the turbulent cascade is a memoryless Markov process in scale. 
IFT is taken as a new tool to prove the optimal functional form of the Fokker-Planck equations
 and subsequently to investigate the question of universality of small scale turbulence in the datasets.
The results of our analysis show that the turbulent cascade contains universal and non-universal features. We identify small scale intermittency as a universality breaking feature. We conclude that specific turbulent flows have their own particular multi-scale cascade, with other words their own stochastic fingerprint.
\end{abstract}


\section{Introduction}
A general description of turbulent flows remains one of the most challenging unsolved scientific problems, e.g. \citet{Nelkin1992,Clay}. Of particular importance is the simplified problem of homogeneous isotropic turbulence (HIT). 
HIT is not only often the starting point for fundamental studies, but also important for applications like CFD-modelling, where the features of small scale turbulence are essential for simplifications, e.g. \citet{Pope2000}.

The idea and the profound understanding of turbulence as a cascade process, cf. \citet{Frisch2001}, plays an important role for such simplifications.  The turbulent cascade can be understood as the evolution of turbulent structures with varying spatial size given by a scale $r$. The evolution can be divided into 3 stages. Firstly, turbulence is initiated at large scales, characterised by the  integral length scale $L$. Secondly, the generated large scale structures decay throughout the cascade process.
It is expected that this decay process is self-similar until, thirdly, it ends at a micro scale, where the viscosity of the fluid starts to significantly affect the decay. The range of non-dissipative scales is called the inertial range and it is this range where we have performed our analysis.

One important question for the understanding of the turbulent cascade is how turbulent structures evolve within the inertial range.
Strongly related to this issue is the question as to whether turbulent structures evolve in a universal way, or whether the evolution depends on the generation process and its specific large scale features.
With other words, how long do large turbulent structures have a significant impact to small scale features?
This central question is frequently discussed in literature since the contributions of Kolmogorov and Landau, cf. \citet[pp. 93-98]{Frisch2001} and it is essential for the concept of universal small scale turbulence.

The overwhelming majority of studies investigate turbulence by statistics of increments $u_r$ like 
structure functions $S^{\kappa}( u_r )$. 
For HIT, scaling features of structure functions are commonly assumed as a function of scale $r$,  characterised by a scaling exponent $\xi_{\kappa}$, cf. \citet{Kolmogorov1941b,Kolmogorov1962,Frisch2001,Davidson2004,Pope2000}.
Universality is assumed to be found in the scaling exponents $\xi_{\kappa}$, whereas pre-factors may change with different flow 
conditions, cf. \citet{Kolmogorov1962}. Within this context, works on finite size effects, \citet{Dubrulle2000}, or the understanding of multi-scale excited turbulent flows and its structures, e.g. \citet{Kuczaj2006, Hurst2007, Valente2012, Nagata2013}, are of particular interest.

Note that the structure function analysis can be expressed equivalently by probability density functions $p( u_r )$ (PDF). From a statistical point of view, structure functions as well as increment PDFs are two point quantities, i.e. the relation of velocities at two points separated by a selected distance, the scale $r$, is investigated.

It was shown, e.g. in \citet{Renner2002}, that 2-point statistics do not completely characterise the 
features of small scale turbulence.
A general multi-scale characterisation of the turbulent cascade can be expressed by the joint PDF $p( u_r ;u_{r+\Delta};... ;$ $u_{r+N\Delta})$ at $N$ scales $r$. Reviews on the stochastic processes in general and in the context of turbulence can be found in \citet{Friedrich2009,Friedrich2011}.
Since velocity increment series show the Markov property, the joint PDF can equally be expressed by a chain of conditional PDFs $p( u_r  | u_{r+\Delta})$, cf. \citet{Friedrich1997,Renner2001,Tutkun2004}. 
Note that this Markov property corresponds to a 2-scale or 3-point closure of the generalised multi-scale  joint probability.
Furthermore, an evolution of this transition probability $p( u_r  | u_{r+\Delta})$ along the scale can be described by the Fokker-Planck equation, where only the first two Kramers-Moyal coefficients $D^{(1)}( u_r )$ and $D^{(2)}( u_r )$ govern the equation.
Thus, these two Kramers-Moyal coefficients describe the stochastic features of the turbulent cascade along the inertial range.
%
Thereby, different cascades, measured in different turbulent flows, can be set into a quantitative comparison with these two coefficients. 
Note, similar to this approach, other approaches contain equivalent joint multi-scale information 
of the turbulent cascade, cf. \citet{Lundgren1967,Monin1967,Procaccia1996,Reza2000,Monin2007}.  
In the discussion of cascade models for multifractal features of turbulence, relations similar to the Markov condition were implicitly discussed. Quantities and kernel functions relating different scales were used \citet{Novikov1994,Castaing1995,She1995,Arneodo1997}, sometimes with the concept of infinitely divisible distributions. These scale relating features can be considered as Chapman-Kolmogorov conditions. The indicated works may therefore be considered as first works in the direction of stochastic cascade processes, however, energy and energy related quantities of the cascades were considered and not the velocity increment itself. Some of these works already put proper scaling behaviour and flow independent universalities for the increment statistics into question.

A different approach using the Langevin equation for the  stochastic description of the cascade was presented in \citet{Marcq2001}. Based on the Langevin equation, the authors reconstructed the noise along the cascade and showed that it is uncorrelated. In their analysis the connection to structure functions were investigated and it was concluded that the odd order structure function could be grasped by the stochastic approach only qualitatively. Furthermore, they posed the question as to what extend the stochastic analysis of the cascade depends on the nature of the turbulent flow, a point which will be discussed in this paper in detail.

\citet{Renner2001} applied such a multi-scale analysis to turbulent data taken in a free jet experiment and the Kramers-Moyal coefficients were related to structure functions. 
In  \citet{Renner2002} it is shown that the multi-scale statistics of free jet generated turbulent cascades change with Taylor-Reynolds number ($\Rey_\lambda$), and the issue of universal features of small scale turbulence was discussed. 
In \citet{Siefert2003,Siefert2006} the common stochastic process of longitudinal and transversal velocity increments were investigated and the relation to the Karman equation was shown.
A similar work addressed the statistics of passive scalars, c.f. \citet{Tutkun2004, Melius2014}, where the features of a turbulent flow was analysed in terms of Kramers-Moyal coefficients from which the stochastic flow features could be reconstructed. 

Based on multi-scale statistics, \citet{Stresing2010B} found that fractal grid generated turbulence exhibits different scaling features and multi-scale statistics than \textit{commonly} generated flows (cylinder wake and free jet), defining a novel class of turbulence.
The work of \citet{Stresing2010,Keylock2015} gives first indications that the turbulent cascade and its Kramers-Moyal coefficients depend on the turbulence generation mechanism. This dependence on turbulence generation contrasts with the universal features found on the basis of 2-point statistics involving structure functions and its scaling exponents.
Therefore, we
revive this analysis with 2-point and joint multi-scale statistics and 
study the question of universality in more detail and in a more general as well as more systematic way. 

Although the joint multi-scale statistics will deliver more details of the complex structure of turbulence, an issue remains with the accuracy in determining the Kramers-Moyal coefficients, since one has to bridge the Einstein-Markov length for their estimation, cf. \citet{Friedrich1998,Renner2001,Lueck2006}. 
\footnote{Note that a general feature of continuous stochastic processes is that it interpolates the real process, such as the deterministic motion of a particle in a diffusion process, or the dissipative structures in turbulent flows considered here.}
Two ways to overcome this technical difficulty have been worked out and applied in this investigation. 
Firstly, the Kramers-Moyal coefficients are estimated by an optimisation procedure. 
Thus, Kramers-Moyal coefficients matches best to the experimental found data, namely $p( u_r  |  u_{r+\Delta})$, cf. \citet{Nawroth,Kleinhans}.
Secondly, an independent way to assess the {adequacy} of the Kramers-Moyal coefficients is based on an integral fluctuation theorem (IFT), cf. \citet{Seifert2005}. 
The IFT is a generalisation of the second law to the non-equilibrium thermodynamics of small systems.
Recently, the formal applicability of the IFT to turbulent cascade processes has been demonstrated by \citet{Nickelsen2013} for a free jet flow.

In our paper, we confirm the applicability of the IFT for a total number of 61 different turbulent flows and explore its implications. As the IFT is based on the knowledge of the underlying stochastic process, we report in this paper in detail the advances which have been worked out for the  reconstruction of the Fokker-Planck equation from given data. Moreover, we use the IFT as a criterion for the adequacy of the estimated stochastic process equations. It is the first time, that the IFT is verified and utilised for such a comprehensive dataset comprising various types of turbulent flows at a wide range of Reynolds numbers, which is our first main result. 

A particular benefit of including the IFT into the estimation procedure is the pronounced sensitivity of the IFT with regard to the proper modelling of small scale intermittency. 
Thus, the IFT can be taken as an independent evidence for the Markov property of the turbulent cascade. 
Utilising this sensitivity, we are further able to pin down the minimal functional form of the first two Kramers-Moyal coefficients and still respect the features of small scale turbulence, 
which is our second main result\footnote{See also a preliminary use of this procedure in a data analysis of turbulence generated by a fractal grid, \citet{Reinke2015b}.}. 

This new approach enables to draw the analogy to non-equilibrium thermodynamic
processes, sheds light on the essential mechanisms of the cascade process and identifies
an energetic term  $u^2$  as the source of intermittent small scale fluctuations.

Finally, the combination of the optimisation procedure and the IFT allows us a profound comparison of stochastic features of the different turbulent flows. We conclude that, within the inertial range and on small scales, universal and non-universal contributions are present. In particular, intermittent features of the cascade can be identified as universality breaking, which is our third main result.

The basis of our analysis forms the longitudinal velocity increment component of 61 datasets, mostly measured by hot-wires, where the turbulence was generated by regular grid, cylinder and free jet, and cover a Taylor-Reynolds number range of  $28\leq \Rey_\lambda \leq996$. 
The entire data was always used, no single dataset has been sorted out for reasons of clarity. 

This paper is organised as follows:  
In section \ref{sec:Databasis} we give more information about the analysed datasets, 
section \ref{sec:Methods} details the used methods regarding structure functions, the IFT and the multi-scale analysis. 
In section \ref{sec:Results} the results are presented, followed by a discussion against the background of universality and the cascade process. Section \ref{sec:Discussion} concludes the paper.

\section{Data basis}
 \label{sec:Databasis}
The data basis used in this study consists of 61 single datasets, 
which correspond to seven subgroups, namely 
 i)  regular grid generated turbulence, with $28\leq \Rey_\lambda \leq153$, cf. \citet{Lueck2006}, 
 ii)  cylinder wake turbulence at the relative distance $\frac{x}{D}$=40 with $261\leq \Rey_\lambda \leq894$, cf. \citet{Lueck2006}, 
 iii)  cylinder wake turbulence at various positions $\frac{x}{D}$=50, 60, 70, 80, 90 and 100 with $\Rey_\lambda \approx 261$, cf. \citet{Lueck2006}, 
 iv) free jet generated turbulence in helium at $\frac{x}{D}$=40 with $208\leq \Rey_\lambda \leq996$, cf. \citet{Chanal2000}, 
 v) free jet turbulence in air at $\frac{x}{D}$=40, 
 vi) free jet turbulence in air at $\frac{x}{D}$=60 and  
 vii) free jet turbulence in air at $\frac{x}{D}$=80.
The free jet data in air cover the Taylor Reynolds number range of $166\leq \Rey_\lambda \leq865$. 
$x$ denotes in all cases the downstream position, $D$ the cylinder diameter or the jet nozzle diameter. 

The subgroups i), ii), v)-vii) are measured with a single hot-wire probe and contain longitudinal velocity component measurements.
The subgroup iii) is measured with a X-wire probe, therefore these data contain the longitudinal and one transversal velocity component, which is not analysed here.
The subgroup iv) is measured with a special micro structured hot-point probe, cf. \citet{Chanal2000}.  
This measurement technique is similar to a single hot-wire probe. 

In the following, results of the single subgroups are indicated by the following markers and notations in legends:
i) open squares, \textit{Grids}; 
ii) open circles, \textit{Cylinders};
iii) solid circles, greyscale indicates $\frac{x}{D}$, the darker the larger $\frac{x}{D}$, \textit{Cylinders (x/D)};
iv) open triangles, \textit{Jets He};
v) solid triangles, \textit{Jets 40D};
vi) open triangles, \textit{Jets 60D} and 
vii) solid triangles, \textit{Jets 80D}. 
Open and solid triangles are differently oriented in order to improve their distinctness.

Four exemplary datasets are used to highlight the results of our analysis. These datasets are chosen either to be comparable in terms of Taylor-Reynolds number $\Rey_\lambda$ (namely $153$ and $166$ or $894$ and $996$) or to be comparable in the kind of turbulence generation.
Characteristic quantities of these datasets are summarised in table \ref{table:data}. 
The listed quantities are, longitudinal integral length $L$, Taylor micro scale length $\lambda$, a newly defined 
Kolmogorov length scale $\Theta$, standard deviation of the velocity $\sigma_v$ and a newly defined standard deviation $\sigma_{\Theta}$ for normalisation.
\begin{table}
\centering
  \begin{tabular}{l|ccccccc} 
  Dataset  &  $\Rey_\lambda$~[-]& $L$~[mm]  & $\lambda$~[mm] & $l_{EM}$~[mm] & $\Theta$~[mm] &  $\sigma_v$~[m/s] & $\sigma_{\Theta}$~[m/s] \\
  \hline
  i)   \ \ \ grid     & 153                   &  25.0          & 2.61               & 5.22                  &     6.57             &  0.845                    &  0.637                  \\
  iii)   \ cylinder & 894                   &  25.2          & 3.35               &6.70                   &     9.78             &  3.94                      &  2.40                  \\
  iv)  \ jet (2)     & 996                   &  1.95          & 0.186             & 0.372                &     0.464           &  0.183                    &  0.119                  \\
  vii) jet (1)       & 166                   &  62.5          & 5.84               & 11.7                  &     15.7              &  0.382                    &   0.278                 \\
\end{tabular}
\caption{Exemplary datasets and their characteristics.}
\label{table:data}
\end{table}

To estimate $\lambda$, the procedure proposed by \citet{Aronson1993} is used, in good agreement with the findings in \citet{Lueck2006} and \citet{Renner2001}. The integral length $L$ is estimated by integrating the autocorrelation function, cf. \citet{Batchelor1953}. In case of a non-monotonous decrease of the autocorrelation function (until first zero crossing) the autocorrelation function is extrapolated by an exponential function, cf. \citet{Reinke2015a}. More information regarding this procedure can be found in appendix A1. Results of $L$ match with findings in \citet{Lueck2006} and \citet{Renner2001}. Note also that the ratio $\frac{L}{\lambda}$ as a function of $\Rey_\lambda$ is shown in appendix A1,
 which demonstrates the consistency of the used datasets.
$\Theta$ and  $\sigma_{\Theta}$ are defined in section  \ref{sec:Methods}.

\section{Methods of analysis}
 \label{sec:Methods}
 This section discusses the methods we use to analyse our data. First the classical method based on structure functions is briefly summarised, before we discuss in some more detail the method used to achieve a multi-scale analysis by means of stochastic processes. In particular, the procedure to obtain from the data a best estimate for a cascade Fokker-Planck equation is discussed. Finally, we show how the estimated cascade equation can be used to determine the entropies for which the integral fluctuation theorem is a rigorous law.
\subsection{Classical analysis}
 \label{sec:Classical}
At first, datasets of velocity time series are transferred to spatial velocity series by the use of Taylor's hypothesis,
\begin{equation} \label{eq:th}
	v(dx) \approx v( -\langle v(t) \rangle dt),
\end{equation}
where $v$ denotes the velocity component in the main flow direction, $x$ is the spatial component in the main flow direction, $t$ is time and $\langle ... \rangle$ denotes the expectation value. The minus sign in eq. (\ref{eq:th}) aligns the orientation of the $x$-axis with the flow direction. 
Taylor's hypothesis of frozen turbulence is commonly known as a good approximation for small velocity fluctuations $\frac{ v'}{\langle v \rangle} \ll 1$, cf. \citet{Taylor1938}, or small turbulence intensities, i.e. smaller than 10-20\%, cf. \citet{Murzyn2005}. 
This limit is fulfilled for almost all datasets. Further discussion of the validity can be found e.g. in \citet{Lin1950,Lumley1965,Pinton1994,Tong1995,Gledzer1997}.  

A standard quantity to analyse the scaling of turbulence is the velocity increment 
\begin{equation} \label{eq:u_r}
	u_r^* := v(x+r)-v(x),
\end{equation}
which is the difference of velocities at two points separated by the distance $r$, therefore its statistics is referred to as 1-scale statistics or, respectively, 2-point statistics on scale $r$.
A common quantity for the analysis of turbulence is the so-called structure function of order $\kappa$,
\begin{equation} \label{eq:sf}
	S^\kappa( r )=\langle (u_r^*)^\kappa \rangle.
\end{equation}
Within the inertial range, it is commonly proposed that for homogeneous isotropic turbulence, structure functions scale according to a power law 
\citep{Kolmogorov1941b} 
\begin{equation} \label{eq:sf2}
	S^\kappa( r )  \propto r^{\xi_\kappa}. 
\end{equation}
\citet{Kolmogorov1962} proposed a refined approach for the structure function exponent (K62),
\begin{equation} \label{eq:se}
	\xi_\kappa=\frac{\kappa}{3} - \frac{\mu}{18}\kappa(\kappa-3),
\end{equation}
with the experimentally measured intermittency factor $\mu\approx0.26$, e.g. \citet{Arneodo1996}.
While Kolmogorov proposed the scaling exponents $\xi_\kappa$ to be constant in $r$, a local scaling exponent $\xi_\kappa( r )$ can be calculated from experimental data by the log-log-derivative of structure functions. Here, the central difference quotient is used, 
\begin{equation}
\label{eq:xi}
\xi_\kappa( r ) = \lim_{\delta r \rightarrow 0}\frac{\ln(\langle (u_{r +\delta r}^*)^\kappa \rangle)-\ln(\langle (u_{r -\delta r}^*)^\kappa \rangle)}
{\ln(r+\delta r)-\ln(r-\delta r)}.
\end{equation}
Experimental results (as shown below) show that the range of constant scaling is rather small or even infinitesimal in case of finite Taylor based Reynolds numbers ($\Rey_\lambda\leq1000$).
This motivates to define a specific scale $\Theta$ at $\xi_2( \Theta) = 0.696$, where the scaling proposed by Kolmogorov is located. Correspondingly, a standard deviation $\sigma_{\Theta}=\sqrt{\langle (u_{r=\Theta}^*) ^2 \rangle}$ is defined. Here, we call $\Theta$ \textit{Kolmogorov length scale} (of second order). 

For the comparison of the statistics of $u_r^*$, it is necessary that velocity increments are normalised in a proper way.
On the one side, we follow the way proposed by \citet{Renner2001}, where a norm based on second order structure function is used,
and on the other side, we are inspired by the work of \citet{Warhaft1996}, where features of the inertial range are investigated.
To account for both, we select $\sigma_{\Theta}$ for the purpose of normalisation,
\begin{equation}
\label{eq:norm}
  u_r:= \frac{ u_r^* }{\sigma_{\Theta}}.
\end{equation}
The benefit of this normalisation becomes more clear below and it should be noted that
this normalisation can be applied without loss of generality in scaling features.
This fact can easily be shown, for instance with eqs. (\ref{eq:sf}) and (\ref{eq:sf2}), where a multiplication with an arbitrary factor does not change the scaling exponent.

\subsection{Joint multi-scale analysis}
\label{Multiscale}
Structure functions can be expressed in terms of unconditional probability density functions  $p( u_r )$,
\begin{equation}
\label{eq:str_pdf}
S^\kappa( r ) = \int^{+\infty}_{-\infty} u_r ^\kappa p( u_r )\mathrm{d} u_r .
\end{equation} 
Thus, the information given by all structure functions is equivalent to the information of $p( u_r )$ at $r$, cf. \citet{Frisch2001}.
Approaches using eq. (\ref{eq:str_pdf}) consider the statistical features for each scale separately. 
As the turbulent cascade is a chain of events on different scales, a more appropriate 
approach sets the statistics of scales in relation to each other. 
A $N$-scale joint PDF $p( u_r ; u_{r +\Delta}; u_{r +2\Delta}; ... ; u_{r +N\Delta})$
captures the entire multi-scale characteristics of the turbulent cascade including its relations over different scales,  cf. \citet{Friedrich1997}.

Earlier studies show that both the turbulent cascade process and its increment series have the Markov property, e.g. \citet{Friedrich1997, Renner2001, Renner2002, Tutkun2004, Lueck2006}.
Thus, general joint PDFs can be expressed by conditional PDFs,
\begin{eqnarray}
	p( u_r ; u_{r +\Delta}; ... ; u_{r +N\Delta})
	=p( u_r |  u_{r +\Delta}) p( u_{r +\Delta}|  u_{r +2\Delta})\times...\nonumber \\ 
	\times p( u_{r +(N-1)\Delta}|  u_{r +N\Delta})p( u_{r +N\Delta}).
\label{eq:j_pdf_fak}
\end{eqnarray}
Motivated by the cascade picture, our notation here implies that an increment $u_{r}$ is conditioned on an increment on a larger scale $u_{r +\Delta}$. 

The Markov property is commonly only valid for finite step sizes, in particular when the 
deterministic microscopic or smallest scale dynamics is coarse-grained to a stochastic dynamics, cf. \citet{Einstein1905}. 
In the case of the turbulent cascade it is valid for $\Delta\geq l_{EM}$, where $l_{EM}$ is the Einstein-Markov coherence length and it is $l_{EM}\propto\lambda$,  cf. \citet{Lueck2006, Stresing2012}. Unless stated otherwise, we take the step width in the cascade as $\Delta = l_{EM} = 2\lambda$. Details of the determination of the Einstein-Markov coherence length are given in appendix A2.
Thus, the stochastic process at scale $r$ is determined by the conditional PDF $p( u_r | u_{r +l_{EM}})$ and the entire cascade with its interscale relations can be described by $p( u_r | u_{r +l_{EM}})$ as function of $r$.

Starting from an initial distribution, e.g. at the integral length scale, $p(u_L)$, or any other starting scale $r'$, the evolution of $p( u_r | u_{L})$ can be described by the Fokker-Planck equation (also known as Kolmogorov equation), cf. \citet{Friedrich1997,Friedrich1997b,Renner2001},
\begin{eqnarray}
	-\frac{\partial}{\partial r} p( u_r | u_{r'}) =
	&-&\ \frac{\partial}{\partial u_r} \left[ D^{(1)}( u_r )p( u_r | u_{r'})\right]\nonumber\\
	&+&\frac{\partial^2}{\partial u_r^2} \left[ D^{(2)}( u_r )p( u_r | u_{r'})\right].
\label{eq:FP}
\end{eqnarray}
Thereby, any $p( u_r | u_{r'})$ with $r<r'$ can be obtained.
The functions $D^{(1)}( u_r )$ and $D^{(2)}( u_r )$  govern the Fokker-Planck equation and are known as drift and diffusion functions or, more generally, first and second Kramers-Moyal coefficients.
The Fokker-Planck equation is a truncated Kramers-Moyal expansion, cf. \citet{Risken1984}, where only the first two terms are considered.
Due to the fact that the first two terms of the Kramers-Moyal expansion strongly dominate the expansion, cf. \citet{Renner2001,Tutkun2004}, the description of the evolution of the conditional PDFs in scale by the Fokker-Planck equation is justified. This dominance can be related to Pawula's theorem, which further corroborates the use of the Fokker-Planck equation, cf. \citet{Risken1984}. 


The Fokker-Planck approach can be related to structure functions by partial integration of eq. (\ref{eq:FP}), e.g. \citet{Friedrich1997c,Renner2001},
\begin{eqnarray} 
\label{eq:momentsFP}
	-r \frac{\partial}{\partial r} S^{\kappa}(r) =  r\kappa \langle u_r ^{(\kappa -1)} D^{(1)}( u_r) \rangle  + r\kappa (\kappa-1) \langle u_r^{(\kappa -2)}  D^{(2)}	(u_r) \rangle.
\end{eqnarray}
It is easily seen that this set of equations for $S^{\kappa}$ becomes unclosed if the drift and diffusion functions have higher order contributions in $u_r$ ($D^{(1)}$ higher than linear, $D^{(2)}$ higher than quadratic).

Besides the Fokker-Planck equation, the two coefficients $D^{(1)}( u_r )$ and $D^{(2)}( u_r )$ define the equivalent Langevin equation (cf. \citet{Risken1984}). The Langevin equation describes the evolution of increments $u_r$ in $r$ as trajectories $u(\cdot)$. The trajectory level illustrates our novel way of investigating turbulence by considering cascade trajectories of increments $u_L \rightarrow u_r$.


\subsection{Estimation of Kramers-Moyal coefficients}

Next, technical aspects are discussed on how to estimate the best Fokker-Planck equation. 
Kramers-Moyal coefficients can be estimated from given data by conditional moments,
\begin{eqnarray} \label{eq:M_k}
	M^{(\kappa)}( u_{r '},\delta) &=& 
	\int^{+\infty}_{-\infty}{( \tilde{u_r}-  u_{r '})^\kappa} \  p(  \tilde{u_r} |  u_{r '}) \ d \tilde{u_r},\\
	&=& \langle \left[  \tilde{u_r}  -  u_{r '}\right]^\kappa \rangle |_{ u_{r'}}\\
\label{eq:D_k}
	D^{(\kappa)}( u_{r '}) &=& \lim_{\delta \to 0} \ \frac{1}{\kappa!\delta} \ M^{(\kappa)}( u_{r '},\delta),
\end{eqnarray}
where $\kappa$ is the order of the coefficient, cf. \citet{Risken1984} and $\delta = r'-r$.
A common method to estimate $\lim_{\delta \to 0}$ in eq. (\ref{eq:D_k}) is an extrapolation. 
The evolution of moments $M^{(\kappa)}( u_r ,\delta)$ between $\delta=l_{EM}$ and $\delta=2l_{EM}$ is linearly extrapolated to $\delta =0$ and
the intersection with the ordinate axis estimates the limit.
%
%
%
Kramers-Moyal coefficients deduced by this approach lead to a rather imprecise description of the turbulent cascade.
Further improvements are possible with the use of higher order corrections, see \citet{Risken1984} and discussions of this issue in \citet{Gottschall2008}; \citet{RennerDiss}, and \citet{Honisch2011}.  
Nevertheless, the limit approximation leads to uncertainties and still leaves room for improvements for their absolute values, whereas the functional form of Kramers-Moyal coefficients are commonly well estimated. 

To determine the Kramers-Moyal coefficients, an optimisation procedure is used to minimise 
possible uncertainties, which was proposed in \citet{Nawroth,Kleinhans}. 
Four steps are necessary for this optimisation: 
i) parametrisation of Kramers-Moyal coefficients, 
ii) calculation of $p( u_r ; u_{r +l_{EM}})$,
iii) estimating the difference between calculated distribution $p( u_r ; u_{r +l_{EM}})$ and measured joint PDF and 
iv) an iterative optimisation procedure that finds best fitting Kramers-Moyal coefficients, ensuring a minimal distance between the distributions.  

We discuss the four steps in detail: \\
{\it Step i)} A common parametrisation ($d_{ij}$) of the Kramers-Moyal coefficients is a linear function for $D^{(1)}$ and a parabolic function for $D^{(2)}$
\begin{eqnarray} \label{eq:D1}
	D^{(1)}( u_r ) &=& d_{11}(r)  \ u_r ,\\
\label{eq:D2}
	D^{(2)}( u_r ) &=& d_{22}(r)   \ u_r^2 + d_{20}(r).
\end{eqnarray}
As we will see later, these functions conform well with the shape of $D^{(1)}$ and $D^{(2)}$ (also denoted as $D^{(1,2)}$) constructed from the measurement data. 

{\it Step ii)} A calculation of $p( u_r | u_{r +l_{EM}})$ is done by the use of the short time propagator 
\citep{Risken1984,Renner2001}
\begin{eqnarray}
p_{STP}( u_r | u_{r +l_{EM}})&\approx& \frac{1}{\sqrt{4\pi  \;D^{(2)}( u_{r +l_{EM}} )\;  l_{EM}  }}\nonumber\\
&\times& \exp\left(-\frac{\lbrack u_{r}-u_{r+l_{EM}} -D^{(1)} ( u_{r +l_{EM}} ) l_{EM} \rbrack^2}{4 D^{(2)}(u_{r +l_{EM}})l_{EM} }\right).
\label{eq:short_time_prop}
\end{eqnarray}
This particular functional form of eq. (\ref{eq:short_time_prop}) is reached by implementing the scale step in units of the Einstein-Markov coherence length $l_{EM}$, which has the advantage that the short time propagator only depends on $D^{(1,2)}$.\footnote{The proper size of a scale step in eq. (\ref{eq:short_time_prop}) is investigated by the use of the Chapman-Kolmogorov equation, cf. \citet{Risken1984}. 
The specified step size $l_{EM}$ leads to consistent results, smaller steps than $l_{EM}$ do not significantly improve the results}   

Since the experimentally found conditional PDFs $ p_{exp}( u_r | u_{r +l_{EM}})$ become noisy at large values of $ u_{r}$ and $ u_{r +l_{EM}}$, it is better to use the joint PDFs for the optimisation procedure. Therefore, conditional PDFs  
$p_{STP}(u_r|u_{r+l_{EM}})$
from eq. (\ref{eq:short_time_prop}) are converted by Bayes' theorem to joint PDFs $p_{STP}( u_r ; u_{r +l_{EM}})$.

{\it Step iii)} Next we set the numerical calculated joint PDF $p_{STP}$ in comparison with the measured reference $p_{exp}$ by the use of a weighted mean square error function in logarithmic space, cf.  \citet{Feller1968}, (analogous to Kullback-Leibler entropy),
\begin{equation} 
\label{eq:KLE}
\epsilon(p_{STP},p_{exp}) =
\frac{\int \limits_{-\infty}^{+\infty}\int \limits_{-\infty}^{+\infty}{(p_{STP}+p_{exp})(\ln(p_{STP})-\ln(p_{exp}))^2 \ du_r^2}} 
       {\int \limits_{-\infty}^{+\infty}\int \limits_{-\infty}^{+\infty}{(p_{STP}+p_{exp})(\ln^2(p_{STP})+\ln^2(p_{exp})) \ du_r^2}},
\end{equation}
with $\ln$ as natural logarithm.
$\epsilon$ is taken as a logarithmic measure for the distance between the distributions. 

{\it Step iv)} The logarithmic distance $\epsilon$ is taken as a cost function for an optimisation procedure. The optimisation procedure systematically changes $D^{(1,2)}$ until $\epsilon$ is minimised.
The optimisation procedure is a 2nd order method based on a Hessian matrix. 
We used for this optimisation the function \textit{fmincon} implemented in MATLAB\textsuperscript{\textregistered} R2012a.
The constraints are set 
in a physically and mathematically meaningful way, 
$-\infty \leq d_{11} \leq0$, 
$0 \leq d_{22} \leq \infty$ and 
$0 < \delta_d \leq d_{20} \leq \infty$, here $\delta_d=10^{-3}$.



\subsection{Integral fluctuation theorem}
\label{ChapIFT}

Next we present the IFT and its thermodynamic origin as a new physical law for the turbulent cascade, and furthermore, as an independent measure to show the validity of the Markov description of the cascade. 
In addition, the functional form of the estimated functions $D^{(1,2)}$ is disclosed and assessed by the IFT. 

The Kramers-Moyal coefficients describe the turbulent system and its increment trajectories from an initial state given by $p(u_L,L)$ to a \textit{final} state $p(u_\lambda,\lambda)$. The corresponding stochastic process that defines the evolution in scale of an initial increment $u_L$ drawn from $p(u_L,L)$ down to the final increment $u_\lambda$ can be interpreted as an analogue of a non-equilibrium thermodynamic process.
Most importantly, drawing the analogy to non-equilibrium thermodynamics entails the notion of irreversible entropy production for a turbulent cascade, with useful implications.
Now, we make a short digression into non-equilibrium thermodynamics to explain the background in more detail.

In non-equilibrium thermodynamics, external forces generate a deterministic drift $D^{(1)}$ and the thermal energy $k_\mathrm{B}T$ defines the diffusion function $D^{(2)}$. 
The drift and the diffusion function are the two ingredients to set up the Fokker-Planck or, equivalently, the Langevin equation of non-equilibrium thermodynamic processes,  cf. \citet{Lemons1997}, \citet{Sekimoto1998}.
For a turbulent cascade, the linear drift $D^{(1)}(u_r)$ in eq. (\ref{eq:D1}) accounts for the tendency of turbulent structures to split into smaller structures, invoked by the non-linearity of the Navier-Stokes equation.

Drawing the analogy further, just as the thermal movement of the particles of an embedding medium is a source of fluctuations for a system of interest, we interpret $u_r^2$ as an energetic term being a source of velocity fluctuations throughout the turbulent cascade $u_r$ itself. 
The functional form of the fluctuation describing coefficient $D^{(2)}(u_r)$ in eq. (\ref{eq:D2}) hence suggests that the $r$-dependency of the $d_{22}(r)$ term constitutes the influence of turbulent energy on small scale intermittency.
This interpretation receives support by the role of the energy transfer rate on the cascade, studied in \citet{Gagne94}, \citet{Naert97}, \citet{Renner2002b}:  Fixing the transferred energy $\varepsilon$ to constant values or, equivalently, looking only at cascade trajectories with constant $\varepsilon$, no intermittency is observed and $p(u_r|\varepsilon)$ becomes Gaussian, i.e. $d_{22}(r)$ vanishes.
In addition to turbulent energy as source of fluctuations, the coefficient $d_{20}(r)$ procures background fluctuations independent from $u_r$. 
We come back to this picture of the turbulent cascade when we discuss our results in section \ref{Par_IFT} and \ref{TuCasKM}.
For more details on the Markov picture of the cascade process we refer the reader to \citet{Nickelsen2017}.

From the energy balance of the thermodynamic Langevin equation follows an expression for the heat exchange with the medium along a single trajectory of the system, cf. \citet{Sekimoto1998}. 
The heat exchange is associated with an entropy production in the medium, $S_\mathrm{m}$, and the change in Gibbs entropy, $\Delta s$, which is associated with the change of system entropy. Thus the total entropy production can be defined as $\Delta S=S_\mathrm{m}+\Delta s$, cf. \citet{Seifert2005}.

As being the entropy production of an isolated system, $\Delta S$ must satisfy the second law of thermodynamics, $\Delta S\geq0$.
However, single fluctuations with a tendency against the average behaviour give rise to entropy consumption, $\Delta S<0$. 
Therefore, it is important to remember that the second law addresses the macroscopic entropy production, which means that on average the total entropy production is non-negative, $\left\langle\Delta S\right\rangle \geq 0$. 
Fluctuating entropy productions are central in the arising field of stochastic thermodynamics, of which one result are the so-called \textit{fluctuation theorems}. 
These theorems express the balance between entropy production and entropy consumption by tightening the second law to an equality, cf. \citet{Seifert2012}. 
Here, we are concerned with the \textit{integral} fluctuation theorem (IFT) for the total entropy production, \citet{Seifert2005}, 
\begin{equation}  \label{eq:FT_pre}
	\left\langle e^{-\Delta S}\right\rangle = 1 .
\end{equation} 
In order to amount to the average of $1$, it is evident that the IFT assigns an exponential weight to the rare entropy consuming fluctuations that contribute with values much larger than $1$ to the average, in opposition to the typical entropy producing fluctuations that contribute with values between $0$ and $1$. 
It is this delicate balance between production and consumption of entropy that makes the IFT so useful for the analysis of non-equilibrium systems.
Typically, the IFT finds applications in small systems on nano-scale, whereas its applicability is not evident for classical turbulent flows with scales in the range of mm and velocity fluctuations of order m/s.


At this point, we turn to turbulent cascades again and refer the interested reader to the comprehensive review by \citet{Seifert2012} for more details on stochastic thermodynamics.
On the basis of single realisations of the cascade process, or, to be explicit, trajectories $u(\cdot)= \{ u_r; r= L\dots\lambda \}$, the formal entropy production $S_\mathrm{m}$ is given by, cf. \citet{Nickelsen2013},
\footnote{Here, we deviate from the previous suppression of the argument $r$ to stress the explicit $r$-dependency of $D^{(1,2)}(u_r,r)$ and $p(u_r,r)$}
\begin{equation} 
\label{eq:S_m}
	S_{\mathrm{m}}[u(\cdot)] = \int^{\lambda}_{L}{\partial_r u_r \,\,\frac{D^{(1)}(u_r,r)-\partial_u D^{(2)}\big(u_r,r\big)}{D^{(2)}\big(u_r,r\big)}\,\mathrm{d}r} .
\end{equation}
The difference in Shannon entropy reads
\begin{equation} 
\label{eq:Ds}
	\Delta s(u_L,u_\lambda) = - \ln \frac{p(u_\lambda,\lambda)}{p(u_L,L)} .
\end{equation}
If the drift and diffusion functions $D^{(1,2)}$ that define the cascade process as well as $p(u_L,L)$ and $p(u_\lambda,\lambda)$ are known, the total entropy production for a single trajectory $u(\cdot)$ can formally be determined as the sum of the two contributions mentioned above,
\begin{equation} \label{eq:S_tot}
	\Delta S[u(\cdot)] = S_{\mathrm{m}}[u(\cdot)] + \Delta s(u_L,u_\lambda) .
\end{equation}

Suppose $\{u(\cdot)\}$ are exact realisations of a cascade process defined by $D^{(1,2)}$ and $p(u_L,L)$, then $\left\langle e^{-\Delta S}\right\rangle$ should converge to exactly $1$ for an unlimited number of $u(\cdot)$ taken into account. In \citet{Nickelsen2013} it has been demonstrated for a single set of data from a free jet experiment that the exponential average in the IFT is indeed converging very fast to $1$. 
The crucial requirement for the convergence
 is the occurrence of rare trajectories with $\Delta S<0$.
\citet{Nickelsen2013} found that the trajectories with $\Delta S<0$ exhibit an \textit{increase} of fluctuations when evolving from $L$ to $\lambda$, against the typical tendency of the cascade.
It is, hence, the phenomenon of small scale intermittency that is responsible for the applicability of the IFT to turbulent cascades and its convergence. 
However, a universal form of $D^{(1,2)}$ and therewith a universal form of the IFT is still missing.
So is, e.g., the exponential average of the IFT deriving from the K62 model eq. (\ref{eq:se}) diverging rapidly for the data of the free jet experiment considered in \citet{Nickelsen2013}, confirming the well known fact that the K62 model does not reflect properly small scale intermittency of real turbulent flows. For the discussion of other models of turbulence see
 \citet{Nickelsen2017}.

Here, we extend the investigations done in \citet{Nickelsen2013} substantially and for the first time employ the utility of the IFT to explore the role large scale structures play for small scale turbulence. 
In doing so, we address three points. 
First of all, we show that the IFT holds for all different types of turbulent flows considered here and thus collect strong evidence that the IFT is a universal feature of turbulence. 
Secondly, the high sensitivity of the IFT regarding realistic modelling of small scale intermittency enables us to verify our estimation of $D^{(1,2)}$ and to quantify the importance and significance of different functional contributions of $D^{(1,2)}$.
And finally, the interpretation as a non-equilibrium thermodynamic process illuminates the role of these functional contributions in terms of small scale intermittency and universality.

To test the quality of the estimated functions $D^{(1,2)}$, we extract $n$ trajectories $\{u(\cdot)\}$ from each of the measured flow fields $v(x)$ by applying the definition eq. (\ref{eq:u_r}) for different values of $x$. 
To ensure the independence of the trajectories, we take equidistant values for $x$ with step size $L$, owing to the fact that $L$ specifies the size of the largest coherent structures in $v(x)$. 
For each set of trajectories, $\{u(\cdot)\}$, we calculate a set of total entropy productions $\{\Delta S_j\}$ from eq. (\ref{eq:S_tot}) in order to employ the IFT by
\begin{equation}  \label{eq:FT}
	1 = \left\langle e^{-\Delta S[u(\cdot)]}\right\rangle \simeq \frac{1}{n}\sum\limits_{j=1}^{n} e^{-\Delta S_j} =: I(n).
\end{equation}

The overall implementation of the IFT is then as follows. For each of the datasets considered here, $D^{(1,2)}$ are estimated and used to define the form of $S_\mathrm{m}$ in eq. (\ref{eq:S_m}), $p(u_L,L)$ and $p(u_\lambda,\lambda)$ are estimated to define $\Delta s$ in eq. (\ref{eq:Ds}), and the very data that was used for this estimation is plugged into the IFT.
If the IFT is confirmed, the Markov process defined by $D^{(1,2)}$ and the initial condition $p(u_L,L)$ correctly captures the cascade process and in particular the small scale features of the turbulent flow. 
In addition to this quality check, the IFT serves as a clear cut criterion to pin down and discuss the minimal functional form of $D^{(1,2)}$ that still captures the cascade process.

\section{Results}
In this section, we show the results obtained from all 61 datasets.
The analysis of structure functions sets our data into the common picture of turbulence and checks if the data obeys a scaling law.  
Subsequently, we demonstrate and discuss the validity of the IFT and our parametrisation of the Kramers-Moyal coefficients, followed by a discussion of their functional form. 
Thereafter, results of multi-scale statistics are presented and discussed in the context of universal and non-universal flow dependencies of turbulence.

\label{sec:Results}
\subsection{2-point scaling features}
\begin{figure}
  \centering    
\includegraphics[width=0.6\linewidth]{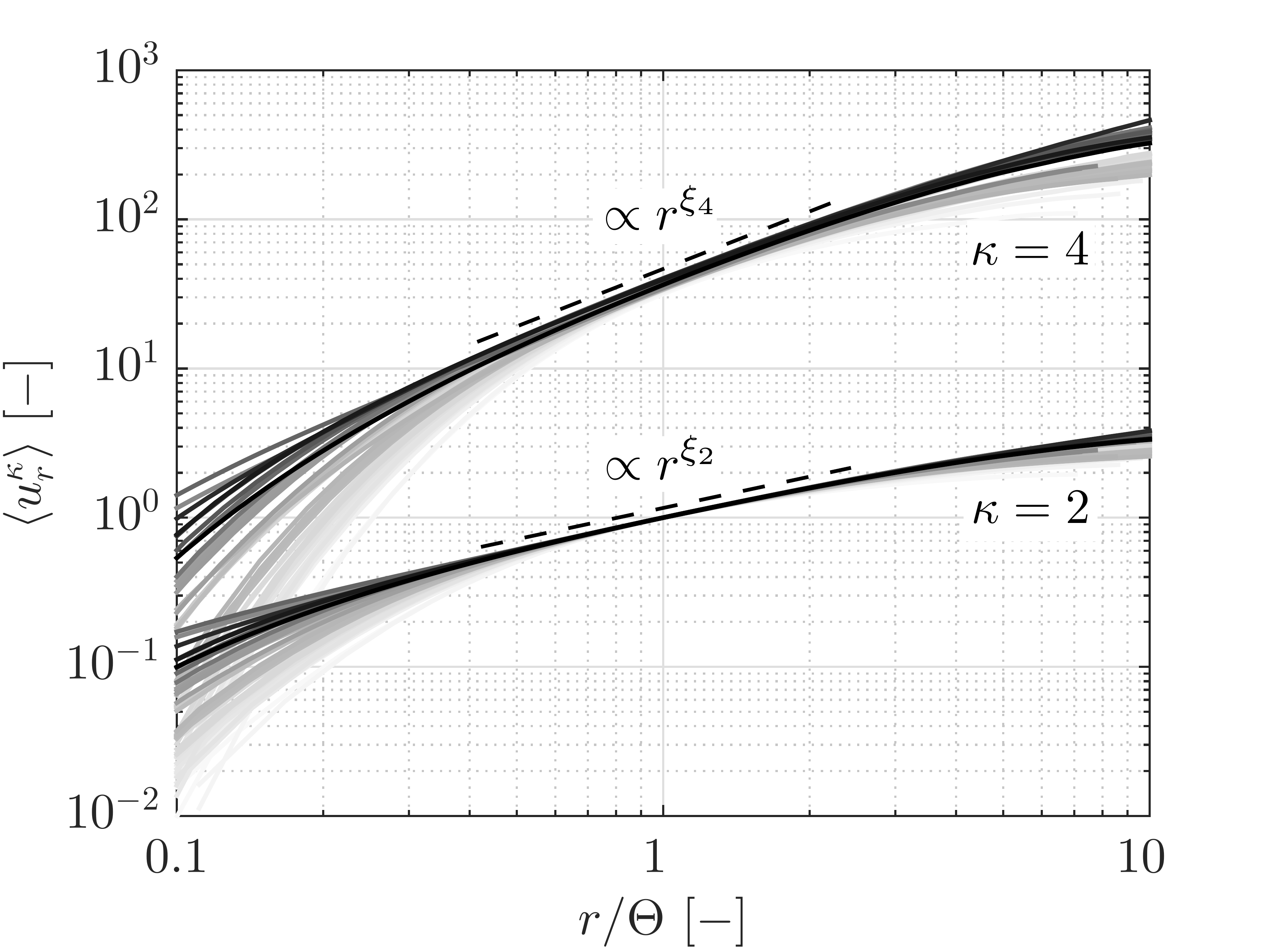}
  \caption{Second ($\kappa = 2$, lower band) and fourth ($\kappa = 4$, higher band) order structure functions versus normalised scale. 
 The whole band of the fourth order structure functions is shifted by a factor of 10 for reasons of presentation. 
 The greyscale used for the curves indicates corresponding Taylor Reynolds numbers, bright grey to black corresponds to increase of $\Rey_\lambda$.
}
\label{fig:sf_2_4}
\end{figure}

We begin with the standard analysis of our data. Figure \ref{fig:sf_2_4} presents second ($\kappa=2$) and fourth order ($\kappa=4$) structure functions according to eq. (\ref{eq:sf}) obtained from all datasets. The lower band corresponds to second and the upper band corresponds to fourth order structure functions. The greyscale of the curves indicates the Taylor Reynolds numbers. It is clearly seen how the \textit{scaling range} increases with $\Rey_\lambda$. Structure functions are normalised in their scale and in their values by $\Theta$ and $\sigma_{\Theta}$, defined in section \ref{sec:Classical}.
The band of the fourth order structure functions is shifted by a factor of $10$ along the ordinate axis for reasons of presentation. The norms $\Theta$ and $\sigma_{\Theta}$ are shown in appendix A3 as functions of $\Rey_\lambda$. These norms are dimensional and show features which depend also on the flow type. 

\begin{figure}
  \centering    
  a)\includegraphics[width=0.47\textwidth]{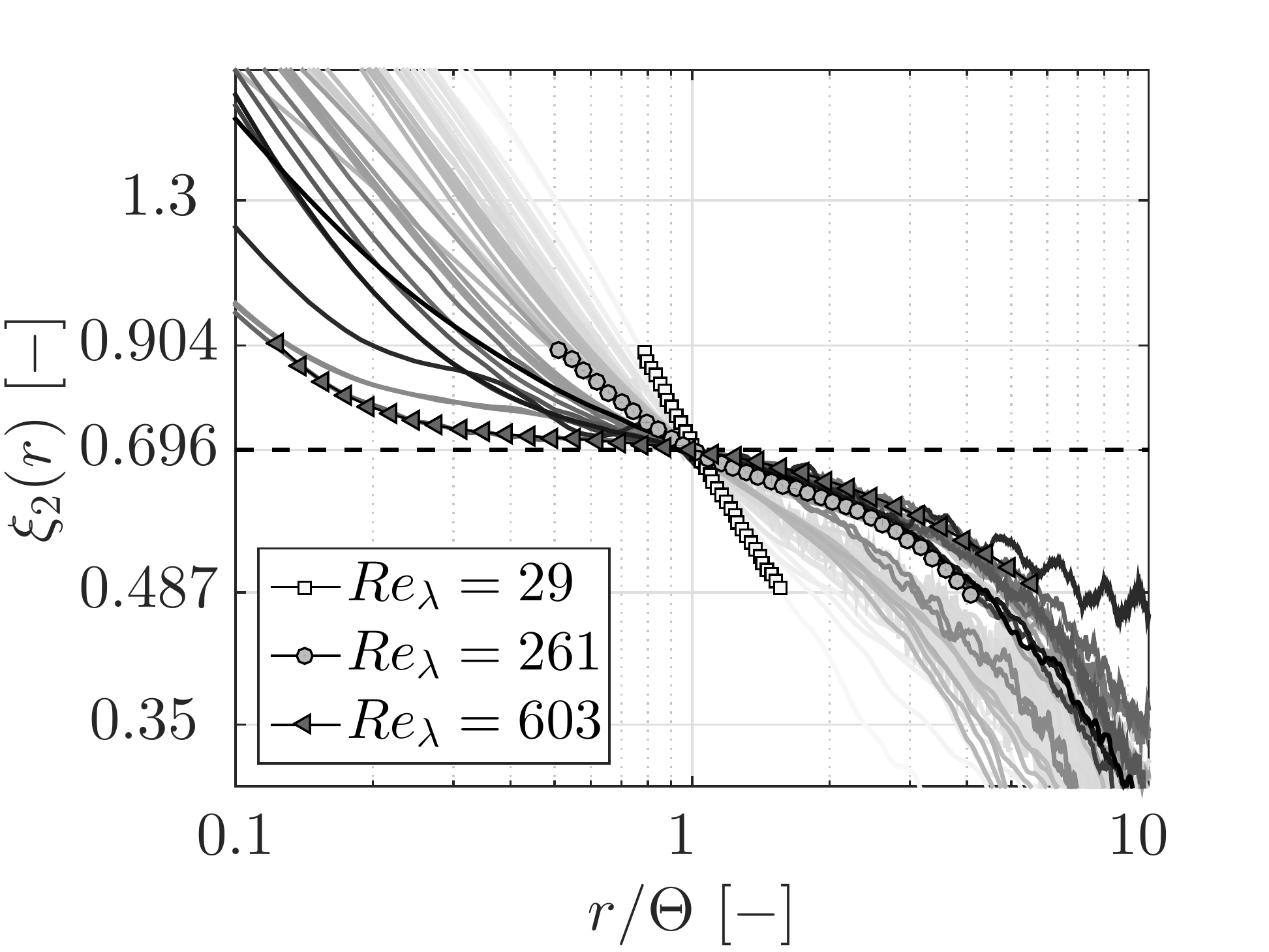}
b)\includegraphics[width=0.47\textwidth]{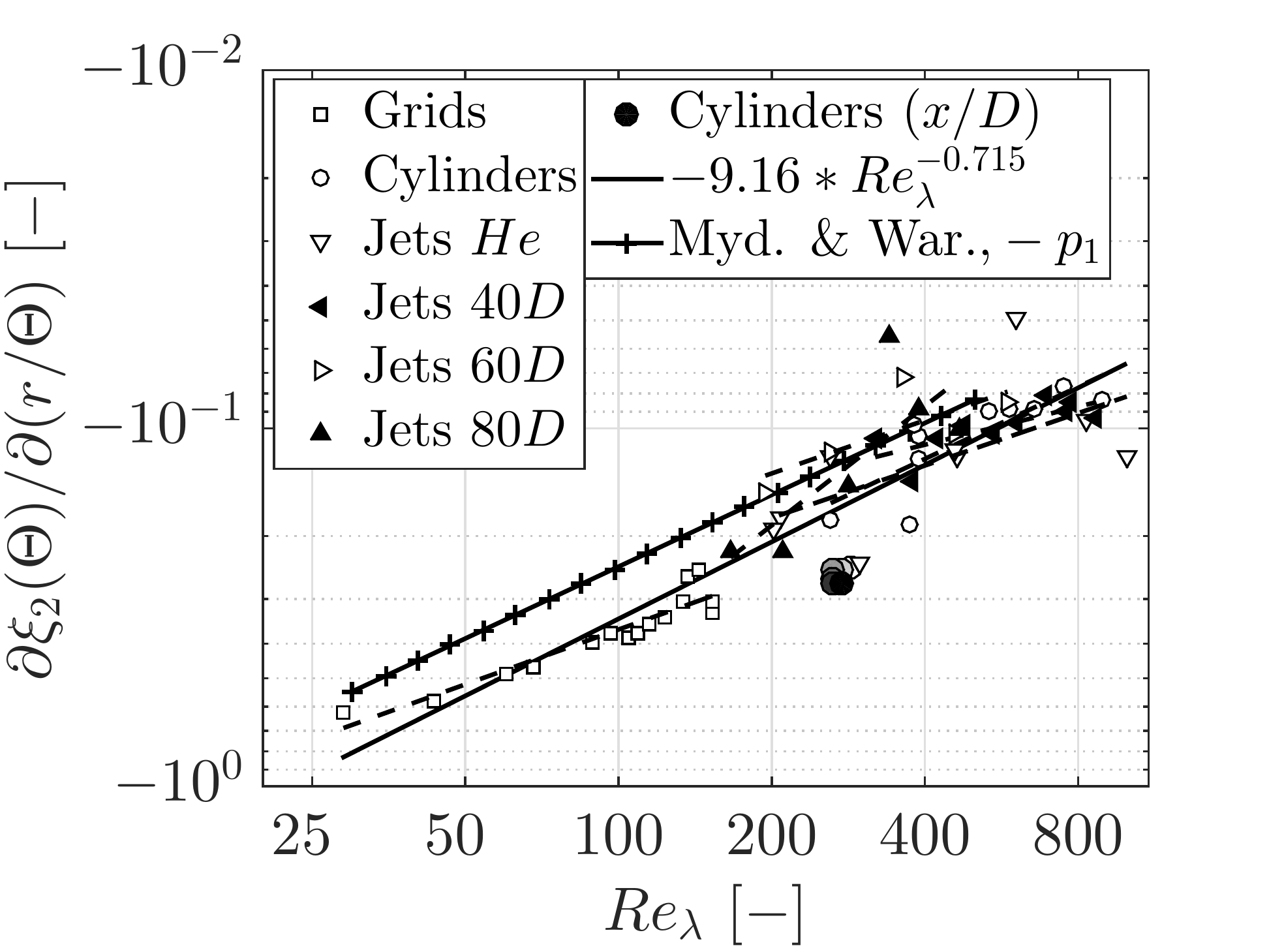}
  \caption{a) Local second order scaling exponent $\xi_2( r )$ according to eq. (\ref{eq:xi}) in a double-logarithmic plot. Presentation alike figure \ref{fig:sf_2_4}. 
b) Slope of local second order scaling exponent $\xi_2( r )$ at $r = \Theta $ in a double-logarithmic presentation together with data from \citet{Warhaft1996}}.
\label{fig:xi2}
\end{figure}

For a more precise analysis of the scaling behaviour of the structure functions, we present in Figure \ref{fig:xi2} a) the local scaling exponent $\xi_2(r)$ according to eq. (\ref{eq:xi}). Due to our normalisation, all datasets pass through $\xi_2=0.696$ at $\frac{r}{\Theta}=1$. 
Three curves 
are marked by symbols to indicate that $\xi_2(r)$ changes from a linear behaviour for low $\Rey_\lambda$ to a power 5 behaviour for high $\Rey_\lambda$ (with respect to the log-log presentation). This shows that the local scaling exponent flattens with increasing $\Rey_\lambda$ around $\xi_2=0.696$. 
Thus, instead of a stretched constant scaling range, only a tendency to usual scaling
\footnote{With usual scaling we refer to the case that no functional deviation from a scaling law is present, i.e. $\xi_{\kappa}(r)$ is rigorously constant in $r$, e.g. eq. (\ref{eq:se})}
behaviour is seen, even for the highest $\Rey_\lambda$ datasets.

For the third and fourth order structure function, similar results are shown in \citet[pp. 121-122, fig. 4.3 and 4.4]{SinhuberDiss} and \citet{Sinhuber2017} and confirm the flattening of structure functions for even higher $\Rey_\lambda$ for grid generated turbulence, also \citet{Grauer2012} used a similar approach. An explanation of this non perfect scaling behaviour of structure functions is that finite size effects play a role, \citet{Dubrulle2000}, which also supports the idea of introducing the specific scale $\Theta$ defined in section \ref{sec:Classical}.

Figure \ref{fig:xi2} b) shows the slope of the local second order scaling exponent $\xi_2( r )$ at $r = \Theta $ as function of $\Rey_\lambda$. 
This slope is obtained by fitting a polynomial of 5th order as shown in figure \ref{fig:xi2} a). The solid line shows the trend of all datasets with the functional form of
\begin{equation}  \label{eq:fit_1}
 	\frac{\partial \xi_2(\Theta)}{\partial (r/\Theta)} \approx -9.16 \ \Rey_\lambda ^{-0.715}.
\end{equation} 
Based on this result, $\partial_{(r/\Theta)} \xi_2(\Theta)$ can be proposed as an alternative to $\Rey_\lambda$. 

Additional data from a similar study done by \citet{Warhaft1996} is added to figure \ref{fig:xi2} b).
This data represents a deviation from the common energy density spectra scaling $E(k) \propto k^{-5/3}$
\footnote{Note that second order structure functions and energy density spectra contain the same information and can be converted into each other by Fourier transformation}.
The deviation is expressed by $p_1$ in the form of  $E(k) \propto k^{-(5/3-p_1)}$.
Turbulent flows and their $E(k)$ are investigated up to $\Rey_\lambda\leq$ 500.
They found that $p_1$ has the form $p_1=5.25 \Rey_\lambda^{-2/3}$, which is very similar to the $\Rey_\lambda$-dependence we found in eq. (\ref{eq:fit_1}). 
The pre-factor differs due to the different approaches.



\subsection{Estimation procedure of Kramers-Moyal coefficients}
\begin{figure}
  \centering    
  a)\includegraphics[width=0.47\textwidth]{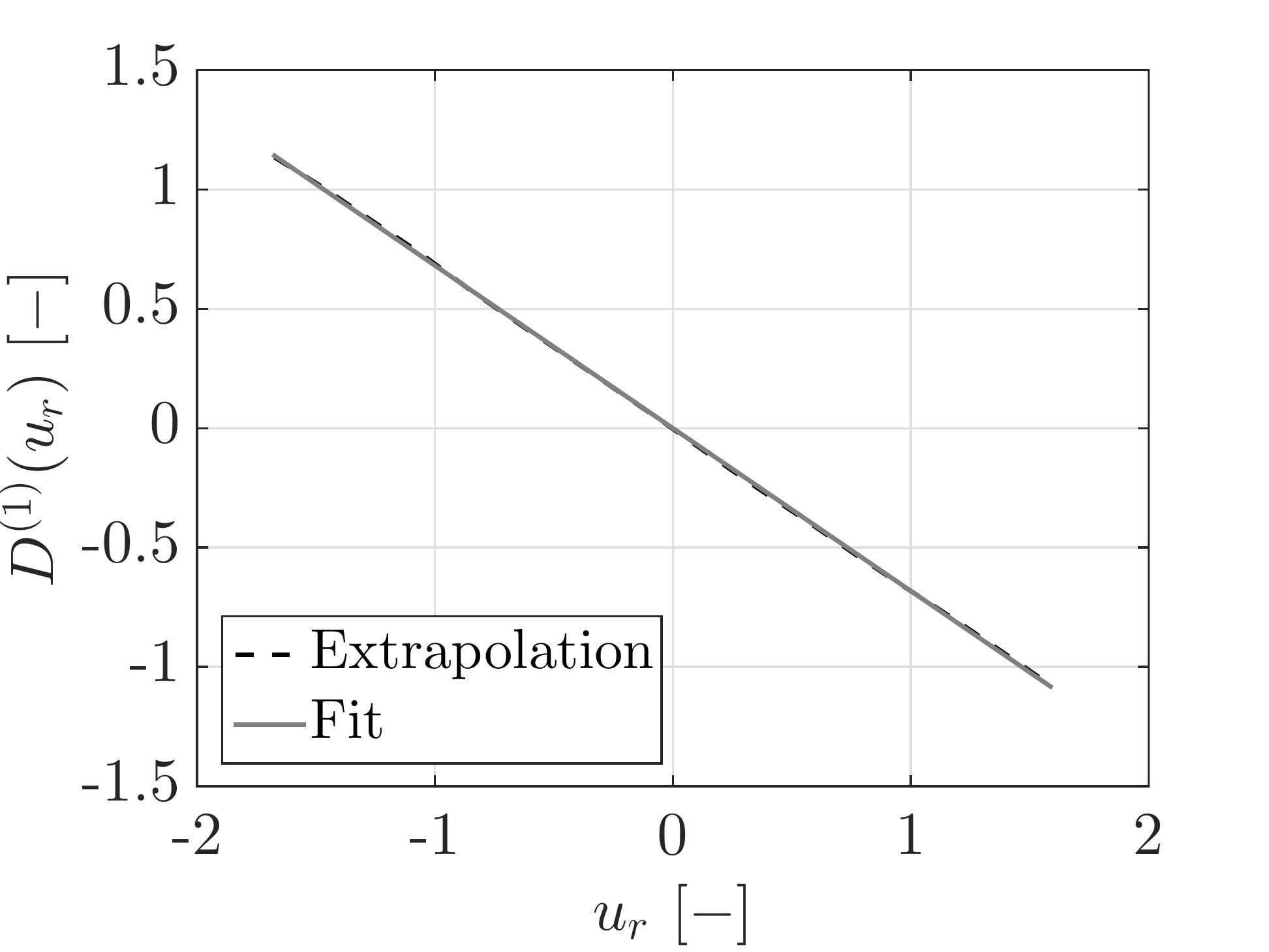}
  b)\includegraphics[width=0.47\textwidth]{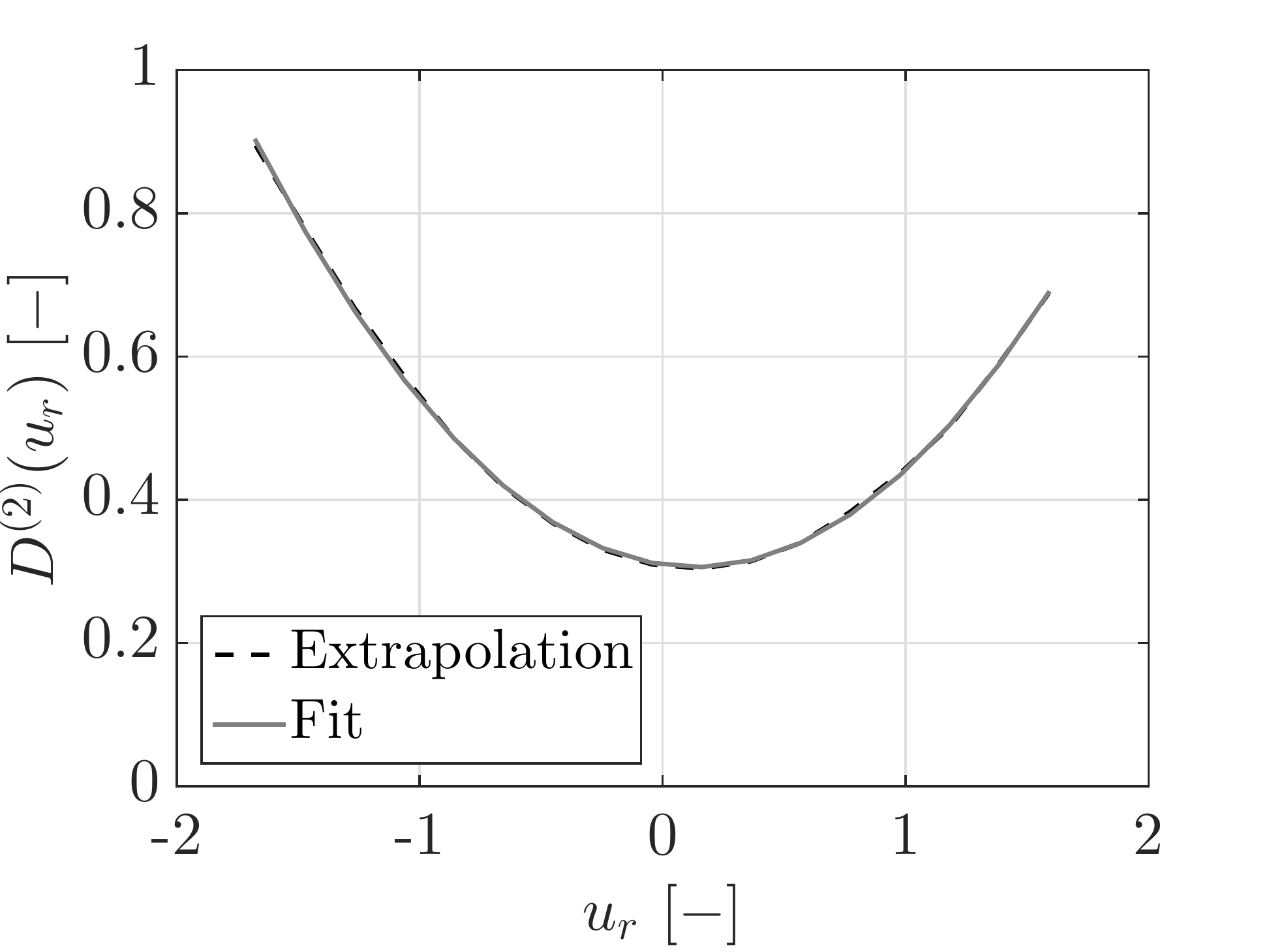}
  \caption{
  $D^{(1,2)}$ versus $u_r$, according to eq. (\ref{eq:D_k}) and higher order corrections, cf. \citet{Gottschall2008}. Linear and parabolic fitting functions illustrate their functional forms. The dataset \textit{jet 1} (table \ref{table:data}) is used at
  $r=1l_{EM}$.
  }
\label{fig:D12_fit}
\end{figure}

We now extend our analysis to joint multi-scale statistics.
The first estimation of Kramers-Moyal coefficients is based on conditional moments and their extrapolation. 
Figure \mbox{\ref{fig:D12_fit} a)} and b) exemplarily show $D^{(1)}$ and $D^{(2)}$ as functions of $ u_r $ according to eq. (\ref{eq:D_k}) and higher order corrections, cf. \citet{Gottschall2008}. 
A linear and a parabolic fitting function reflect the common functional forms of $D^{(1,2)}$, which are used for the subsequent optimisation. The used fit range is indicated, $-1.5\le u_r \le1.5$.
Here one might argue that higher order contributions can be seen in $D^{(1,2)}$ (compared to the approach in eq. (\ref{eq:D1}) and (\ref{eq:D2})). We come back to this point when we discuss the results regarding the IFT.

\begin{figure}
  \centering    
  a)\includegraphics[width=0.47\textwidth]{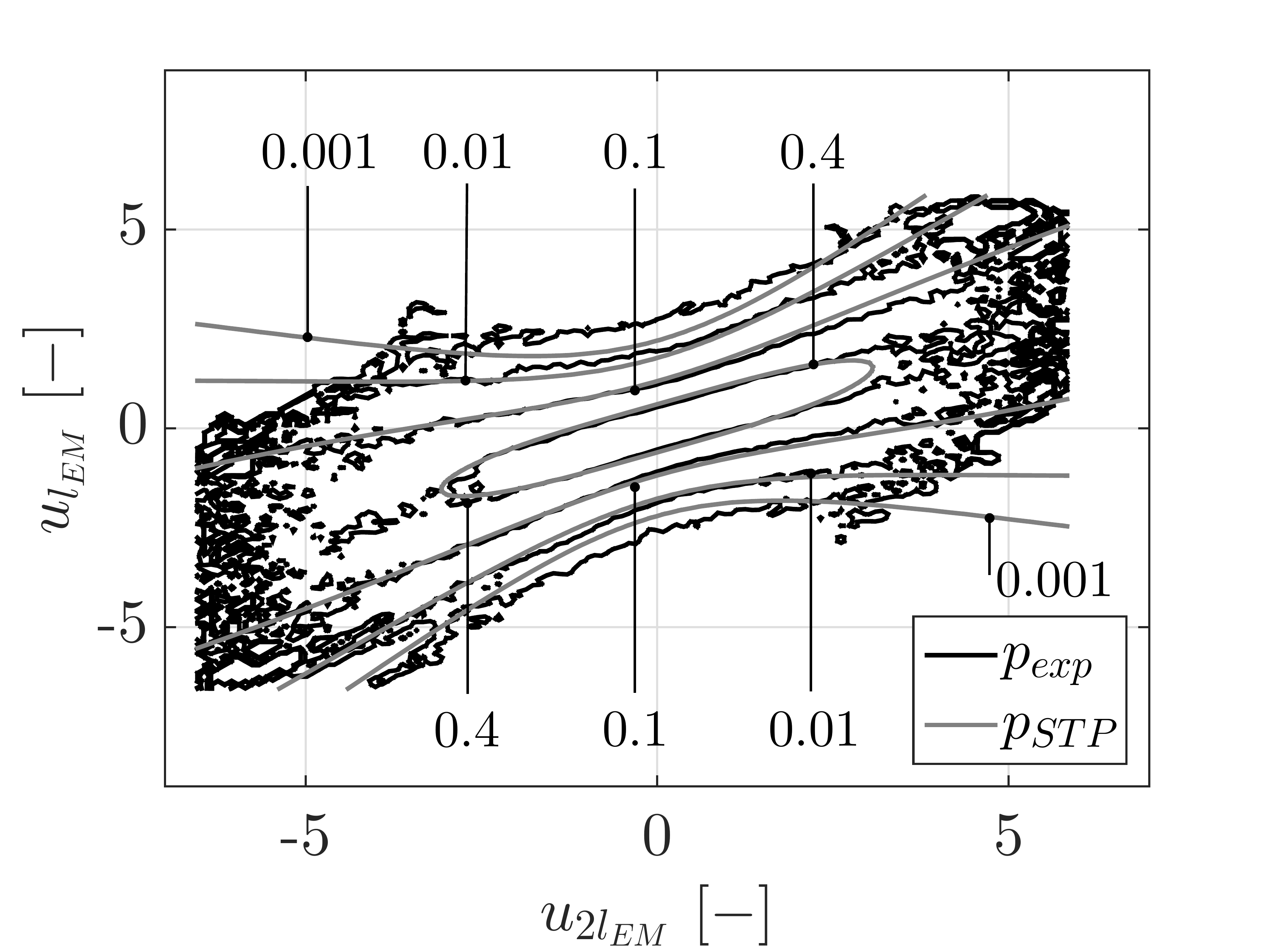}
  b)\includegraphics[width=0.47\textwidth]{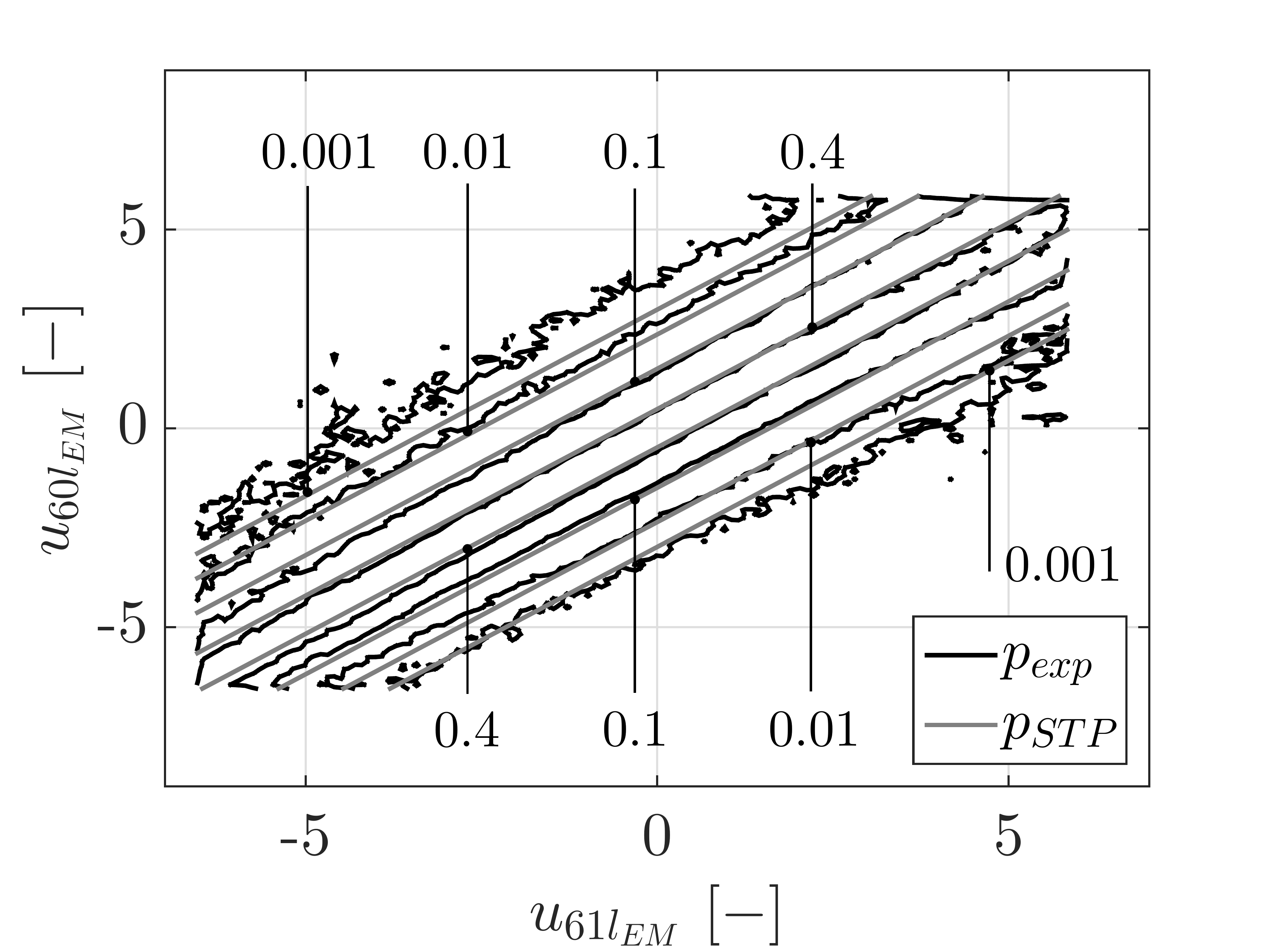}
  \caption{
  Contour plots show experimental and reconstructed conditional PDFs at two different scales, a) $r=l_{EM}$ and b) $r=60l_{EM}$.
  A dataset of subgroup (v) is used with $\Rey_\lambda=475$.
  }
\label{fig:p_1}
\end{figure}

As a second step for the estimation of Kramers-Moyal coefficients, the optimisation procedure of Chapter 3.3 is applied.
This procedure is based on the reconstruction of PDFs according to eq. (\ref{eq:short_time_prop}). 
Figure \ref{fig:p_1} shows the quality of reconstructed conditional PDFs by the optimised Kramers-Moyal coefficients at two different scales, $r=l_{EM}$ and $r=60l_{EM}$.
To pronounce the differences, we have chosen $r=60l_{EM}$ which corresponds to a scale larger than $L$. The agreement of experimental and reconstructed conditional PDFs is good.


\begin{figure}
  \centering    
  a)\includegraphics[width=0.47\textwidth]{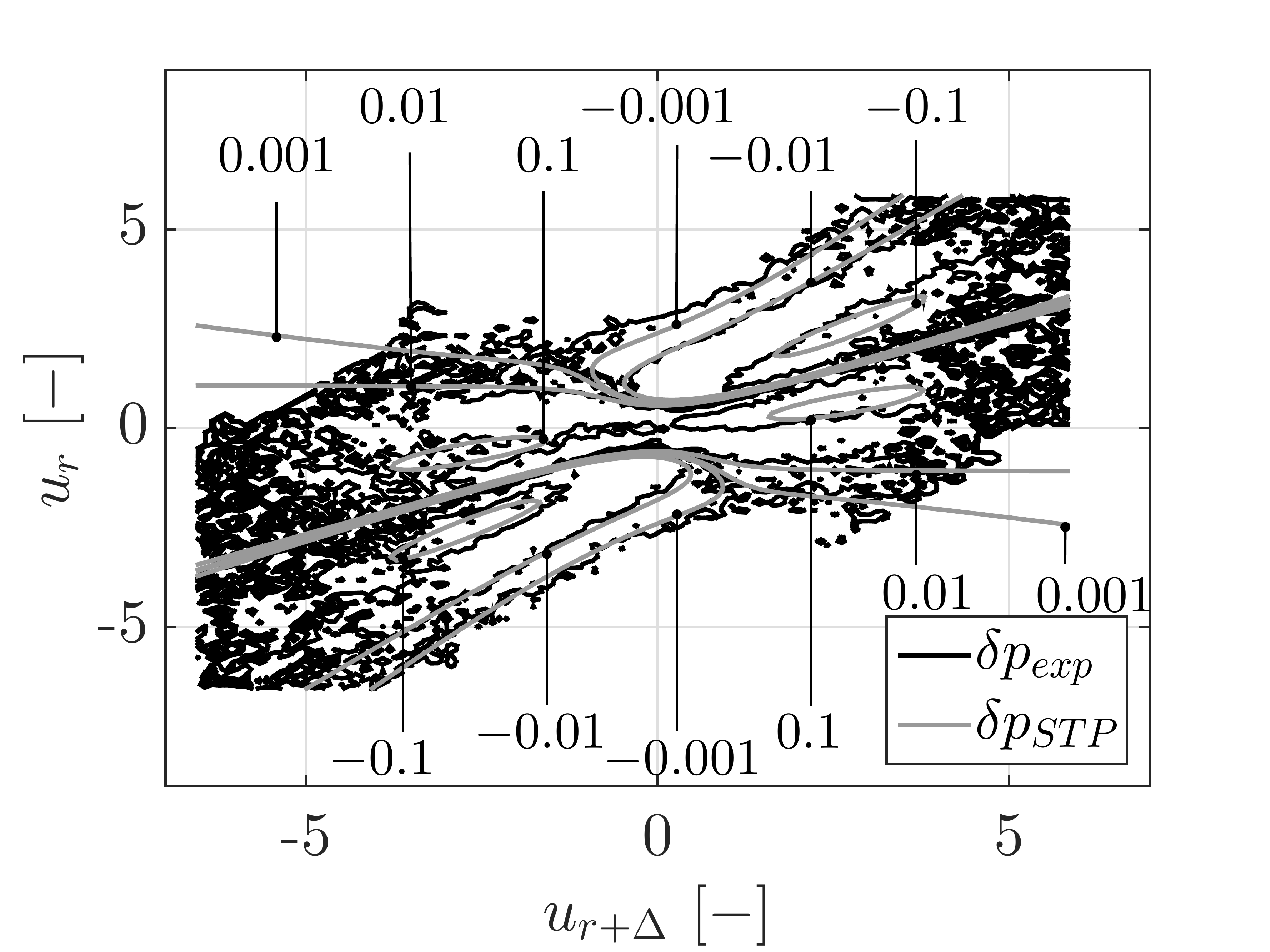}
  b)\includegraphics[width=0.47\textwidth]{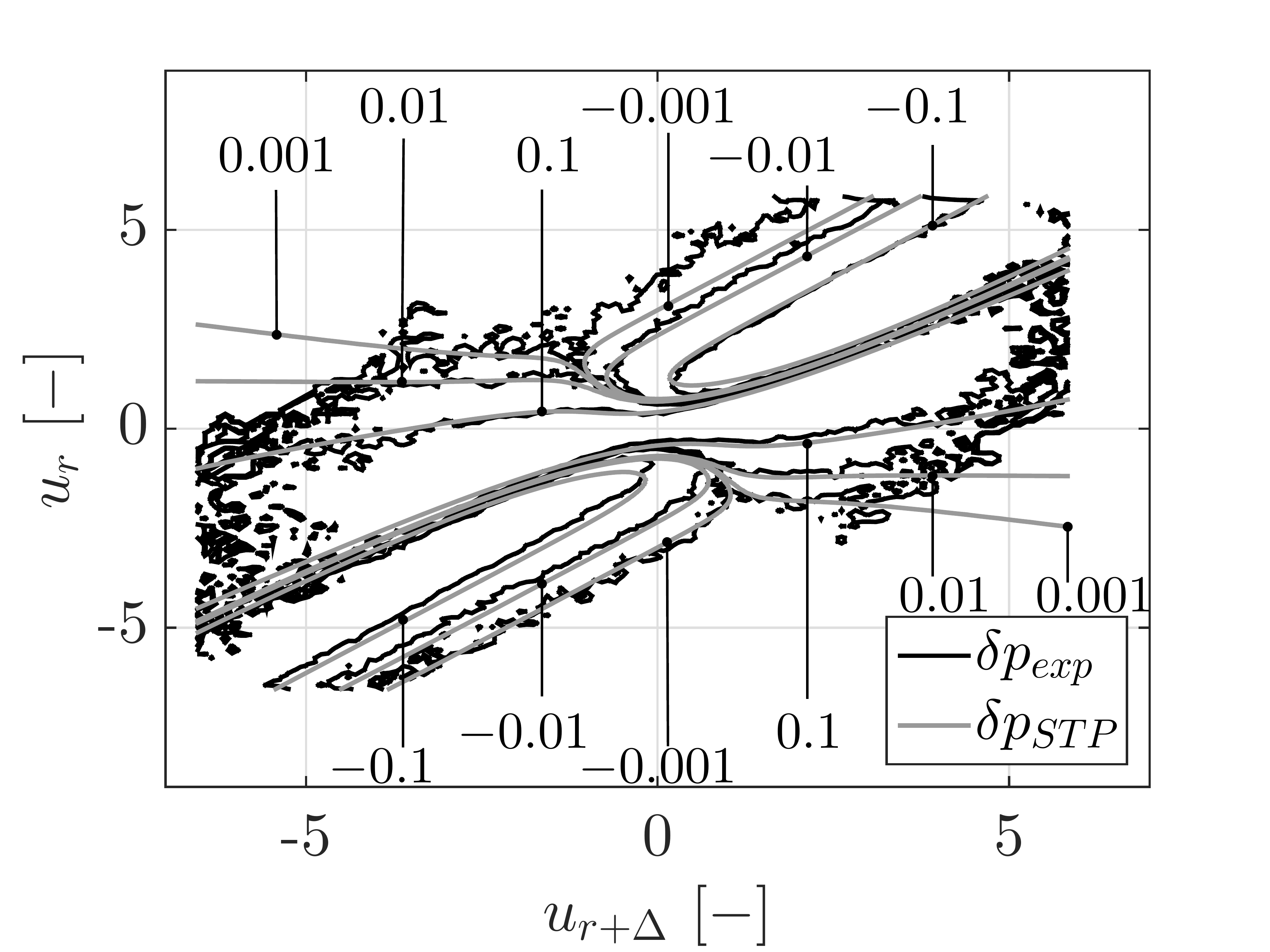}
  \caption{Change of conditional PDF $\delta p$ along the scale. a) $\delta p$ for $N=2$, which corresponds to small scale change and b) for $N=60$, which corresponds to the entire change within inertial range.
A dataset of subgroup (v) is used with $\Rey_\lambda=475$.
}
\label{fig:p_w}
\end{figure}

In order to further show the quality of the reconstructed conditional PDF, we analyse the change of conditional PDFs along the scale, cf. eq. (\ref{eq:j_pdf_fak}) and (\ref{eq:FP}).
Figure \ref{fig:p_w} shows the cumulative change of conditional PDF, expressed as 
\begin{eqnarray}
\label{eq:dpdf}
\delta p \ 
=&& \left[p(u_{l_{EM}} | u_{2l_{EM}}) - p(u_{2l_{EM}} | u_{3l_{EM}})\right]  
\nonumber\\
+ && \left[p(u_{2l_{EM}} | u_{3l_{EM}}) - p(u_{3l_{EM}} | u_{4l_{EM}}) \right] 
\nonumber\\
+ &&... 
\nonumber\\
+ && \left[p(u_{(N-1)l_{EM}} | u_{N l_{EM}}) - p(u_{Nl_{EM}} | u_{(N+1)l_{EM}}) \right]
\\
\label{eq:dpdf2}
=  &&p(u_{l_{EM}} | u_{2l_{EM}})- p(u_{Nl_{EM}} | u_{(N+1)l_{EM}}) .
\end{eqnarray}

In figure \ref{fig:p_w}, $\delta p_{exp}$ corresponds to the cumulative change of experimental data and $\delta p_{STP}$ 
is deduced by means of the short time propagator, eq. (\ref{eq:short_time_prop}), and the optimised Kramers-Moyal coefficients.
Due to the scale-wise optimisation procedure, eqs. (\ref{eq:dpdf}) and (\ref{eq:dpdf2}) are not necessarily equal 
for the reconstructed PDFs, thus we use eq. (\ref{eq:dpdf}).
\mbox{Figure \ref{fig:p_w} a)} shows $\delta p$ for $N=2$ (the change of the cPDFs at small scales) and 
figure \ref{fig:p_w} b) presents $\delta p$ for $N=60$ (the change of the cPDFs over a broad range of scales).
We see in both figures that the distributions of $\delta p_{exp}$ and $\delta p_{STP}$ are in good agreement. 
Thus, the optimisation procedure works well from the smallest scale changes $\delta r = l_{EM}$  up to the cumulative change of the entire inertial range. 
Consequently, we assume that the description of the turbulent cascade process is properly characterised by optimised Kramers-Moyal coefficients. 

\begin{figure}
  \centering    
 \includegraphics[width=0.47\textwidth]{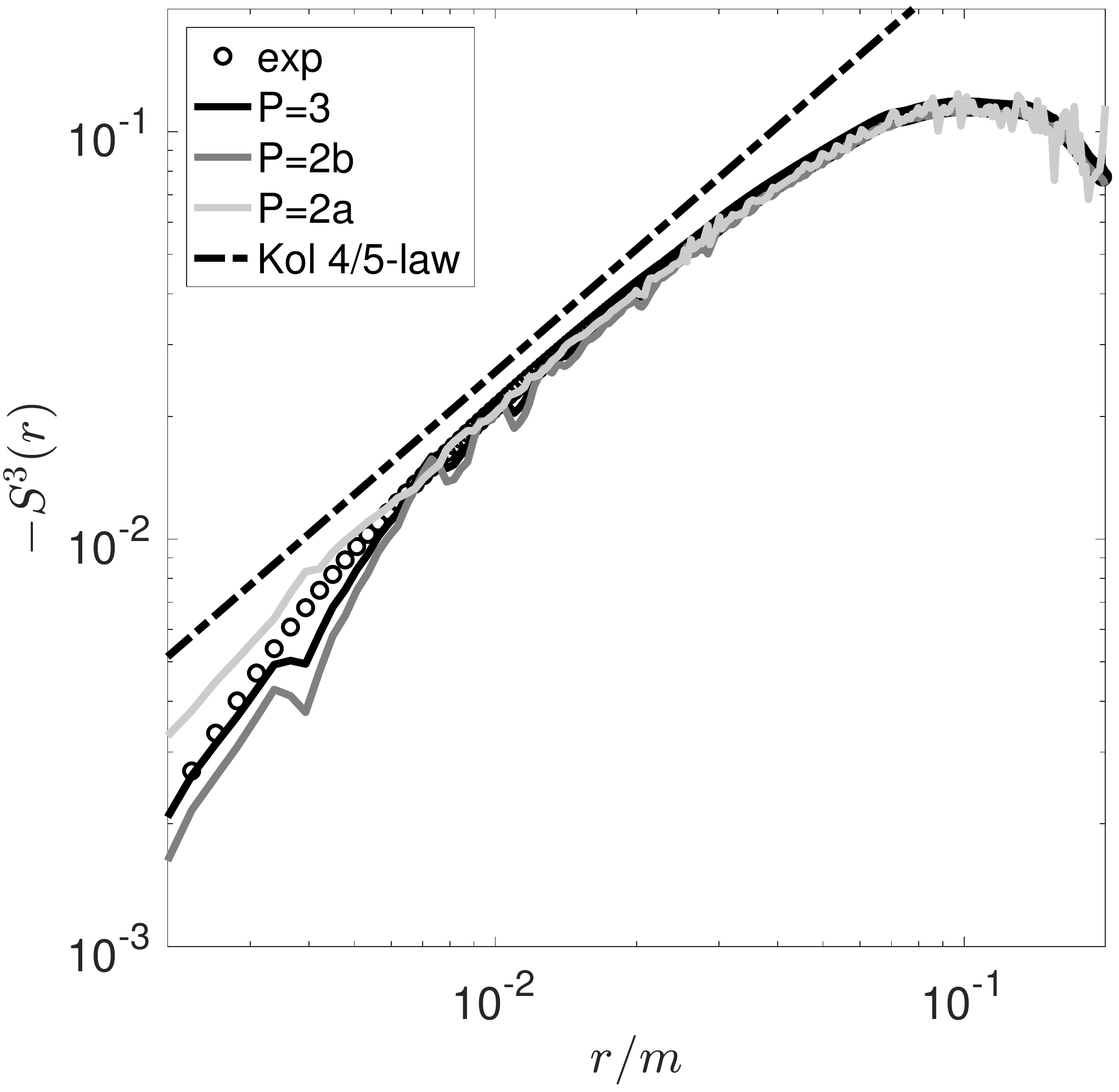}
  \caption{Third order structure function $-S^3$ as a function of scale $r/\Theta$. Shown are the results obtained from the experimental data (exp) and for different functional forms of the Kramers-Moyal coefficients corresponding to the cases of table \ref{table:IFT}. In addition the four-fifth law is shown. The dataset of subgroup vii  is used with $\Rey_\lambda=166$.}
\label{struc3}
\end{figure}

It is a very strong demand on the quality of the reconstructed Kramers-Moyal coefficients to show that the cPDFs are well reproduced. Naturally, also the commonly investigated increment PDFs $p(u_r)$ and the structure functions are reproduced well by these optimised stochastic equations. As the third order structure function $S^3$ with its four-fifth law plays a key role for turbulence, we show exemplarily for one dataset in figure \ref{struc3} the comparison of $S^3$  between the result of the Fokker-Planck equation and the measured one. For further discussion see later on.

\subsection{General validity of the integral fluctuation theorem and parametrisation of $D^{(1,2)}$}
\label{Par_IFT}
As a next step, we first demonstrate the validity of the IFT for our datasets and then employ the IFT to assess the quality of the optimised Kramers-Moyal coefficients and discuss their functional contributions. 

\begin{figure}
\centering  
\includegraphics[width=0.6\textwidth]{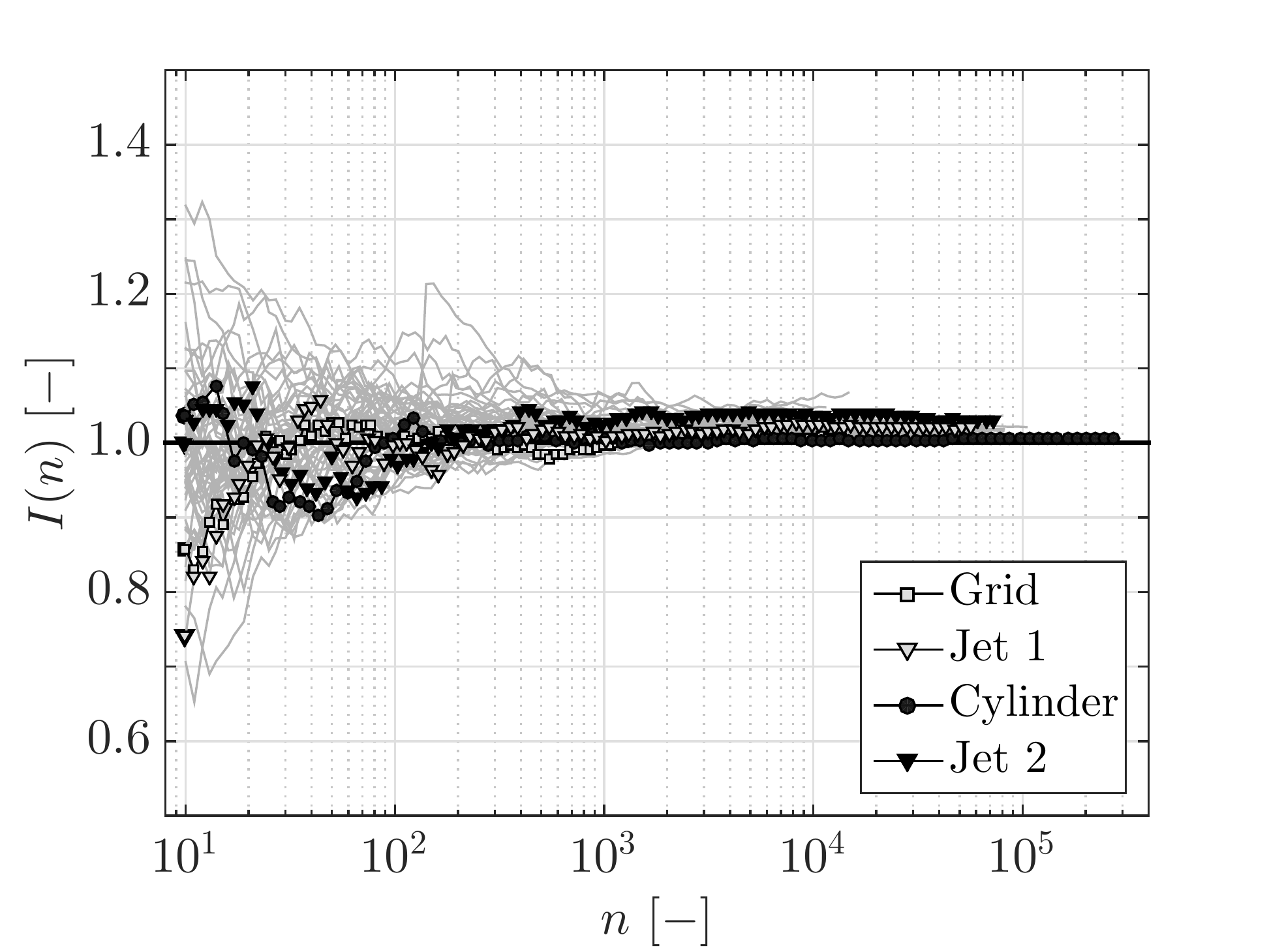}
\caption{IFT according to eq. (\ref{eq:FT}) as a function of the number of trajectories $n$. 
Results from all data; the selected datasets from table \ref{table:data} are highlighted by symbols, all other datasets are shown as grey lines.
}
\label{fig:11}
\end{figure}

Figure \mbox{\ref{fig:11}} presents the IFT for all  datasets as a function of the number $n$ of considered velocity increment trajectories $u(\cdot)$ as stated in eq. (\ref{eq:FT}). 
Here we used again the optimised functions $D^{(1,2)}$. 
Due to different numbers of samples in the datasets and varying Taylor micro-scale as well as integral length scales, the maximal number of independent trajectories differ from dataset to dataset. 
The four selected datasets from table \ref{table:data} are highlighted by symbols, other datasets are shown in grey. 
At $n\approx1000$, $I(n)$ has already largely converged to $1$, eq. (\ref{eq:FT}), and only slight improvements of convergence are present at larger number of trajectories, $n>1000$.  
The average of the 61 IFT values 
taken at $n=\mathrm{maximal}$ is 
$\left\langle I_\mathrm{max}\right\rangle_{\textrm{{\scriptsize datasets}}} \approx 1.01$
and the standard deviation of the 61 $I_\mathrm{max}$ is  
$ \sigma ( I_\mathrm{max} ) \approx0.01$, 
which we take as strong evidence that the IFT is a universal feature of turbulent cascades.

\begin{figure}  
\centering  
\includegraphics[width=0.6\textwidth]{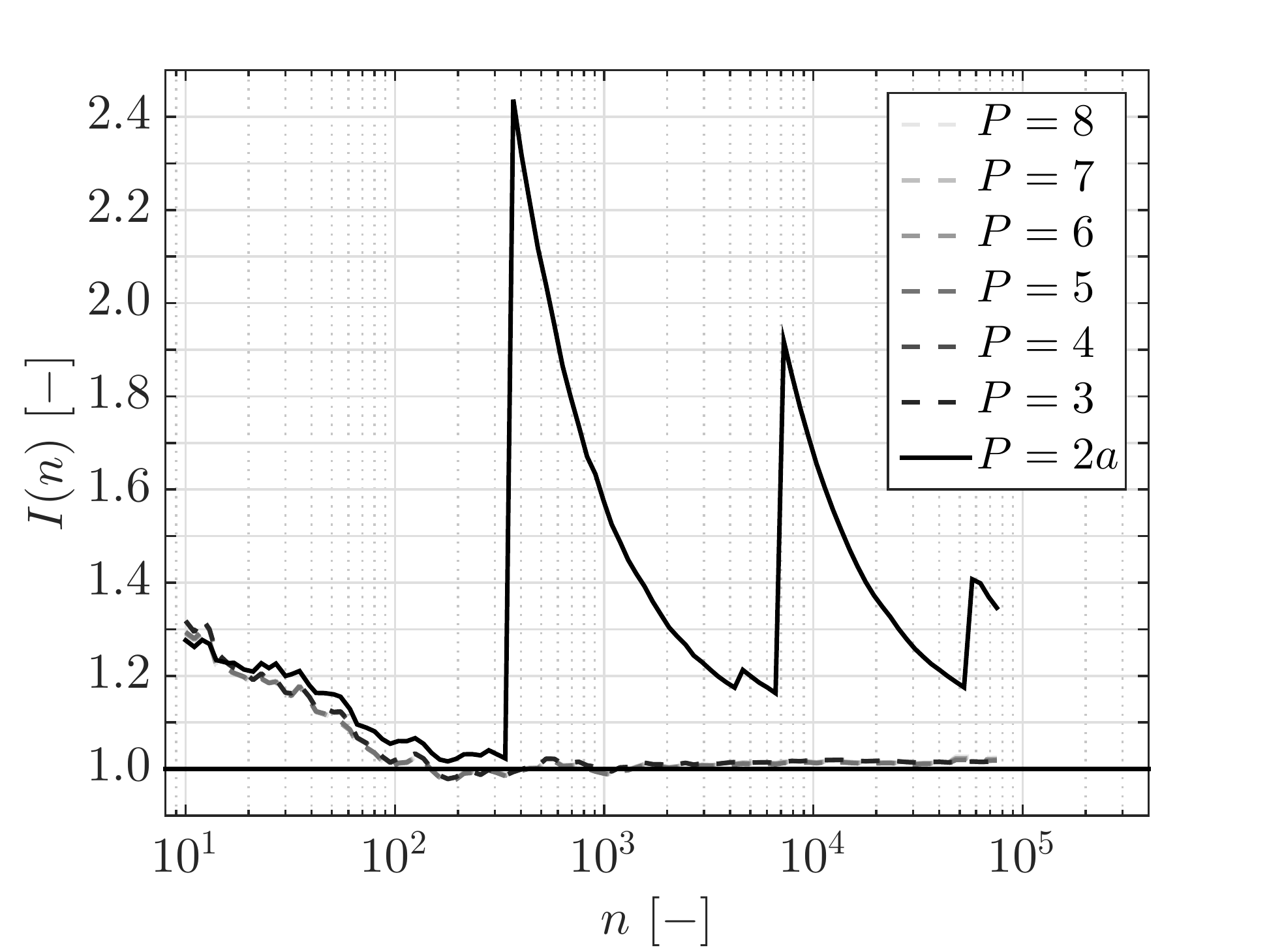}
\caption{ IFT, according to eq. (\ref{eq:FT}), as function of the number of trajectories $n$ for different parametrisation of $D^{(1,2)}$ according to table \ref{table:IFT}. Note that $P = 2b$ is not shown in figure \ref{fig:11_2}, since the IFT diverges for this case.}
\label{fig:11_2}
\end{figure}

Next we reverse the argumentation and take the IFT as a given fundamental law
for the turbulent cascade. Thus we can verify our parametrisation and the significance of functional contributions of $D^{(1,2)}$ can be worked out.
To test these aspects we parametrise $D^{(1,2)}$ up to third order in $u_r$
\footnote{Preliminary investigations with even higher order parametrisations do not show any improvements or other features to this third order approach}, 
i.e. with 8 parameters,
\begin{eqnarray} \label{eq:D1e}
	D^{(1)}( u_r ) &=& d_{13}(r) u_r^3 \  + \ d_{12}(r)  u_r^2 \  + \ d_{11}(r)  u_r  \ +  \ d_{10}(r),\\
\label{eq:D2e}
	D^{(2)}( u_r ) &=& d_{23}(r) u_r^3 \  + \ d_{22}(r)  u_r^2 \  + \ d_{21}(r)  u_r  \ +  \ d_{20}(r).
\end{eqnarray}
The parameters $d_{ij}$ are continuous functions of $r$ which is advantageous for the determination of $I(n)$. 
To single out the important parameters among the $d_{ij}$ above, the number of parameters has successively been reduced from 8 to 2 as shown in table \ref{table:IFT}.
Figure \ref{fig:11_2} presents this investigation for one exemplary dataset (subgroup (v), $\Rey_\lambda = 475$). 
For each selection of parameters the above mentioned optimisation procedure to determine $D^{(1,2)}$ was repeated and the IFT was tested. 
The resulting values  $I_\mathrm{max}$ are shown in \mbox{table \ref{table:IFT}.} 
We estimate an error of the $I_\mathrm{max}$ by considering the convergence of the IFT ($1000\leq n \leq \mathrm{maximal}$) and its fluctuations around the saturation value $I_\mathrm{max}$, figure \ref{fig:11_2}. In accordance with the mentioned standard deviation ($\sigma (I_\mathrm{max}) \approx$ 0.01), the estimated error
 is $0.016$.
From the results presented in table \ref{table:IFT} and in figure \ref{fig:11_2} we conclude that different functional representations of $D^{(1,2)}$ are possible to fulfil the IFT in good quality.
\begin{table}
\centering
  \begin{tabular}{c|ccccccccc} 
  P  &  $d_{13}$ &  $d_{12}$ &  $d_{11}$ &  $d_{10}$ &  $d_{23}$ &  $d_{22}$  &  $d_{21}$  &  $d_{20}$ & $I_\mathrm{max}$ \\
  \hline
 2a             & -                & -               & X               & -               & -               & -               & -               & X        &      1.3(46) $\pm$ 0.016\\
 2b             & -                & -               & X               & -               & -               & X               & -               & -        &      $\infty$\\
 3               & -                & -               & X               & -               & -               & X              & -               & X        &     1.0(17) $\pm$ 0.016 \\
 4               & -                & -               & X               & X              & -               & X              & -               & X        &     1.0(17) $\pm$ 0.016 \\
 5               & -                & -               & X               & X              & -               & X              & X              & X        &     1.0(18) $\pm$ 0.016  \\
 6               & -                & X              & X               & X              & -               & X              & X              & X        &     1.0(21) $\pm$ 0.016 \\
 7               & X               & X              & X               & X              & -               & X              & X              & X        &     1.0(21) $\pm$ 0.016 \\
 8               & X               & X              & X               & X              & X              & X              & X              & X        &     1.0(24) $\pm$ 0.016   \\
\end{tabular}
\caption{Scheme of the chosen parameters  $d_{ij}$  for $D^{(1,2)}$ according to eqs. (\ref{eq:D1e}) and (\ref{eq:D2e}). On the left row the number of used parameters are given. Crosses indicate non-zero parameters. The corresponding 
 $I_\mathrm{max}$
values are given with their uncertainties.  }
\label{table:IFT}
\end{table}

Next we discuss the meaning of the single cases. 
The case $P=2a$ in table \ref{table:IFT} corresponds to pure additive noise in the cascade process. This implies only Gaussian increment PDFs, i.e. no intermittency is taken into account.
The value of  $I_\mathrm{max} = 1,3(46)$ and the large jumps with $n$ evidently suggest that the IFT cannot be fulfilled with this functional ansatz, see figure \ref{fig:11_2}. The jumps in the convergence are a consequence of an unbalanced weight on entropy consuming trajectories in the exponential average of the IFT. We consequently reject the case $P=2a$  as a functional form of $D^{(1,2)}$.

The case $P=2b$ corresponds to pure multiplicative noise in the cascade process which includes a quadratic term in the diffusion function $D^{(2)}$.
Note that this parametrisation corresponds to a generalisation of the scaling proposed by \citet{Kolmogorov1962}, \mbox{see eq. (\ref{eq:se})} and \citet{Friedrich1997}, \citet{Nickelsen2013}.
As the IFT diverges, this functional form is rejected as well.

For all the other choices with three and more parameters ($P\ge3$), we see good fulfilment of the IFT.
More coefficients do not lead to a significantly better fulfilment of the IFT and its convergence.
We hence conclude that the case $P = 3$ is sufficient for a good stochastic model of the turbulent cascade that captures the features of small scale intermittency. Higher order terms as used in \citet{Renner2001} and \citet{Nickelsen2013} seem not to be necessary. Similar stochastic models have been discussed before, see e.g. \citet{Dubrulle2000} and \citet{Laval2001}. Starting with the Navier-Stokes equation, Laval et al. derive after certain simplifications and approximations a Langevin equation with correlated additive and multiplicative noise, which is of similar but more involved form as our case $P=3$. The amplitude of the additive noise is related to our $d_{20}$ term, and the amplitude of the multiplicative noise to the $d_{22}$ term. The conceptual difference, however, is that in \citet{Laval2001} the skewness is generated by the correlation between the two noise sources, which would imply a linear term $d_{21}$ (case $P=5$), whereas in our model the skewness already present at integral length scales is taken and transported down the cascade by the deterministic $d_{11}(r)$ term. This can also be seen in figure \ref{struc3} where already the additive or the multiplicative noise alone are enough to reproduce the measured skewness (cases $P=2a$ and $P=2b$), as long as the linear drift term $d_{11}$ is part of the model.

To conclude, we see that the minimal functional form to fulfil the IFT is for $D^{(1)}$ a linear term, $d_{10}(r) u_r$, and for $D^{(2)}$ an additive term $d_{20}(r)$ as well as a quadratic term $d_{22}(r) u_r^2$, which is in accordance with eq. (\ref{eq:D1}) and (\ref{eq:D2})
\footnote{\citet{Renner2002b} interpreted $d_{20} \neq 0$ and $d_{22} \neq 0$ as a necessary combination for turbulence and its energy cascade}.
We also recognised that $d_{22}$ becomes increasingly important for the fulfilment of the IFT with increasing $\Rey_\lambda$, indicating that small scale intermittency becomes more dominant at higher $\Rey_\lambda$.

The need of the three parameters can also be discussed in the context of structure functions,
see Section \ref{Multiscale} and eq. (\ref{eq:momentsFP}).
Inserting the three parameters and a multiplication by $-\frac{1}{S^\kappa}$ (from the left) lead to
\begin{eqnarray} 
\label{eq:momentsFP2}
\xi_\kappa = \frac{r}{S^\kappa} \frac{\partial S^{\kappa}}{\partial r}
     &=&
- r \kappa \biggl( d_{11}(r)   + (\kappa-1) \Bigl(  d_{22}(r)+  \frac{S^{(\kappa -2)}(r)}{S^{\kappa}(r)}d_{20}(r) \Bigr)\biggr) .
\end{eqnarray}
Here, we used the substitution $\frac{r}{S}\frac{\partial S}{\partial{r}} = \frac{\partial \ln(S)}{\partial{\ln(r)}} =\xi_\kappa(r)$.

%
Note that the coupled set of differential equations (\ref{eq:momentsFP2}) is closed: All structure functions can be determined successively from  $S^{0} = 1$ and $S^{1} = 0$.
It is obvious that if more parameters were needed, eq. (\ref{eq:momentsFP2}) becomes more complicated.
In the case of negligible $d_{20}$, eq. (\ref{eq:momentsFP2}) is an analytic expression for $\xi_\kappa(r)$ and can be transformed to a form comparable to Kolmogorov's expression of the scaling exponent, eq. (\ref{eq:se}), if $d_{11}(r)$ and $d_{22}(r)$ depend reciprocally on $r$, cf. \citet{Renner2001}, \citet{Nickelsen2017}. 
This special case also recovers the four-fifth law $\xi_3=1$, cf. \citet{Kolmogorov1941b}. Since the four-fifth law is only exact for the \textit{ideal} case of homogeneous and isotropic turbulence at infinite Reynolds numbers, we do not expect that the four-fifth law is identically reproduced by our analysis of \textit{real} experimental data. The deviation from the four-fifth law rather accounts for corrections that are mandatory for realistic turbulent flows, similar to the widely accepted approach of extended self-similarity introduced by \citet{Benzi1993,Benzi1996}. For one exemplarily chosen dataset we have shown in figure \ref{struc3} the resulting third order structure functions. Even for the cases $P=2a$ and $P=2b$ the third order structure function is well reproduced. This result does not put the importance of the negative skewed statistics of the velocity increment expressed by $S^3$ into question, but we see quite different Kramers-Moyal coefficients can be used to reproduce $S^3$. This is easily understood as structure functions (as well as the PDFs $p(u_r)$) are only two point or one scale quantities and result from a projection of the multi-scale statistics. Thus there are whole families of Fokker-Planck equations that can reproduce these structure functions. Note that depending on the $d_{ij}$ terms, other structure functions, like $S^2$ via $d_{21}$, may contribute to the $r$-dependence of $S^3$, otherwise $S^3(r)$ is a relaxation process of the initial large scale skewness $S^3(L)$. This is easily derived from eq. (\ref{eq:momentsFP}) and \ref{eq:momentsFP2}).

The consistency of the determined coefficients with the measured structure functions is shown in appendix A4.

\subsection{The turbulent cascade in terms of Kramers-Moyal coefficients}
\label{TuCasKM}
Based on the results in section \ref{Par_IFT},
we take the \textit{simplest} functional form of the Kramers-Moyal coefficients, eq. (\ref{eq:D1}) and (\ref{eq:D2}), and discuss 
to what extent the parameters $d_{11}(r)$, $d_{20}(r)$ and $ d_{22}(r)$ behave in a universal way with respect to Taylor-Reynolds number and flow type.

\begin{figure}  
\centering  
a)\includegraphics[width=0.47\textwidth]{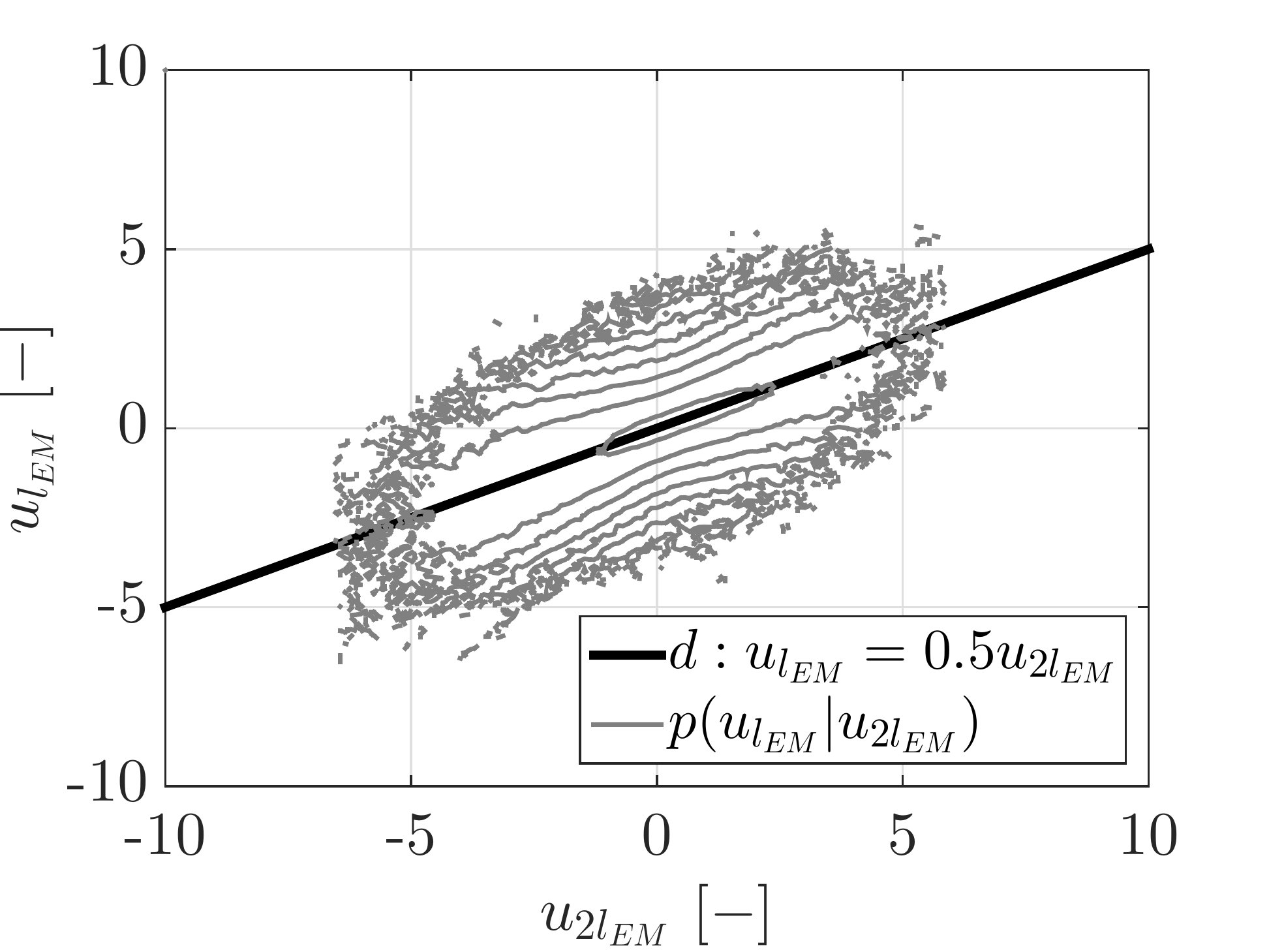}
b)\includegraphics[width=0.47\textwidth]{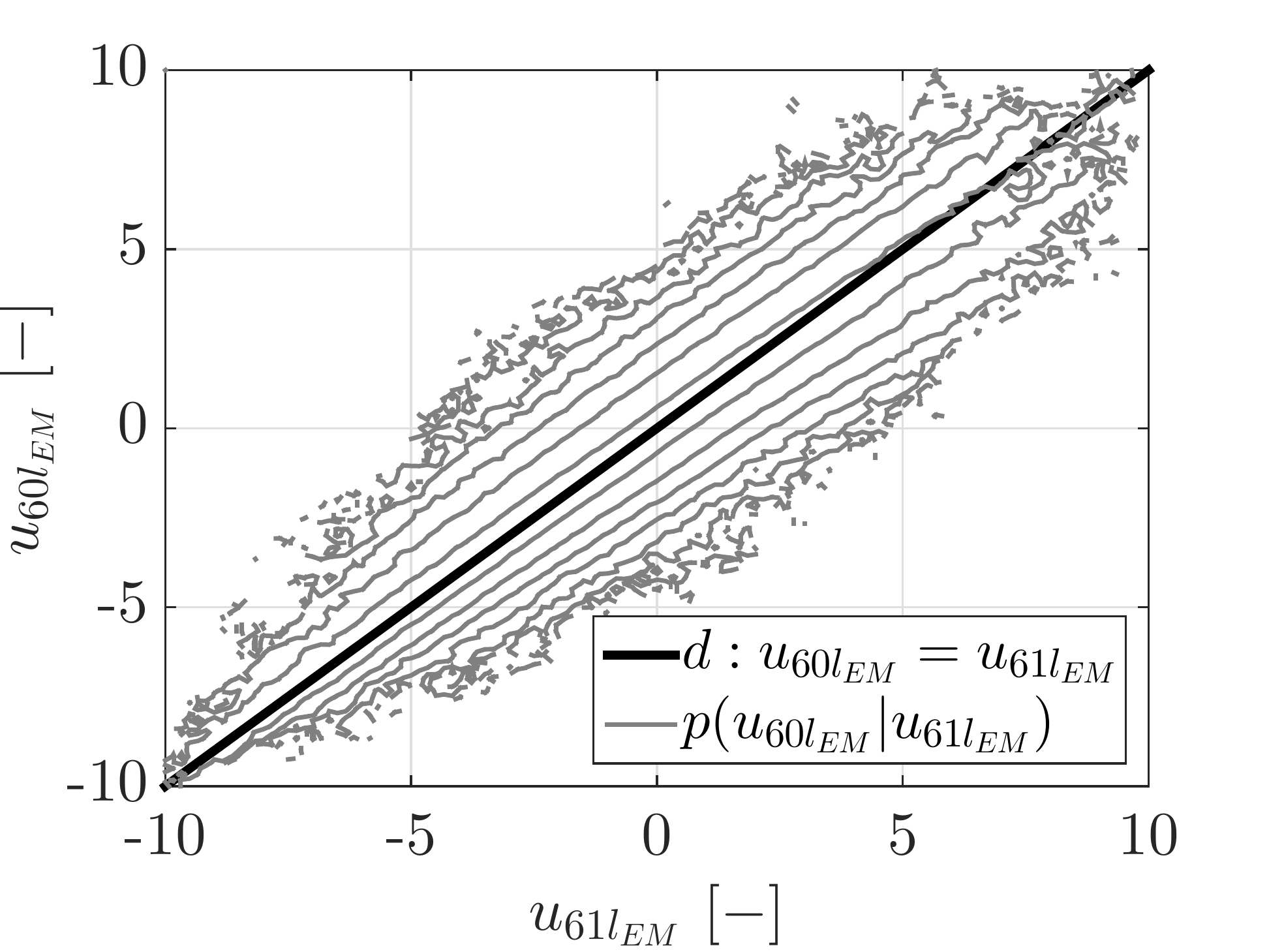}
\caption{a) Conditional PDF at $r=l_{EM}$. The location of the diagonal of the cPDF is indicated by the function
 $d \; : \; u_{l_{EM}}=0.5u_{2l_{EM}}$.
b) Conditional PDF at $r=60l_{EM}$, the diagonal function is
 $d \; : \; u_{60l_{EM}}=u_{61l_{EM}}$.
 For a) and b) a dataset of subgroup (v) is used with $\Rey_\lambda=475$.
}
\label{fig:bsp}
\end{figure}

Before we discuss the results in detail, we analyse in figure \ref{fig:bsp} the topological changes of the conditional PDFs with varying scale, which will serve us as a basis for the interpretation of the parameters $d_{11}(r), \ d_{20}(r)$ and $ d_{22}(r)$.
Figure \ref{fig:bsp} a) shows an exemplary conditional PDF at a small scale, $r=l_{EM}$. 
The maximum of the distribution is located at a diagonal of the form $d \; : \; u_{r} =c u_{r'}$, which is referred to as \textit{distribution diagonal} or simply \textit{diagonal}. Figure \ref{fig:bsp} a) shows that at scale $r=l_{EM}$ the distribution diagonal is $d \; : \; u_{l_{EM}}=0.5u_{2l_{EM}}$.
The contour isolines are curved beneath and above this diagonal.

Figure \ref{fig:bsp} b) shows for the same dataset the conditional PDF at larger scale, $r=60l_{EM}$.\footnote{To pronounce the differences between small and large scale, $r=60l_{EM}$ is chosen larger than $L$}
The distribution diagonal is now roughly at $d:u_{60l_{EM}}=u_{61l_{EM}}$. 
The contour isolines are not curved but almost parallel to the diagonal. 

These changes in conditional PDF shape can be related to the parameters by considering the short time propagator, eq. (\ref{eq:short_time_prop}). 
The leading term for the maximum is given by $D^{(1)}$ and $d_{11}$. The smaller the magnitude of $d_{11}$, the more the diagonal approaches $d \; : \;  u_{r} =u_{r+l_{EM}}$. The width $w$ of the distribution around the diagonal is proportional to the denominator in the exponential term of eq. (\ref{eq:short_time_prop}), $w  \propto d_{20} + d_{22} \ u^2_{r+l_{EM}}$. 
Thus, the coefficient $d_{22}$ expresses the non-linearity in the standard deviation and
gives rise to the curvature of the contour isolines:
The lower $d_{22}$, the smaller the curvature will be. 
In terms of unconditional PDF, a large $d_{22}$ is equivalent to fat-tailed increment distributions, confirming the previous interpretation that intermittent fluctuations are caused by $d_{22}$.
At $u_{r+l_{EM}}=0$, only 
$d_{20}$ characterises the width of the distribution. Further details are discussed in
appendix A5.

\begin{figure}  
\centering  
a)\includegraphics[width=0.6\textwidth]{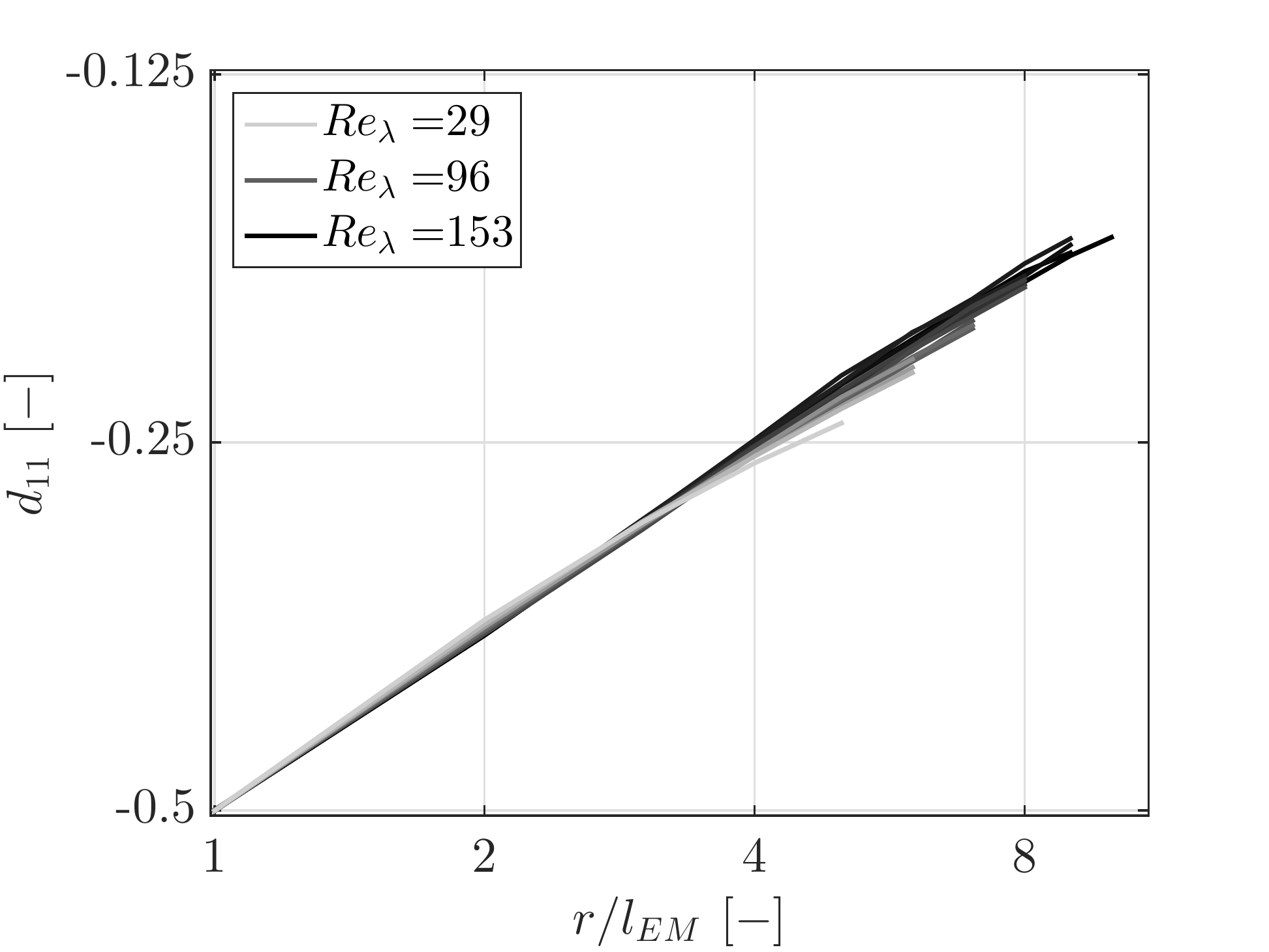}\\
b)\includegraphics[width=0.6\textwidth]{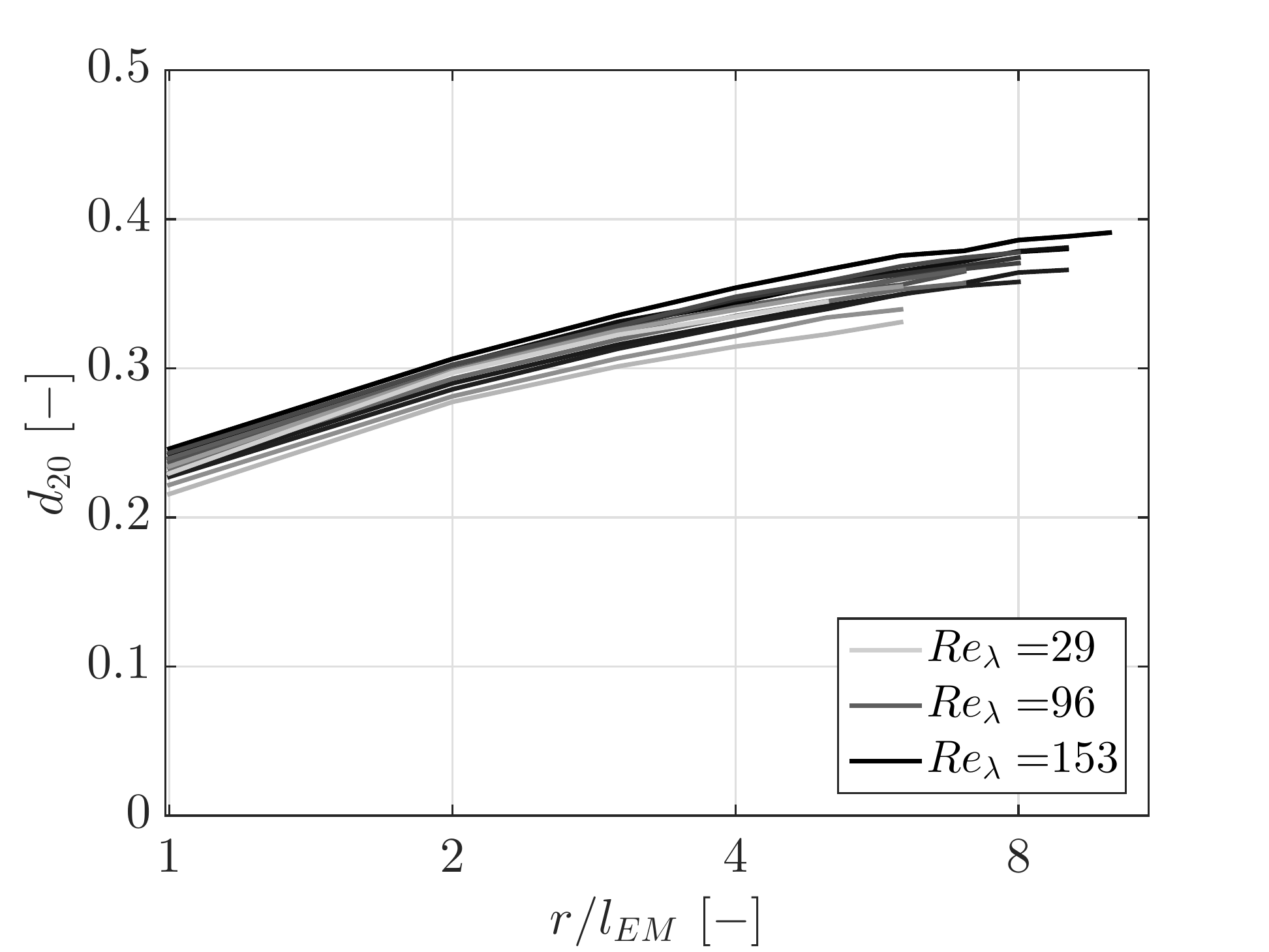}\\
c)\includegraphics[width=0.6\textwidth]{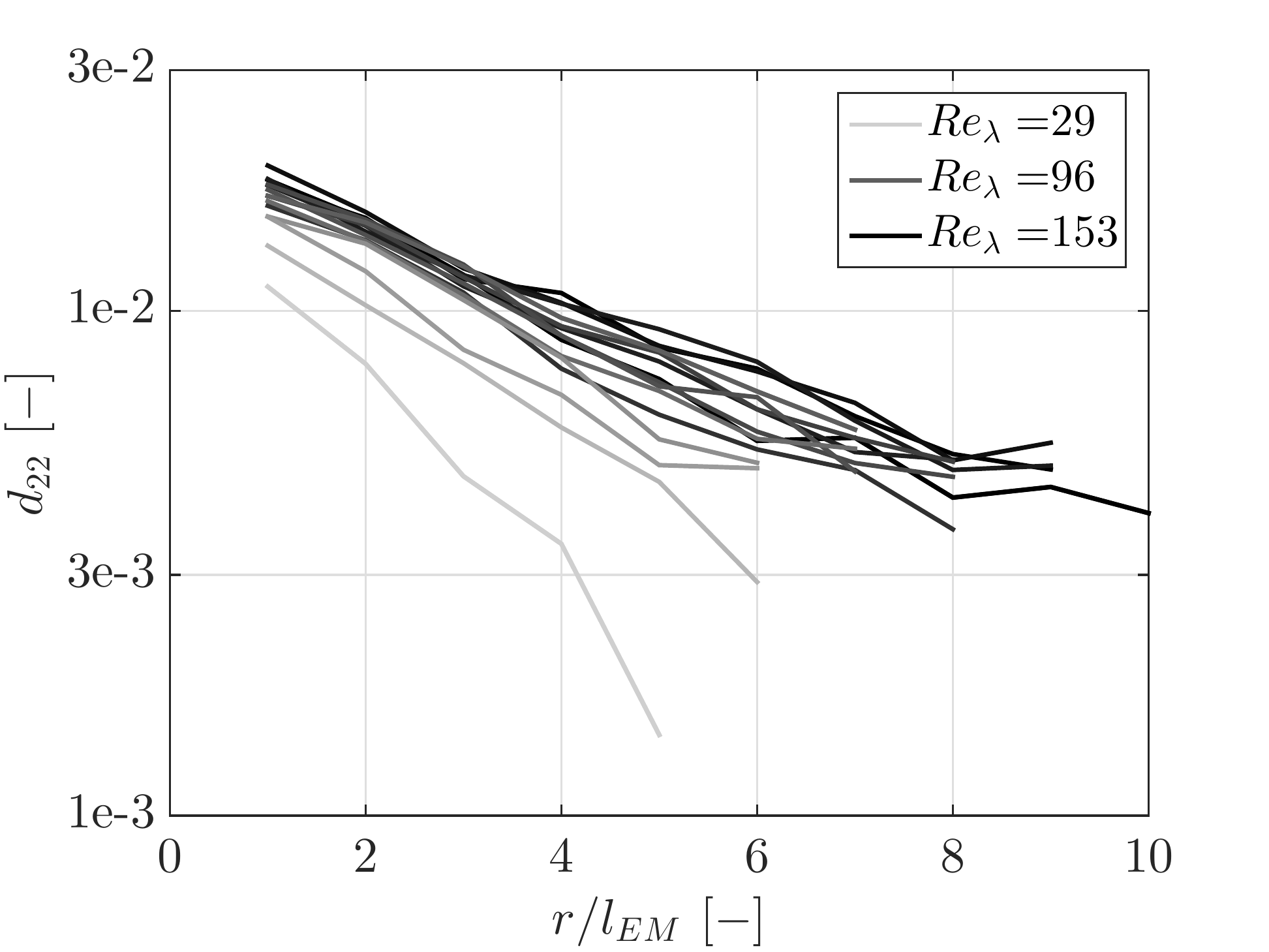}
\caption{
Parameters of grid generated turbulence as function of scale. 
The greyscale indicates  $\Rey_\lambda$.
a) shows $d_{11}$ in a double-logarithmic plot.
b) shows $d_{20}$ in a semi-logarithmic plot.
c) shows $d_{22}$ in a semi-logarithmic plot.
}
\label{fig:5_grid}
\end{figure}

Now, we return to the results of the three parameters $d_{11}, \ d_{22}$ and $d_{20}$ and their scale dependence.
In figure \ref{fig:5_grid}, these parameters are shown as a function of the scale $r$ and the Taylor-Reynolds number $\Rey_\lambda$ 
for grid generated turbulence.
Grid data was selected, because for this data the smallest scattering was observed. 
The same presentation of the parameters for cylinder and free jet flows are shown in appendix A6.

Figure \ref{fig:5_grid} a) shows the convergence of all $d_{11}$ to the fixed value $d_{11}=-0.5$ at $r=l_{EM}$,
which corresponds to a distribution diagonal $d \; : \;  u_r =0.5u_{r+l_{EM}}$, like figure \ref{fig:bsp} a).
Furthermore,  all $d_{11}$ are approximately linear at small scales in the log-log presentation.
Thus, $d_{11}$ can be expressed according to
\begin{equation}
\label{d11_para}
d_{11} \approx -0.5 \bigl( \frac{r}{l_{EM}}\bigr)^{-\frac{1}{2}}.
\end{equation}
Eq. (\ref{d11_para}) is found to hold universally for all types of flows, as shown in appendix A6.
The exponent $-\frac{1}{2}$ is in accordance with an earlier indication of \citet{Renner2002} for free jet flows.
At larger scales, $r \geq 4 l_{EM}$, the slope of $d_{11}$ decreases slightly and deviations from the power law become visible.
The higher $\Rey_\lambda$, the larger the scale where the deviation from the power law sets in.
Thus,  eq. (\ref{d11_para}) indicates that the case $d \; : \;  u_r=u_{r+l_{EM}}$ is reached only at very large scales and in the limit of very large $\Rey_\lambda$.

Furthermore, the power law with negative exponent eq. (\ref{d11_para}) implies an accelerating decay of turbulent structures along the cascade. 
From a uniform cascade process, as in the K62 model eq. (\ref{eq:se}), one would expect the power law exponent for $d_{11}$ to be $-1$ with a prefactor $\approx-0.36$, cf. \citet{Friedrich1997}, \citet{Nickelsen2017}. Compared to the K62 process, the smaller exponent $-1/2$ found in our analysis together with the prefactor $-0.5$ constitutes a decelerated decay of turbulent structures towards smaller scales.

Figure \ref{fig:5_grid} b) presents $d_{20}$ in a semi-logarithmic plot. 
All $d_{20}$ show a similar functional form with an increase in scale and with a small monotone curvature. The  $\Rey_\lambda$-dependency is rather weak. 
The functional form can be approximated by 
\begin{equation}
\label{d20_para}
d_{20}\approx (\alpha_{20}\ln(\frac{r}{l_{EM}}) + \beta_{20})^{\frac{1}{3}},
\end{equation}
where $\alpha_{20}$ and $\beta_{20}$ are fit parameters, see also appendix A4 and figure \ref{fig:fits} c).
Accordingly,  background fluctuations decrease moderately to smaller scales.

Figure \ref{fig:5_grid} c) presents $d_{22}$ in a semi-logarithmic plot. 
All $d_{22}$ decrease strongly with increasing scale, with a slower decrease for increasing $\Rey_\lambda$.
This functional behaviour can be approximated by 
\begin{equation}
\label{d22_para}
d_{22}\approx \alpha_{22}\ln(\frac{r}{l_{EM}}) + \beta_{22},
\end{equation}
where $\alpha_{22}$ and $\beta_{22}$ are fit parameters, see also appendix A4 and figure \ref{fig:fits} b).
In contrast to $d_{20}$, $d_{22}$ increases strongly to smaller scales. 
Since $d_{22}$ is the magnitude of the $u^2$ term in $D^{(2)}$, intermittent features also increase towards smaller scales, establishing the link to small scale intermittency.

The above parametrisation also allows a qualitative discussion of the infinite Reynolds number limit. From figures \ref{fig:5_grid}, \ref{fig:6_klein} and \ref{fig:cylinder_jet_d11_all} - \ref{fig:cylinder_jet_d22_all} it can be concluded that $d_{11}$ and $d_{20}$ appear to be rather independent from $\Rey_\lambda$. For $d_{11}$, this independence is apparent, for $d_{20}$ one may also see a decreasing tendency with $\Rey_\lambda$ which was also found in \citet{Renner2002b}. In contrast, $d_{22}$ clearly increases with $\Rey_\lambda$, confirming that small scale intermittency intensifies with $\Rey_\lambda$. A direct quantitative analysis in terms of the intermittency factor $\mu$ is questionable due to non-scaling behaviour, \citet[figure 4b]{Renner2002b}, but qualitatively it follows from (\ref{eq:momentsFP}) that $\mu(r) = 18 d_{22}(r)$, neglecting $d_{20}$ and taking $d_{11}$ and $d_{22}$ as “K62 contribution”. The resulting local intermittency factor, $\mu(r)$, indeed increases with $\Rey_\lambda$ from $\mu=0.2$ to $\mu=0.4$ at $r=l_{EM}$, and from $\mu=0.1$ to $\mu=0.3$ at $r=3 l_{EM}$, in qualitative agreement with the universal value of $\mu\approx0.26$.


%
%

\begin{figure}  
\centering  
a)\includegraphics[width=0.6\textwidth]{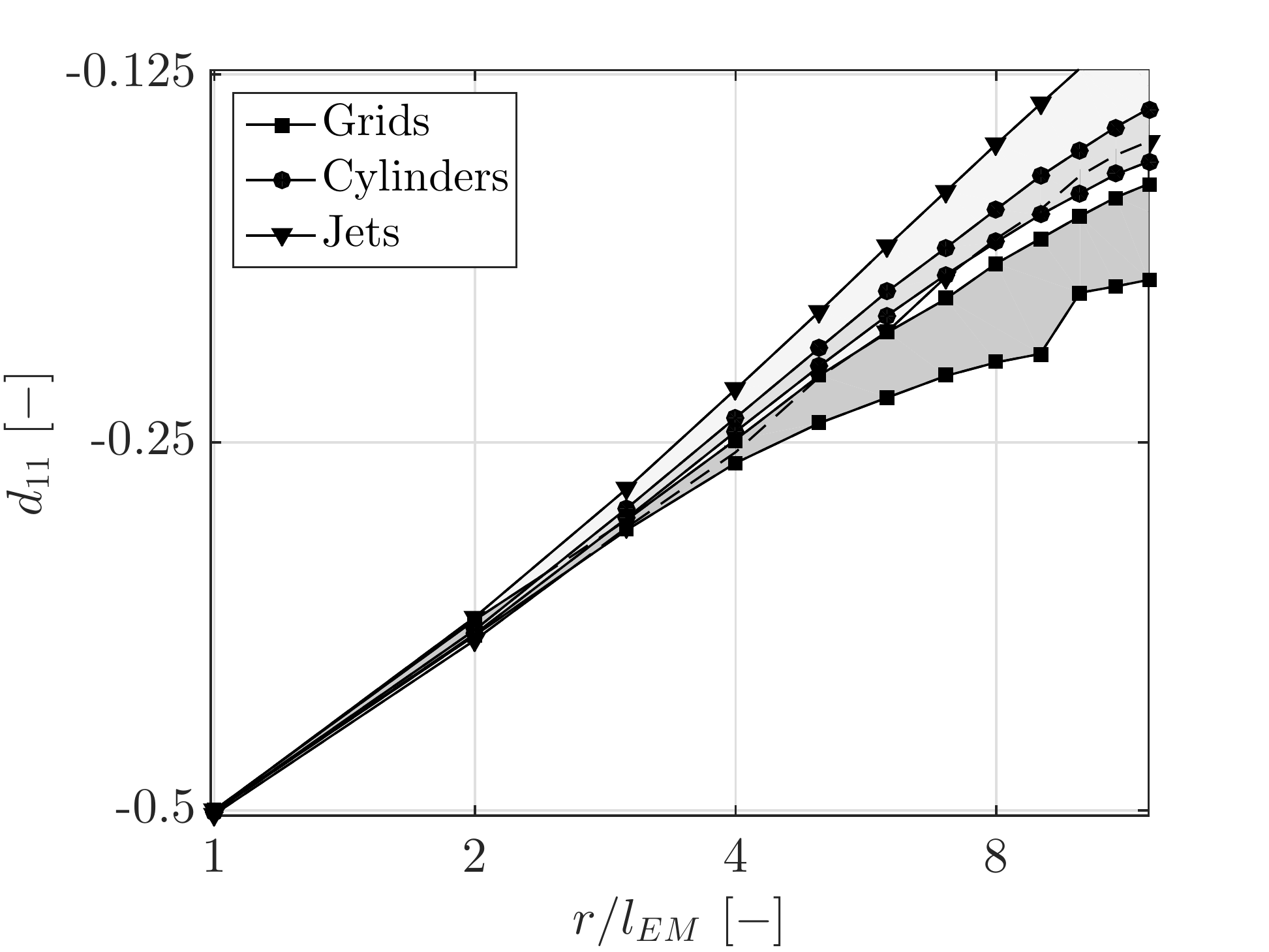}\\
b)\includegraphics[width=0.6\textwidth]{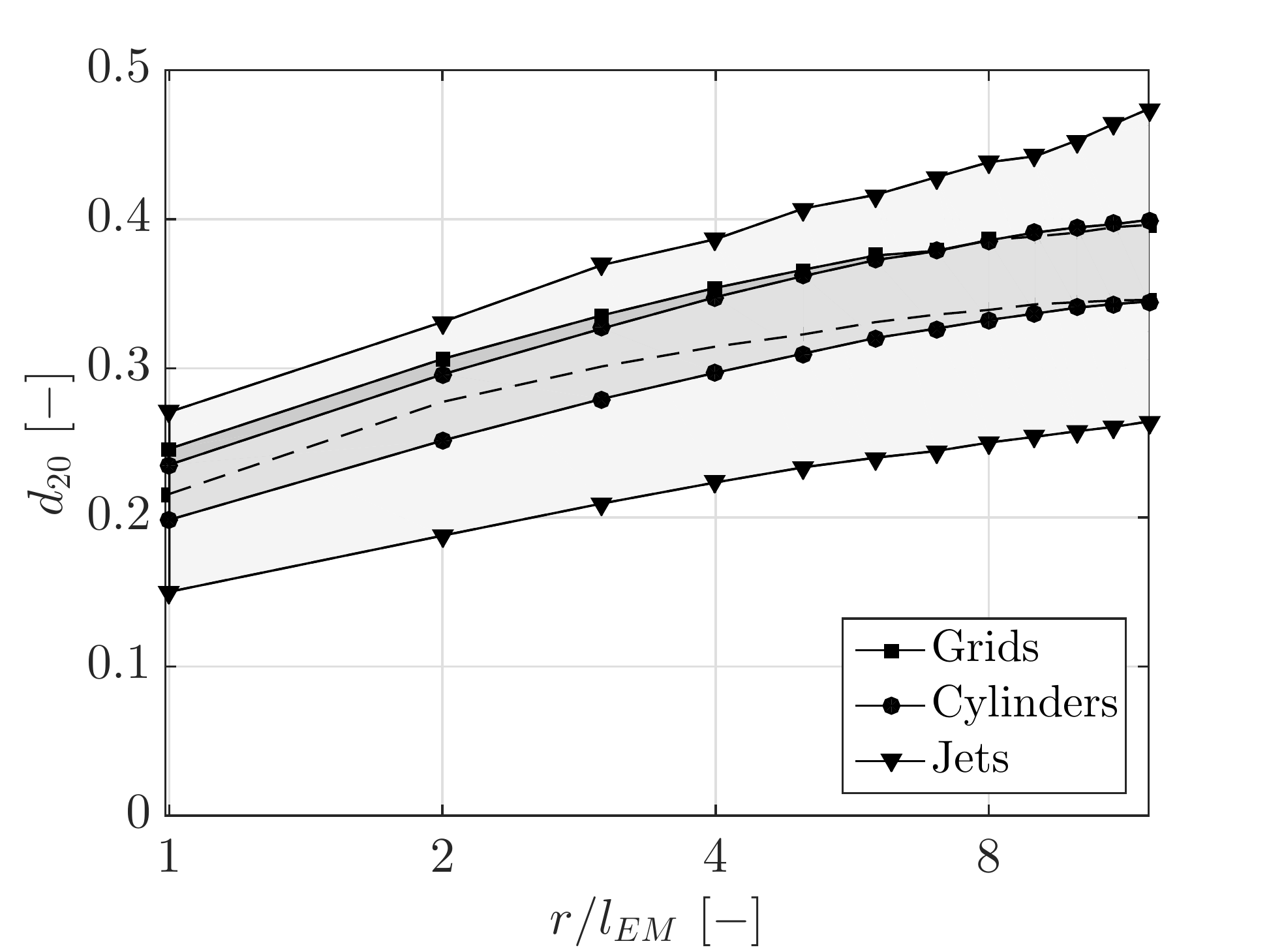}\\  
c)\includegraphics[width=0.6\textwidth]{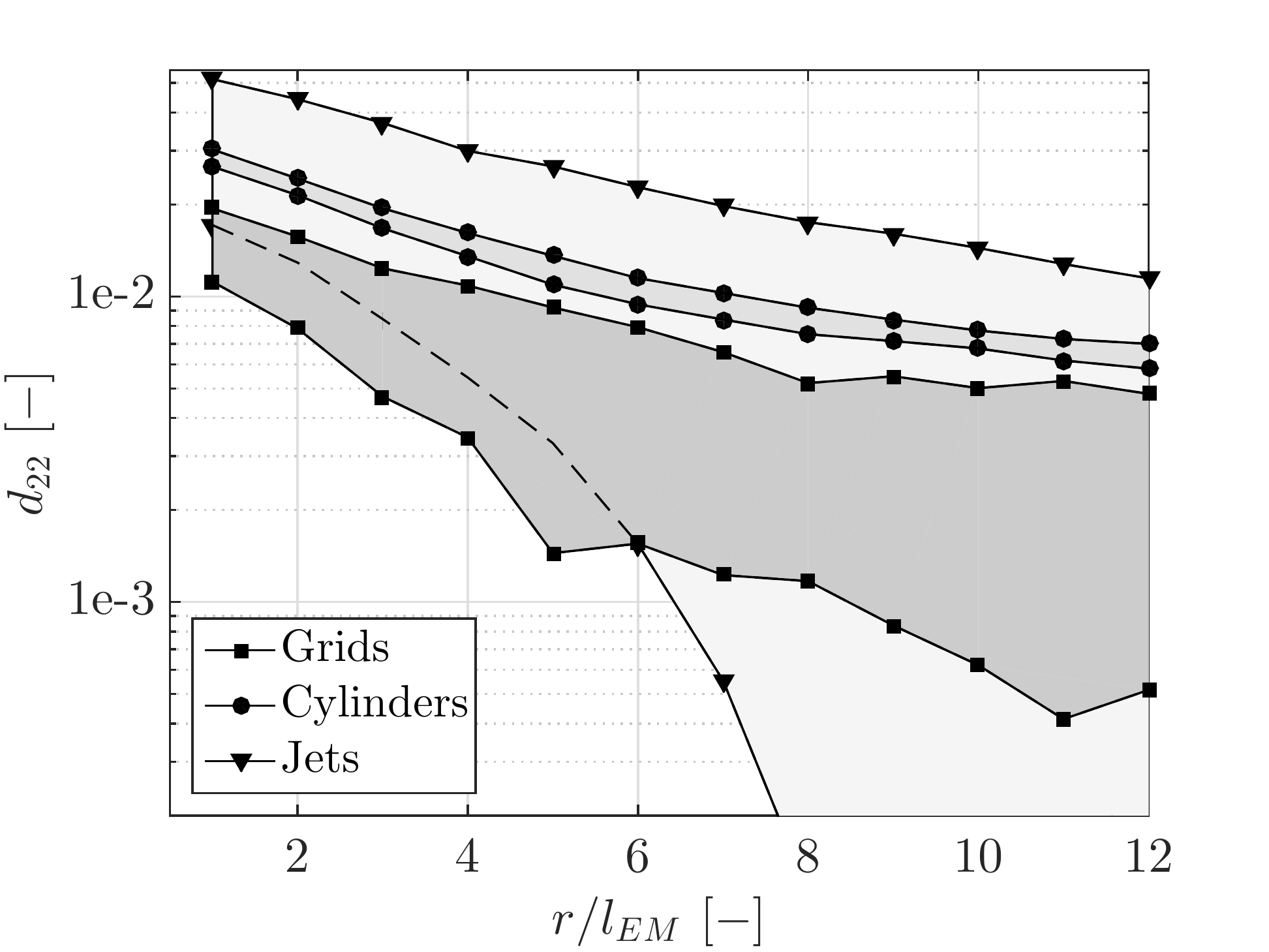}
\caption{ Parameters ($d_{11}$, $d_{20}$ and $d_{22}$) as function of scale. Surfaces indicate areas where the parameters of specific turbulent flow are located.
a) shows $d_{11}$ in a double-logarithmic plot.
b) shows $d_{20}$ in a semi-logarithmic plot.
c) shows $d_{22}$ in a semi-logarithmic plot.
}
\label{fig:5}
\end{figure}

\begin{figure}
\centering  
\vspace{0.2cm}
\includegraphics[width=0.7\textwidth]{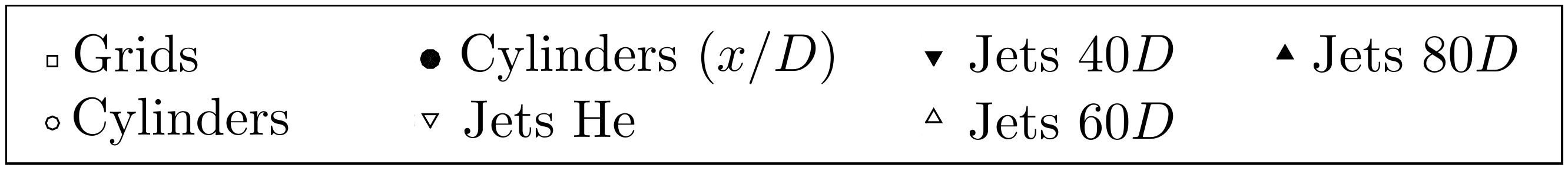}
\vspace{0.1cm}
\\
a)\includegraphics[width=0.45\textwidth]{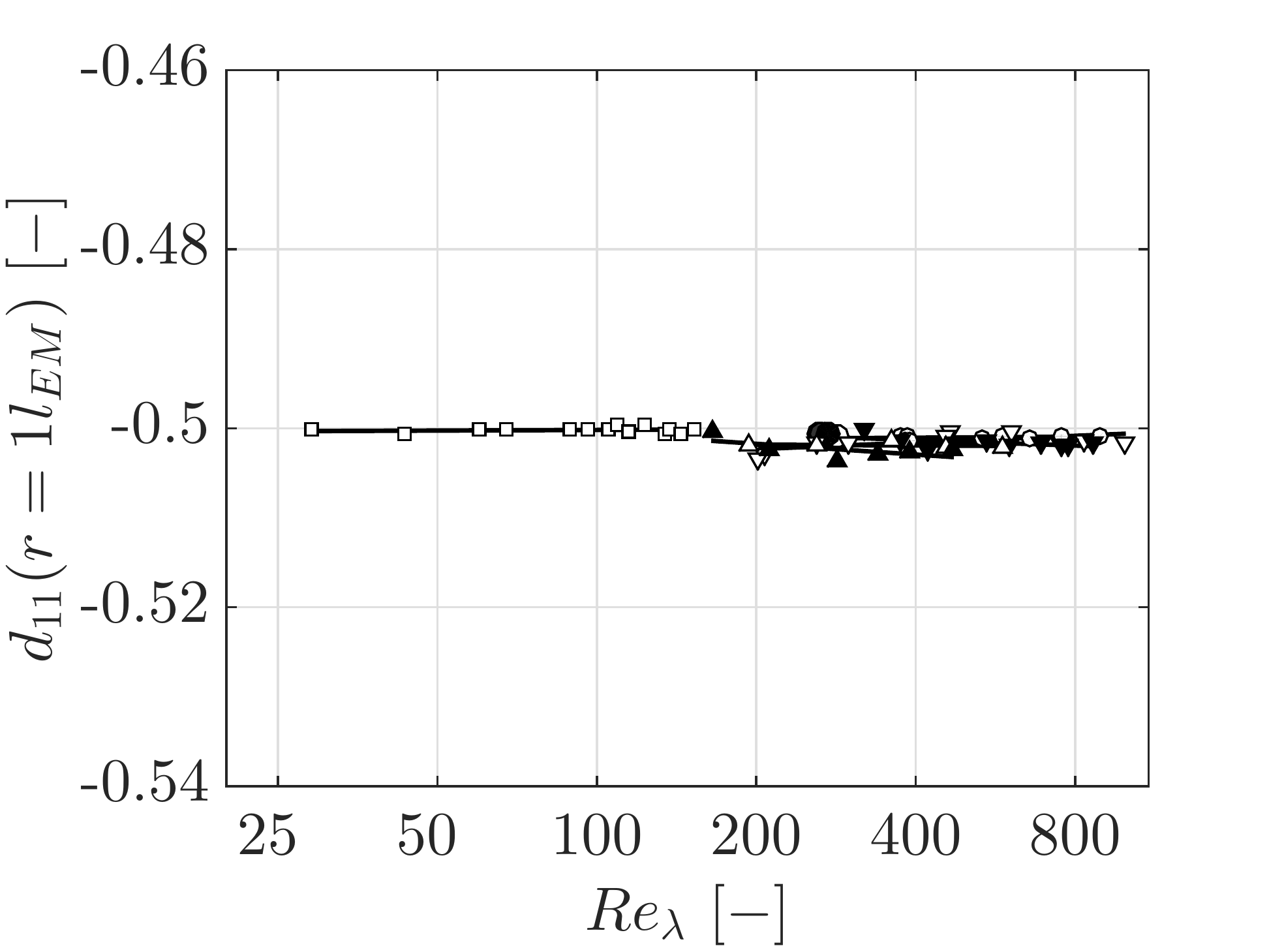}
b)\includegraphics[width=0.45\textwidth]{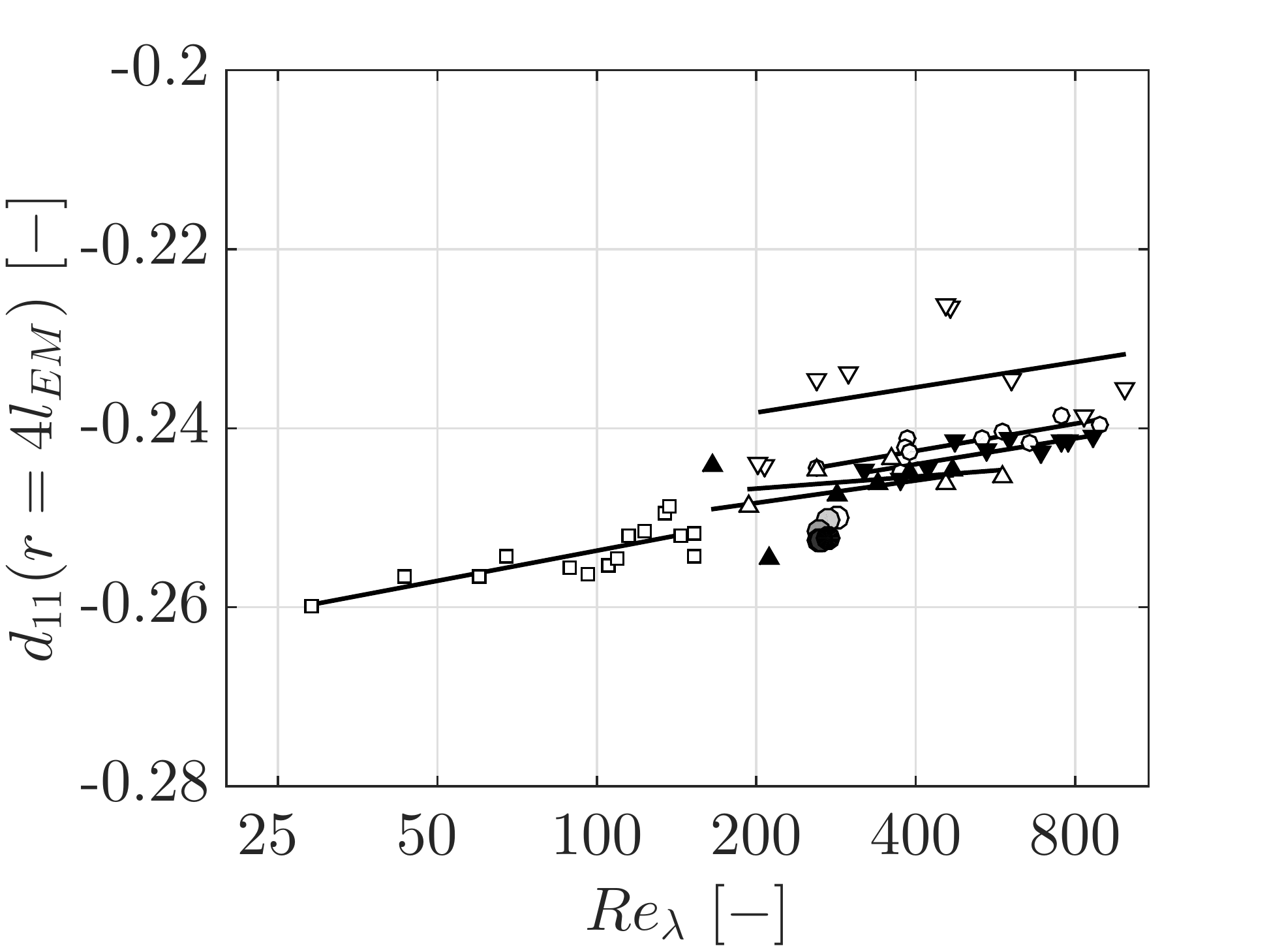}\\
c)\includegraphics[width=0.45\textwidth]{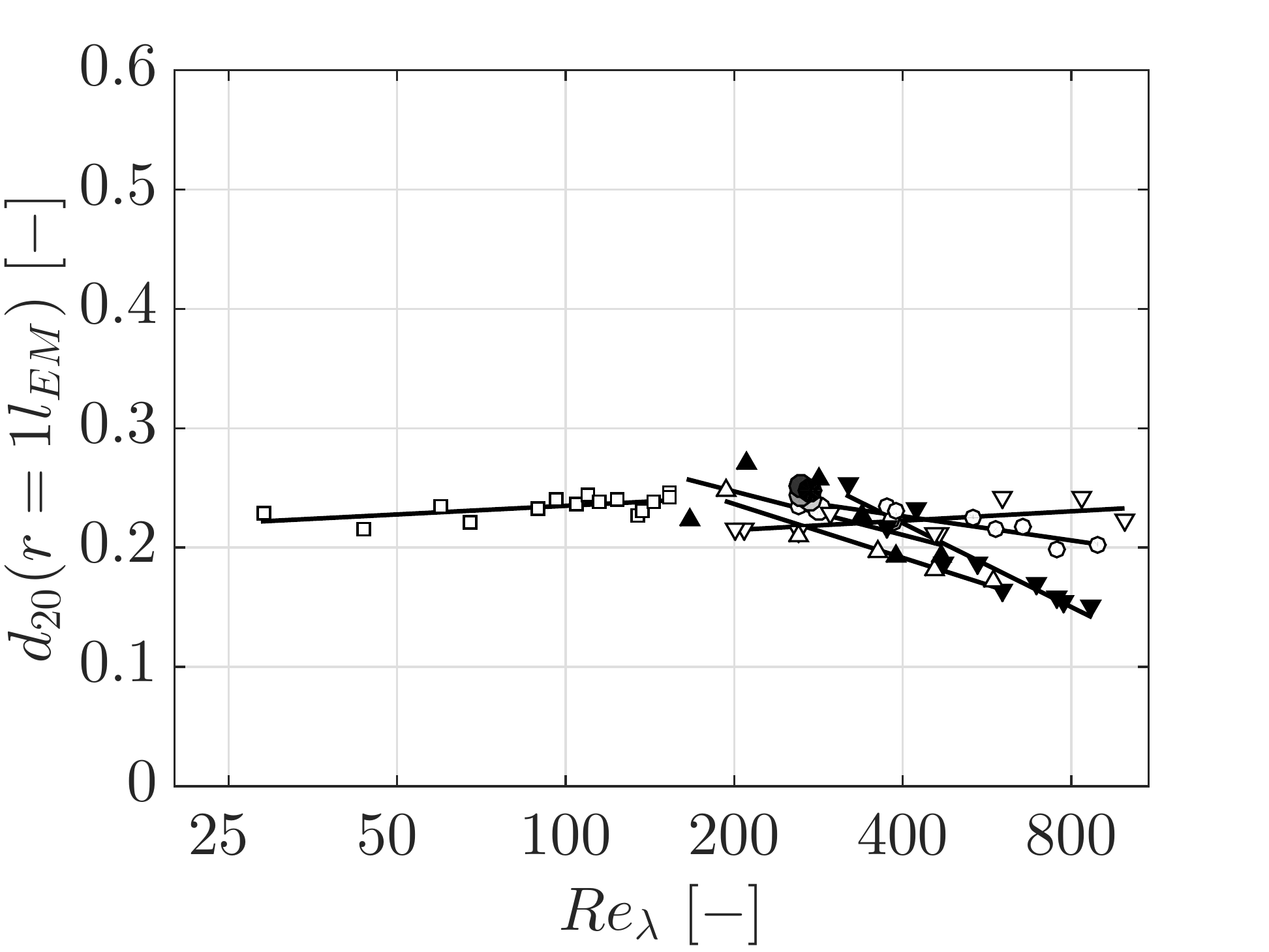}
d)\includegraphics[width=0.45\textwidth]{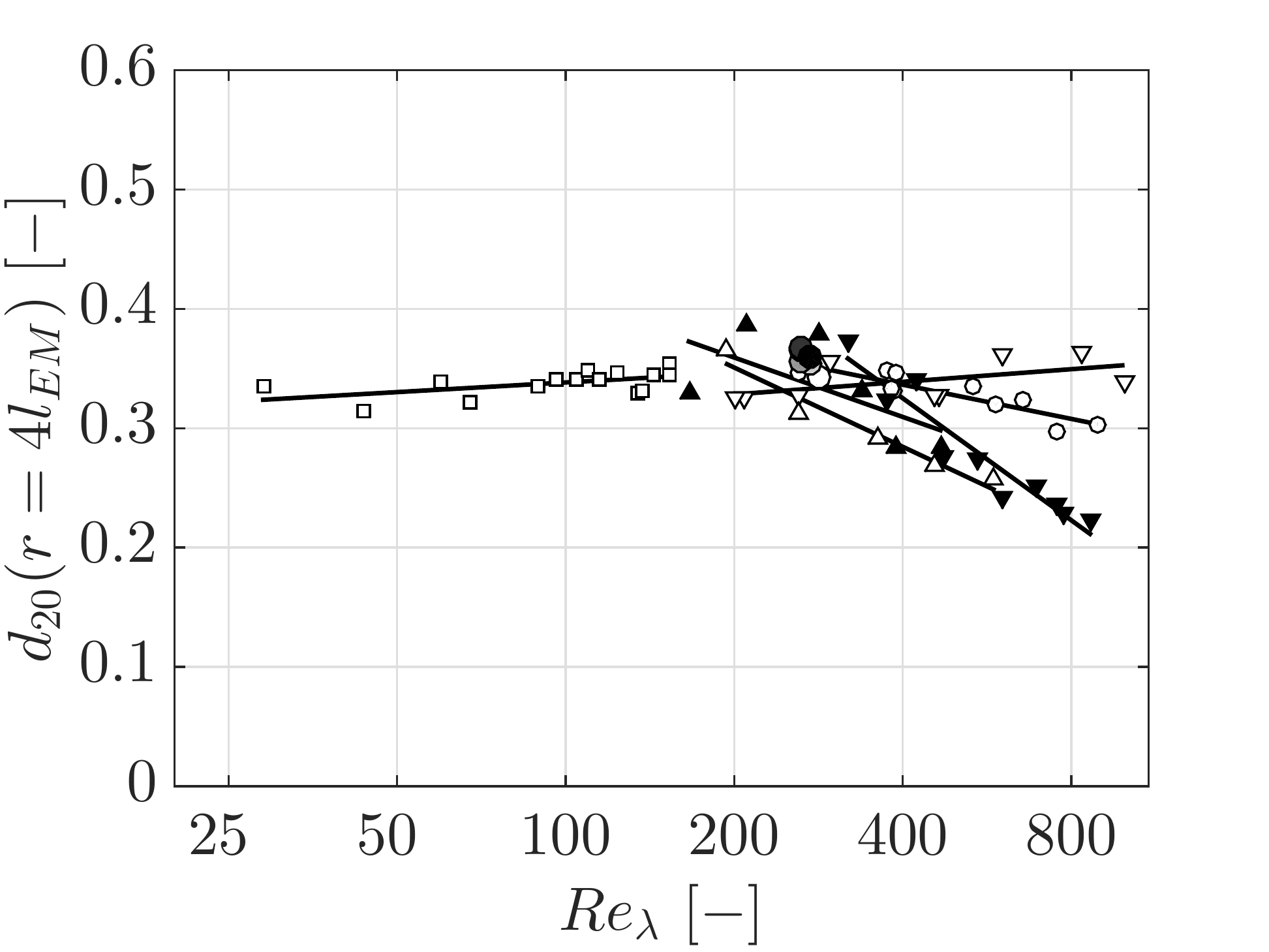}\\
e)\includegraphics[width=0.45\textwidth]{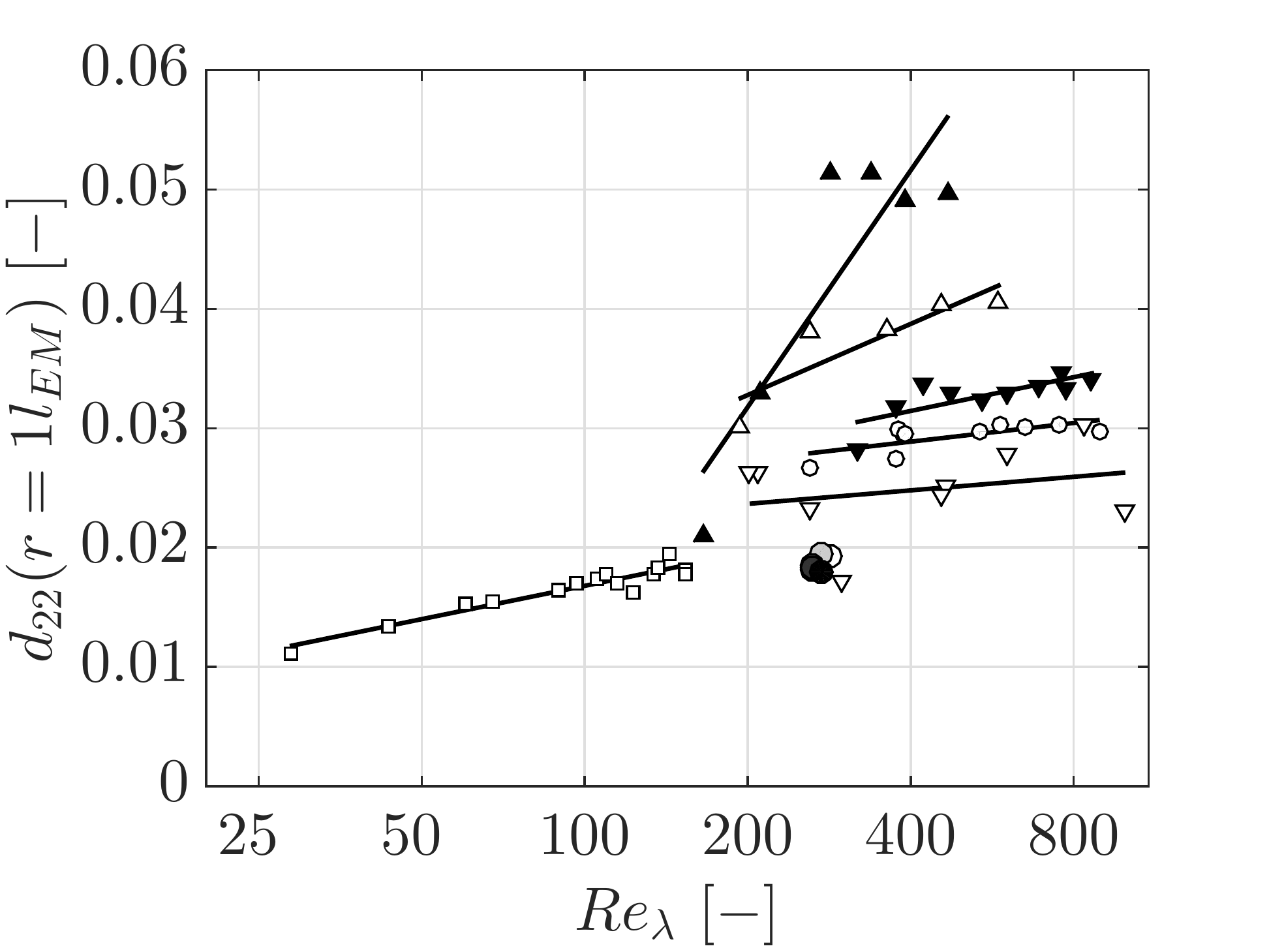}
f)\includegraphics[width=0.45\textwidth]{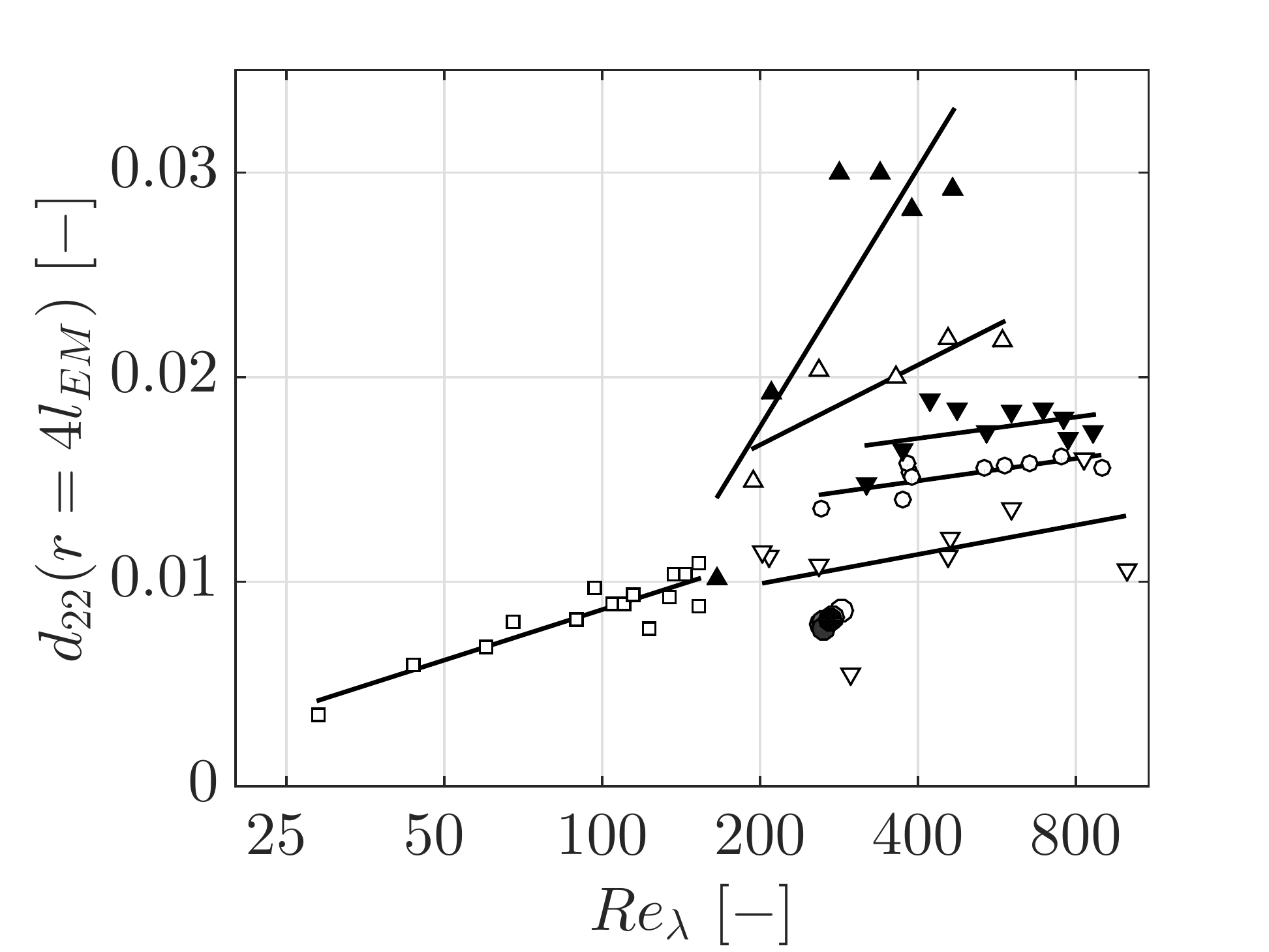}
\caption{Parameters as function of $\Rey_\lambda$ at two fixed scales ($r= l_{EM}$ and $4l_{EM}$).
Additionally, \textit{linear} fit functions indicate trends. 
All plots are semi-logarithmically plotted.
}
\label{fig:6_klein}
\end{figure}

We continue with an analysis regarding the universality of the functional forms of $d_{11}$, $d_{20}$ and $d_{22}$. 
Figure \ref{fig:5} shows the three parameters for all datasets in logarithmic plots. 
The parameters are sorted in three groups which reflect the different turbulent flow types. All parameters which belong to grid flows, to cylinder flows or to free jet flows are covered by a surface with different greyscale and are enclosed by different symbols.  
Dashed lines indicate where edges are overlaid. 
In addition, figure \ref{fig:6_klein} shows the parameters as function of the Taylor-Reynolds number at the fixed scales $r=l_{EM}$ and $r=4l_{EM}$.

At small scales, all $d_{11}$ converge to an approximate power law with the exponent $-0.5$ at $r\gtrsim l_{EM}$, as shown in figure \ref{fig:5} a). 
Therefore, universal scaling features of $d_{11}$ are found for our datasets at small scales, which is confirmed by plotting the $\Rey_{\lambda}$-dependence of $d_{11}(r=l_{EM})$, see figure \ref{fig:6_klein} a).
At larger scales, the three groups broaden differently and start to deviate from the power law behaviour.
%
%
Figure \ref{fig:6_klein} b) indicates that the different behaviour in $d_{11}$ is mainly a $\Rey_{\lambda}$ effect and does not show significant differences for the different flows. More specifically, at small scales $d_{11}$ acquires a universal form, which shows neither a dependence on the kind of turbulent flow nor on $\Rey_\lambda$, and at larger scales, $\Rey_\lambda$ becomes important, but the kind of flow does not have a strong effect on $d_{11}$ features.

Figure \ref{fig:5} b) shows $d_{20}$.
The three groups of flows increase with scale in a similar manner as for $d_{11}$. 
As a further characterisation, we show the $\Rey_{\lambda}$-dependence of $d_{20}$ again for the same two scales in figures \ref{fig:6_klein} c) and d).
The overall trend of the entire datasets is rather constant and subgroups overlap.
Thus, a quite uniform band of data is found.
Note that $d_{20}$ is sensitive to the normalisation of the velocity increment series, eq. (\ref{eq:norm}).
On the one hand, a more sophisticated norm based on $d_{20}$ might result in a perfect collapse of the $d_{20}(r)$, at least at one scale.
On the other hand, a norm $\sigma_\infty = \sqrt{2}\sigma(v)$ would lead to a distinct dependence of $d_{20}$ on $\Rey_\lambda$.
We hence stick to the normalisation using $\Theta$ and $\sigma_{\Theta}$ to deliver a uniform picture in the 2-point and multi-point analysis, which entails the displayed weak dependence of $d_{20}$ on $\Rey_\lambda$, and which therefore supports our choice to use the new quantities $\Theta$ and $\sigma_{\Theta}$ for normalisation.

Regarding $d_{22}$, the turbulent flows divide into distinct groups, as shown in \mbox{figure \ref{fig:5} c).}
These groups are different in shape and broadness,
where, from bottom to top, the functional form of the group's borders changes in a quite 
systematic way.
To investigate whether this behaviour can be explained by a $\Rey_{\lambda}$-dependence or not,
we proceed as above by plotting $d_{22}(r = l_{EM})$ and $d_{22}(r = 4l_{EM})$ as a function of $\Rey_{\lambda}$, see figures \ref{fig:6_klein} e) and f).
In contrast to the findings of $d_{11}$ and $d_{20}$, the data does not collapse on a
uniform $\Rey_{\lambda}$ behaviour, but clearly distinct dependencies can be seen.  
Each dataset seems to follow a different evolution with $\Rey_{\lambda}$. 
Most interestingly, this also holds for the case of small scales $r = l_{EM}$, for which a convergence to a flow type independent behaviour should be expected in the common understanding of universal turbulence. 

We further extend our research and try to combine the various subgroups of $d_{22}$.
Plotting the parameter $d_{22}$ as a function of the dimensionless length $\frac{L}{\lambda}$ instead of $\Rey_{\lambda}$ may allows to combine the subgroups within the flow types, i.e. free jet data becomes more aligned, see figure \ref{fig:6_klein2} in appendix A7. 
However, the different flow types itself cannot be combined to one universal functional form.


\section{Conclusion}
\label{sec:Discussion}
In the common understanding of turbulence, large scale structures, where energy is fed into a cascade process, are known to be dependent on the generating process. 
For small structures, it is commonly believed that they have flow independent features. 
This assumption of universal small scale structures is utilised in turbulence modelling.
It is the subjective of this paper to discuss to what degree such a universal behaviour is present.
Therefore, we performed a comprehensive investigation based on 2-point statistics (i.e. structure functions), multi-scale statistics (i.e. Markov analysis), and the analogue thermodynamic process (i.e. integral fluctuation theorem, IFT), as well as a multitude of 61 different turbulent flows with various kinds of turbulence generation.

Our first analysis step of the extensive dataset confirms the common 2-point statistics' scaling picture of homogeneous isotropic turbulence, cf. \citet{Kolmogorov1962}, figure \ref{fig:sf_2_4}.
Furthermore, a closer investigation of local scaling exponents of structure functions and its features is also set in a universal dependence on $\Rey_\lambda$, which is supported by results of \citet{Warhaft1996}, figure \ref{fig:xi2}.
The shown results do not allow a distinction between the different turbulent flows and no significant clustering with regard to the generation of turbulence could be detected. 
These results and their scattering are in agreement with other studies, e.g. \citet{Gylfason2004}, and confirm the common picture of universal small scale turbulence.

For this traditional scaling analysis of structure functions, we introduced two new quantities, $\Theta$ and $\sigma_{\Theta}$, for the purpose of normalisation of scale and magnitudes of structure functions as in eq. (\ref{eq:norm}). 
As these quantities are dimensional and contain information on large scale structures, we achieved a comparable basis and a collapse of the data of all considered turbulent flows.

After we have put our results in the context of the common understanding of turbulence, we continued with an advanced study which includes joint multi-scale statistics and the utilisation of the IFT as a consequence of the interpretation of the turbulent cascade as a non-equilibrium thermodynamic process.

As a first main result, we were able to show the validity of the IFT for all datasets, eq. (\ref{eq:FT}) and figure \ref{fig:11}. The high quality of the validity of the IFT leads us to propose that the IFT can be taken as a new general law for the turbulent cascade.
Furthermore, utilising the IFT in combination with a self consistent optimisation procedure, we were able to work out the significant functional contributions to the stochastic cascade process, see eqs. (\ref{eq:D1}) and (\ref{eq:D2}), and figure \ref{fig:5_grid}. 
For all data, we see that these contributions are given by three aspects and parameters:

A deterministic part, expressed by $d_{11}(r)$ in eq. (\ref{d11_para}), corresponds to a returning force which pulls increments to the equilibrium state $u=0$ while the cascade proceeds towards smaller scales. Thus, the deterministic part describes the reduction of the magnitude of the increments $u_r$ with decreasing scales and fixes the average tendency of the cascade. 
The functional form of $d_{11}$ in scale follows a power law, where the scaling exponent indicates, 
compared to the K62 model eq. (\ref{eq:se}), a decelerated decay of turbulent structures towards smaller scales.

Stochastic forces are represented by an additive and a multiplicative noise term, given by $d_{20}(r)$ and $d_{22}(r)$ in eqs. (\ref{d20_para}) and (\ref{d22_para}) respectively. 
Both stochastic forces add fluctuations to the deterministic evolution of increments and 
counteract the equilibrium tendency of the deterministic part.
%
However, both are quite different in their significance and nature.
The additive noise term $d_{20}$ adds background fluctuations independent from $u_r$ with a noticeable decrease with decreasing scale.
In contrast, the multiplicative noise term $d_{22}$ incorporates the energetic term $u^2$ as a source of fluctuations, just as the thermal energy of a reservoir is a source of fluctuations in a thermodynamic process. As $d_{22}$ increases with decreasing scale, heavy tails of PDFs grow and extreme events get more likely towards smaller scales.
%
%
%
%
This result is in agreement with the evolution of the energy transfer rate along the cascade, 
which starts off with a non-fluctuating average large value and becomes more and more fluctuating towards smaller scales.
The \textit{increasing} occurrence of extreme events is responsible for the phenomenon of small scale intermittency, and
it is important to note that these fluctuations do not occur haphazardly but are well balanced with average realisations of the cascade process as checked by the IFT.

Our analysis demonstrated in particular that these three contributions are sufficient to also reproduce the measured evolution of skewness along the cascade in compliance with the four-fifth law, cf. figure \ref{struc3}. In other words, due to the absence of a term $d_{21}$, the two stochastic forces do not need to be correlated for the correct inclusion of skewness, contrary to many other stochastic models. Instead, with the additive noise acting predominantly on large scales and the multiplicative noise on small scales, an intuitive fluid dynamic picture of the turbulent cascade is conveyed: The additive noise injects energy on large scales, the linear drift is responsible for the transfer down the cascade, and the multiplicative noise effectively damps the fluctuations (weak noise for small $u$), superimposed by intermittent outbursts (strong noise for large $u$). In that picture, it is the linear drift term $d_{11}$ that is responsible for the correct evolution of initial large scale skewness along the cascade.

The second main result is the reduction of the stochastic model to these three influences and their interpretation. 
Adding to that, the derived stochastic cascade model (based on the three parameters) leads to a closed set of differential equations for structure functions, eq. (\ref{eq:momentsFP2}), which allows a decent reproduction of structure functions, and includes correct deviations from usual scaling behaviour, e.g. \citep{Kolmogorov1962}.

Owing to our finding that the stochastic cascade process of turbulence can be represented by scale dependent parameters $d_{ij}$, we were able to investigated the aspects of universality in detail, figures \ref{fig:5} and \ref{fig:6_klein}.
We found that the evolution in scale of $d_{11}$ as well as $d_{20}$ are universal in the sense that they are flow independent. Thus, their functional forms can be expressed by scale- and/or $\Rey_\lambda$-dependencies.

Our third main result is that in strong contrast to the findings of structure functions and parameters $d_{11}$ and $d_{20}$, the parameter $d_{22}$ as function of $\Rey_\lambda$ splits up into distinct clusters for every considered dataset subgroup, which are characterised by specific large scale flow structures and the turbulence generation mechanism.
Thus, we conclude that specific turbulent flows have their own particular multi-scale cascade, with other words, their own stochastic fingerprint. 
\\
\\
\\
\\
We like to acknowledge the funding from DFG and the cooperations 
as well as discussions with  D. Bastine,  A. Engel,  A. Fuchs, A. Hadjihosseini,  P. Lind, P. Milan, R. Stresing, M. R. R. Tabar and M. W\"achter as well as for the experimental datasets provided by O. Chanal, St. L\"uck and Ch. Renner.


\section{Appendix}

\subsection*{A1: Determination of integral length }
\begin{figure}
  \centering    
a)\includegraphics[width=0.47\textwidth]{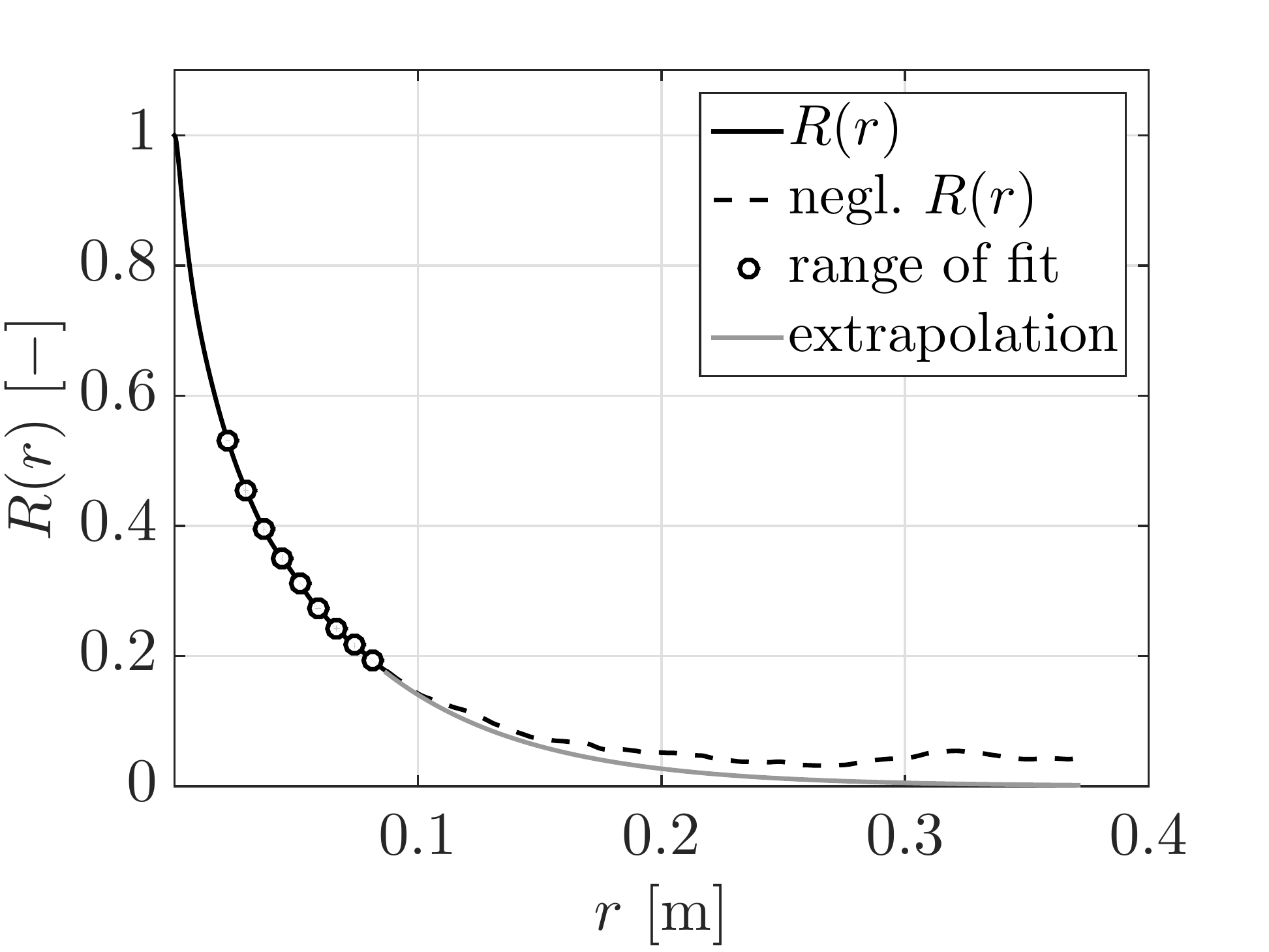}
b)\includegraphics[width=0.47\textwidth]{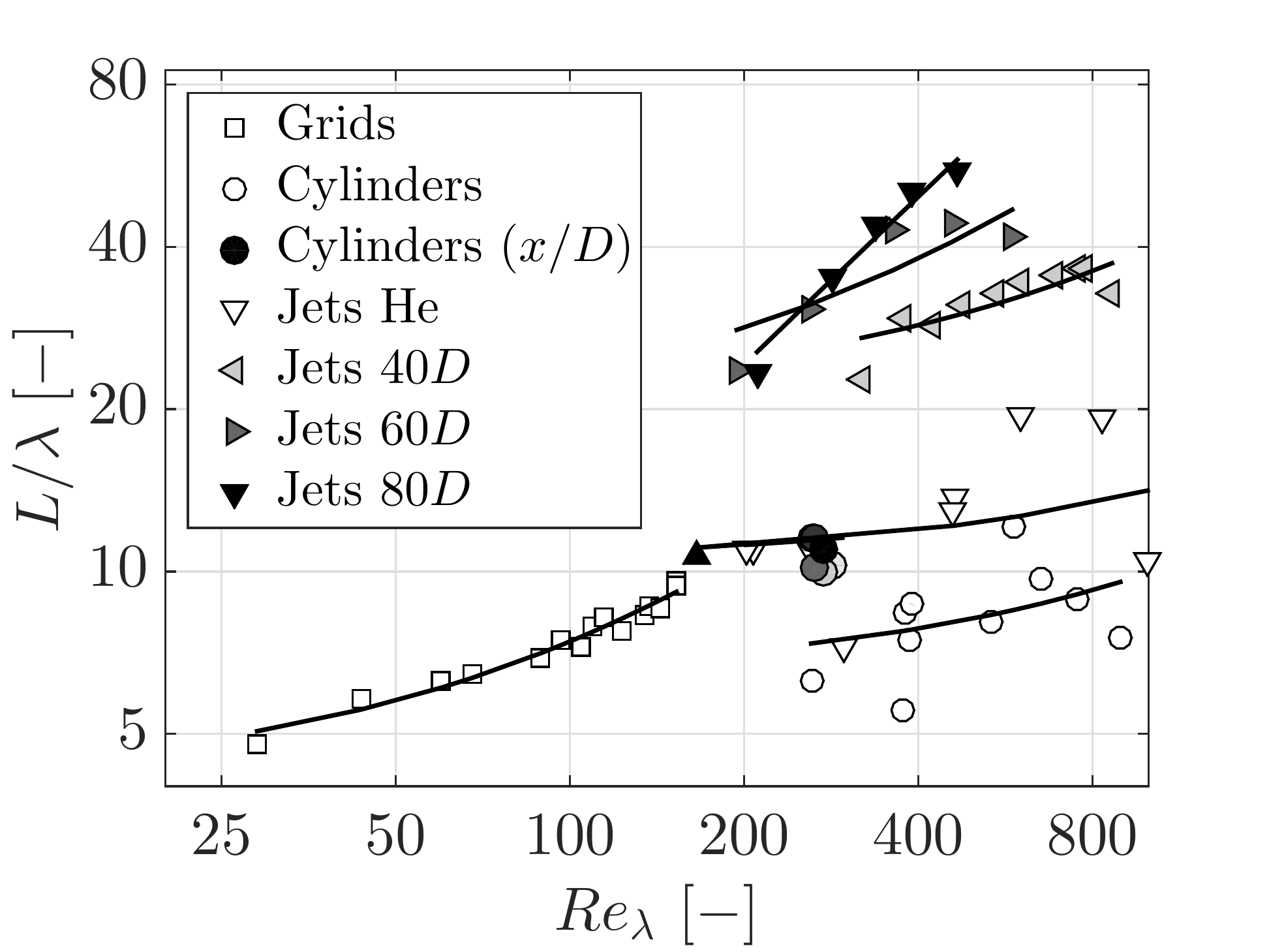}
  \caption{a) Illustration of the procedure of determining the integral length. Shown is the autocorrelation function, the considered part, solid black line and the neglected part, dashed black line, the range of fit, open circles and the extrapolation of $R(r)$, grey solid line.
b) The \textit{length} of the inertial range $\frac{L}{\lambda}$ as function of  $\Rey_\lambda$ for all datasets.
}
\label{fig:L_appendix_0}
\end{figure}

The determination of the integral length is based on the definition in \citet[p. 47]{Batchelor1953}, where the autocorrelation $R(r)$ is integrated from zero to infinity.
From a particle point of view the integration is stopped before infinity, thus, an upper limit $r_{up}$ is needed. For our data we investigate, going from zero to larger scales, if the autocorrelation decreases monotonically with $\partial_r R(r) \le 0$ and $\partial_{rr} R(r) \ge 0$.  For this case, the integration is done up to $R(r)\leq0.01$.
If this is not the case, a $r_{up}$ is defined at the location where 
the monotonous decrease is violated. 
After $r=r_{up}$, the autocorrelation function becomes extrapolated as inspired by  \citet[pp. 92-96, fig. 5.2]{Batchelor1953},
\begin{equation}
\label{eq:def_int_laenge2}
L=\int_{0}^{r_{up}}{R(r)dr} + \frac{e^{a\cdot r_{up} + b}}{a}.
\end{equation}
The parameters $a$ and $b$ are obtained by a fit routine.
The fitting range is between $r_{low}$ and  $r_{up}$. 
Low border $r_{low}$ is chosen as the larger value of $R(r_{low})=\frac{2}{e}$ or $3R(r_{up})$. 
\mbox{Figure \ref{fig:L_appendix_0} a)} exemplarily illustrates this procedure. 
Figure \ref{fig:L_appendix_0} b) shows the integral length $L$ in units of the Taylor microscale $\lambda$ versus the Taylor based Reynolds number. 
For all subgroups of datasets, the ratio $\frac{L}{\lambda}$ increases with $\Rey_\lambda$, as highlighted by \textit{linear} fits. $\frac{L}{\lambda}\propto \Rey_\lambda$ can be derived from $L\approx$ constant, $\lambda\propto Re^{-\frac{1}{2}}$ and $\Rey_\lambda \propto Re^{\frac{1}{2}}$  cf. \citet{Lueck2006}. 
Note, in figure \ref{fig:L_appendix_0} b) a double-logarithmic presentation is used for reasons of clarity. 

\subsection*{A2: Einstein-Markov coherence length}
\begin{figure}
  \centering    
\includegraphics[width=0.6\textwidth]{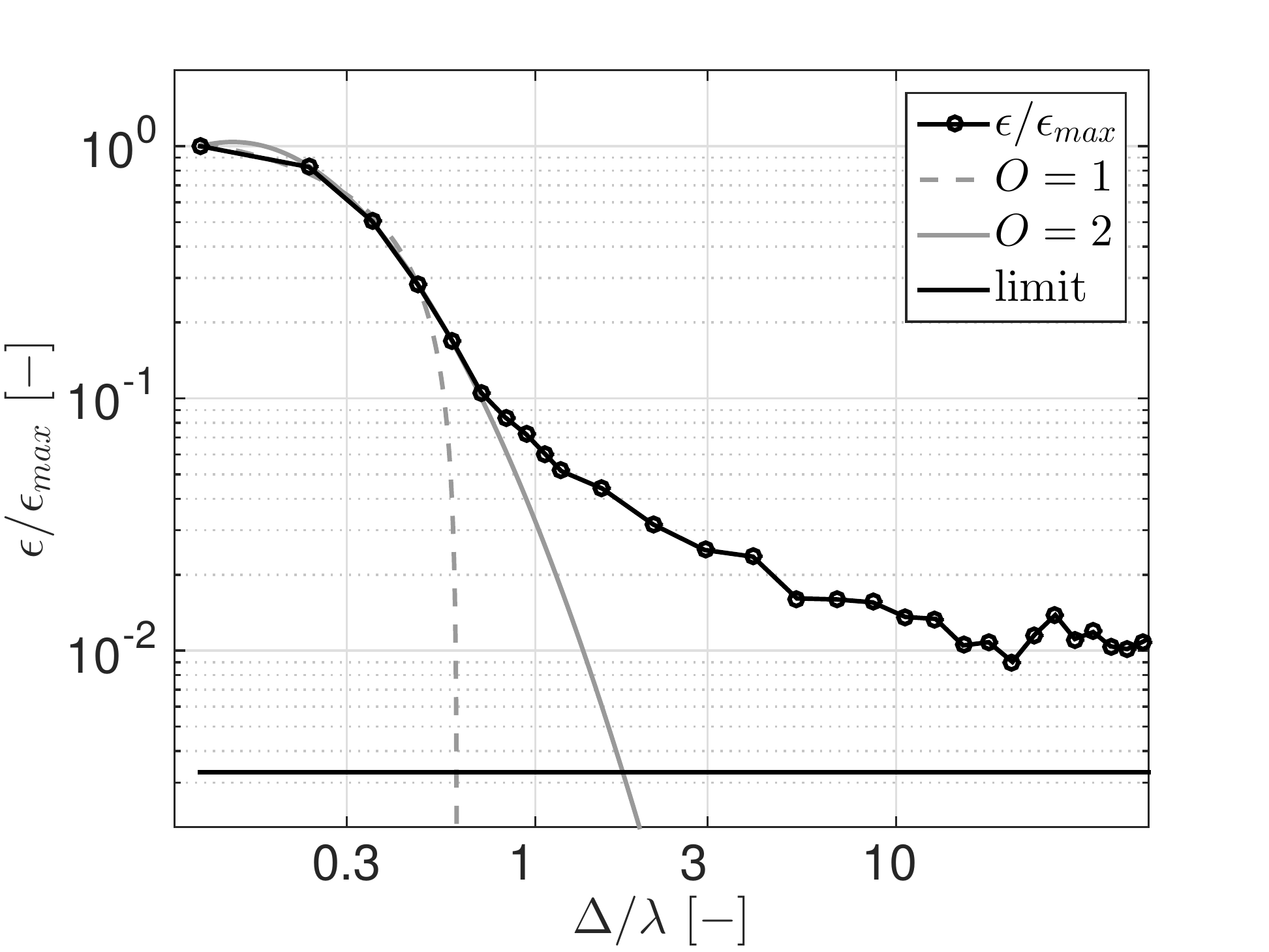}
  \caption{Test of the Markov property by normalised $\epsilon$, according to eq. (\ref{eq:KLE}), with PDFs from the left and right hand side of eq. (\ref{eq:EM_lenght}). The limit indicates the best expected $\epsilon$. Fits of first and second order estimates $l_{EM}$.
}
\label{fig:EM_test}
\end{figure}

The Einstein-Markov coherence length $l_{EM}$ is a length at which 
\begin{eqnarray}
\label{eq:EM_lenght}
p(u_r|u_{r+l_{\Delta}}) &=& p(u_r|u_{r+l_{\Delta}},u_{r+2l_{\Delta}}
=0\pm\chi),\\
\chi&=&0.1\sigma_v
\end{eqnarray}
is fulfilled. \citet{Lueck2006} used the Wilcoxon test to verify this length, which can be considered as a \textit{linear measure}.
Here a test is used which is based on eq. (\ref{eq:KLE}). Since eq. (\ref{eq:KLE}) is a logarithmic measure, rare events are more strongly weighted.
Figure \ref{fig:EM_test} shows an exemplary examination of eq. (\ref{eq:EM_lenght}) under variation of $\Delta$, which is given in units of $\lambda$.
The first steep part of $\epsilon$ is extrapolate to estimate $l_{EM}$.
A first order polynomial ($O=1$) leads to a similar result as published by \citet{Lueck2006} and a second order polynomial ($O=2$) leads to a higher estimation of $l_{EM}\approx 2\lambda$. 
For both estimations ($O=1$ and $O=2$) of $l_{EM}$ a non-Markovian share remain. 
Note, the relation between $l_{EM}$ and $\lambda$  slightly changes the IFT convergence, $l_{EM}=\lambda \ \rightarrow \langle I_{max} \rangle \approx 1.02$ and $l_{EM}=2\lambda \ \rightarrow \langle I_{max} \rangle \approx 1.01$.

\subsection*{A3: Characterisation of norms}
\begin{figure}
  \centering    
  \vspace{0.2cm}
\includegraphics[width=0.7\textwidth]{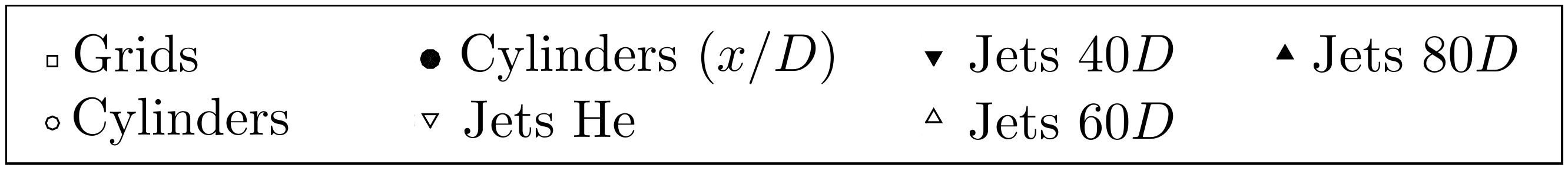}
\vspace{0.1cm}
\\
a)\includegraphics[width=0.47\textwidth]{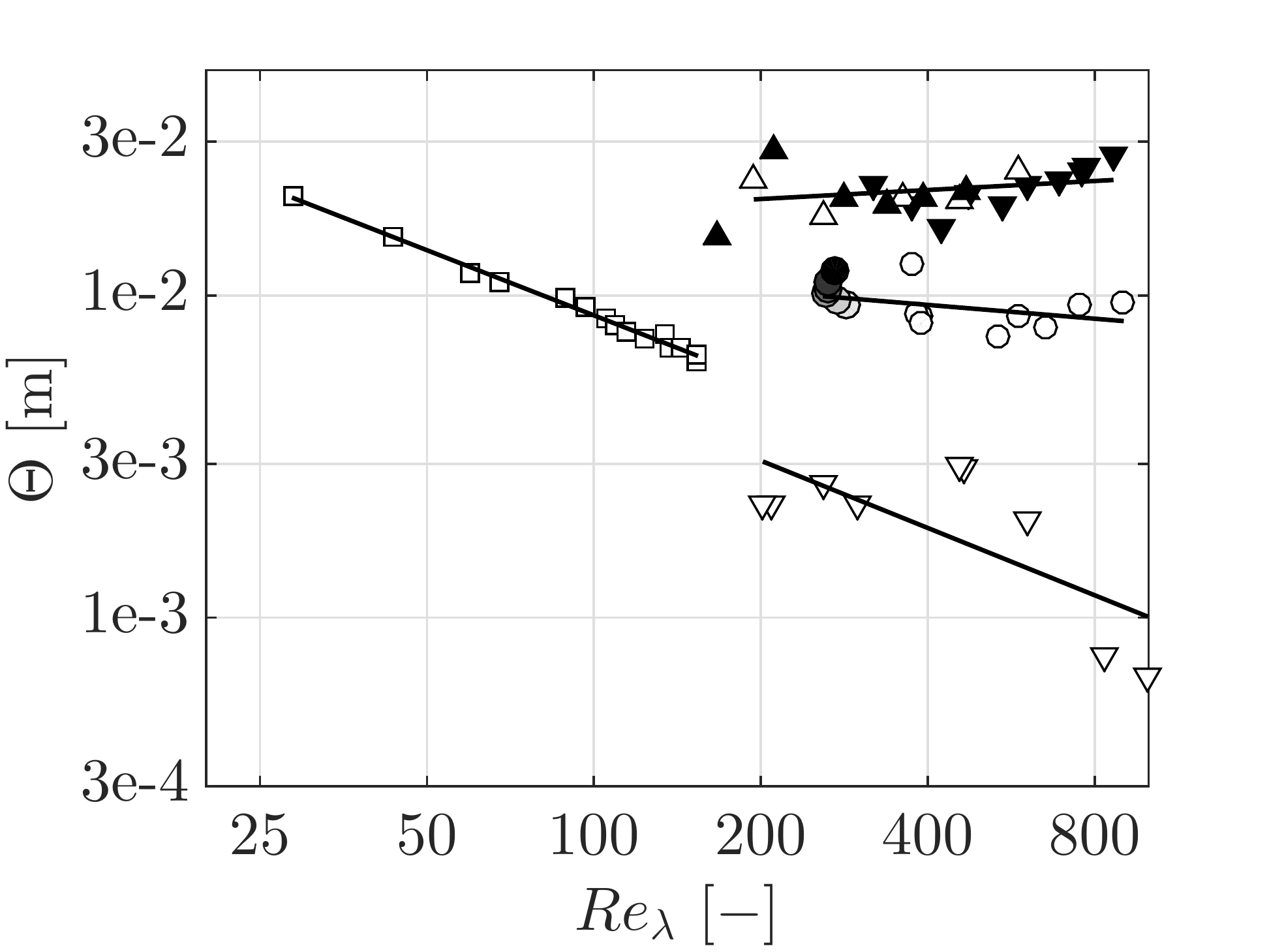}
b)\includegraphics[width=0.47\textwidth]{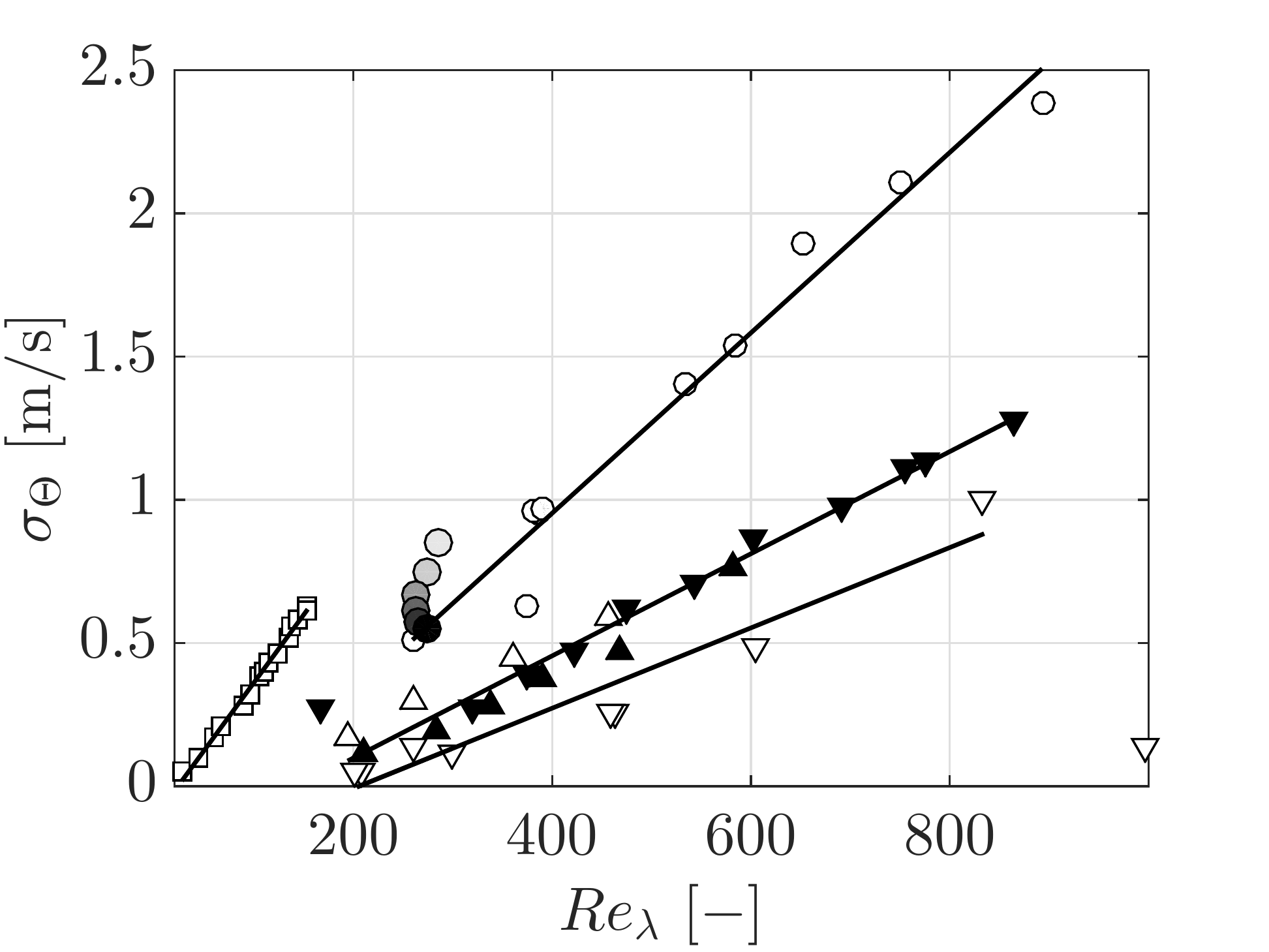}
  \caption{a) shows the norm $\Theta$ versus $\Rey_\lambda$ in a double-logarithmic plot. b) shows the norm $\sigma_{\Theta}$ versus $\Rey_\lambda$. Both norms are shown for all datasets. Trends are highlighted by linear fit functions.}
\label{fig:norm_appendix_1}
\end{figure}

Figure \ref{fig:norm_appendix_1} shows the dimensional norms $\Theta$ and $\sigma_{\Theta}$ as function of $\Rey_\lambda$.
$\Theta$ is plotted in a double-logarithmic fashion, thereby a power law dependence from $\Rey_\lambda$ becomes evident. 
$\sigma_{\Theta}$ depends linearly on $\Rey_\lambda$, figure \ref{fig:norm_appendix_1} b). 
For both quantities the different generated flows show up in individual clusters, which are, interestingly, arranged in different ways.
Although, $\Theta$ and $\sigma_{\Theta}$ are connected by their definition, the different arrangement and the different functional dependence on $\Rey_\lambda$ show that they are not trivially linked to each other. The ratio $\frac{\Theta}{\lambda}$ (e.g. as function of $\Rey_\lambda$) shows a rather scattered distribution, values are between $2$ and $\sim4$, however, some ratios go up to $\sim7$. 

\subsection*{A4: Structure functions deduced by $D^{(1,2)}$ }
\begin{figure}
\centering  
a)\includegraphics[width=0.6\textwidth]{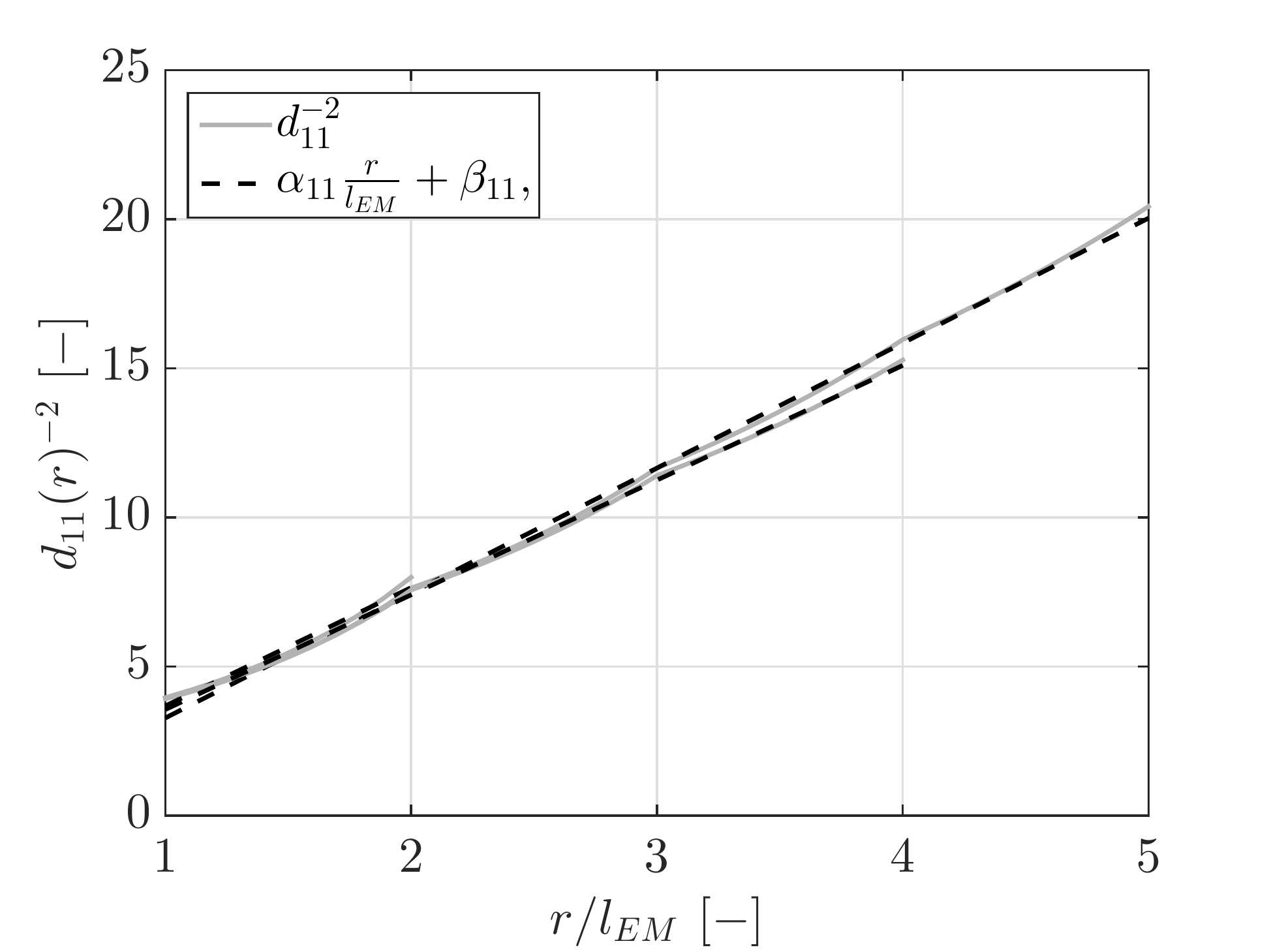}
b)\includegraphics[width=0.6\textwidth]{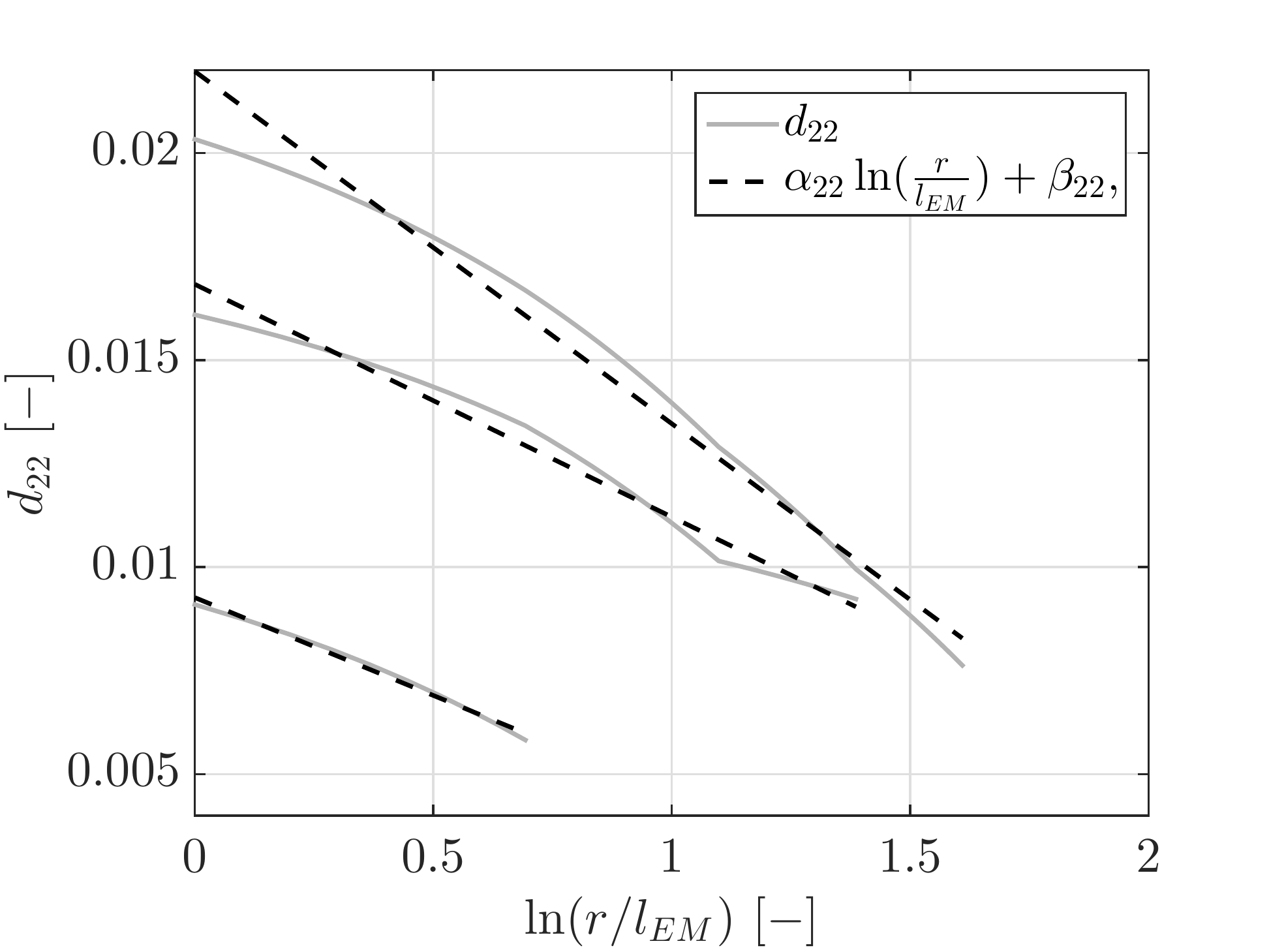}
c)\includegraphics[width=0.6\textwidth]{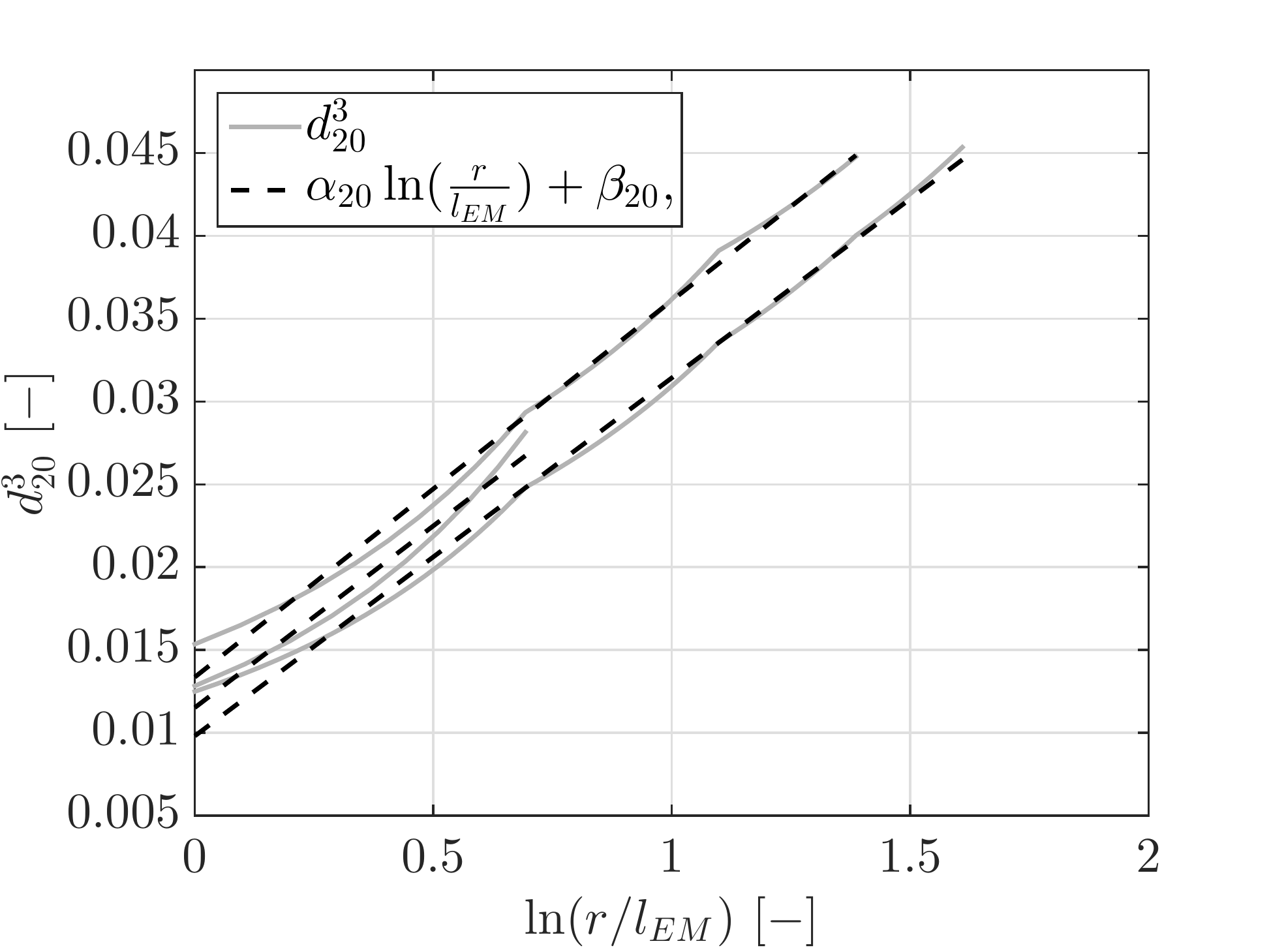}
\caption{$d_{ij}$ in a linearised form as well as fit functions. Length of $d_{ij}$ corresponds to $Re_\lambda$, see table \ref{table:dij_parameter}.}
\label{fig:fits}
\end{figure}

\begin{figure}
\centering  
a)\includegraphics[width=0.6\textwidth]{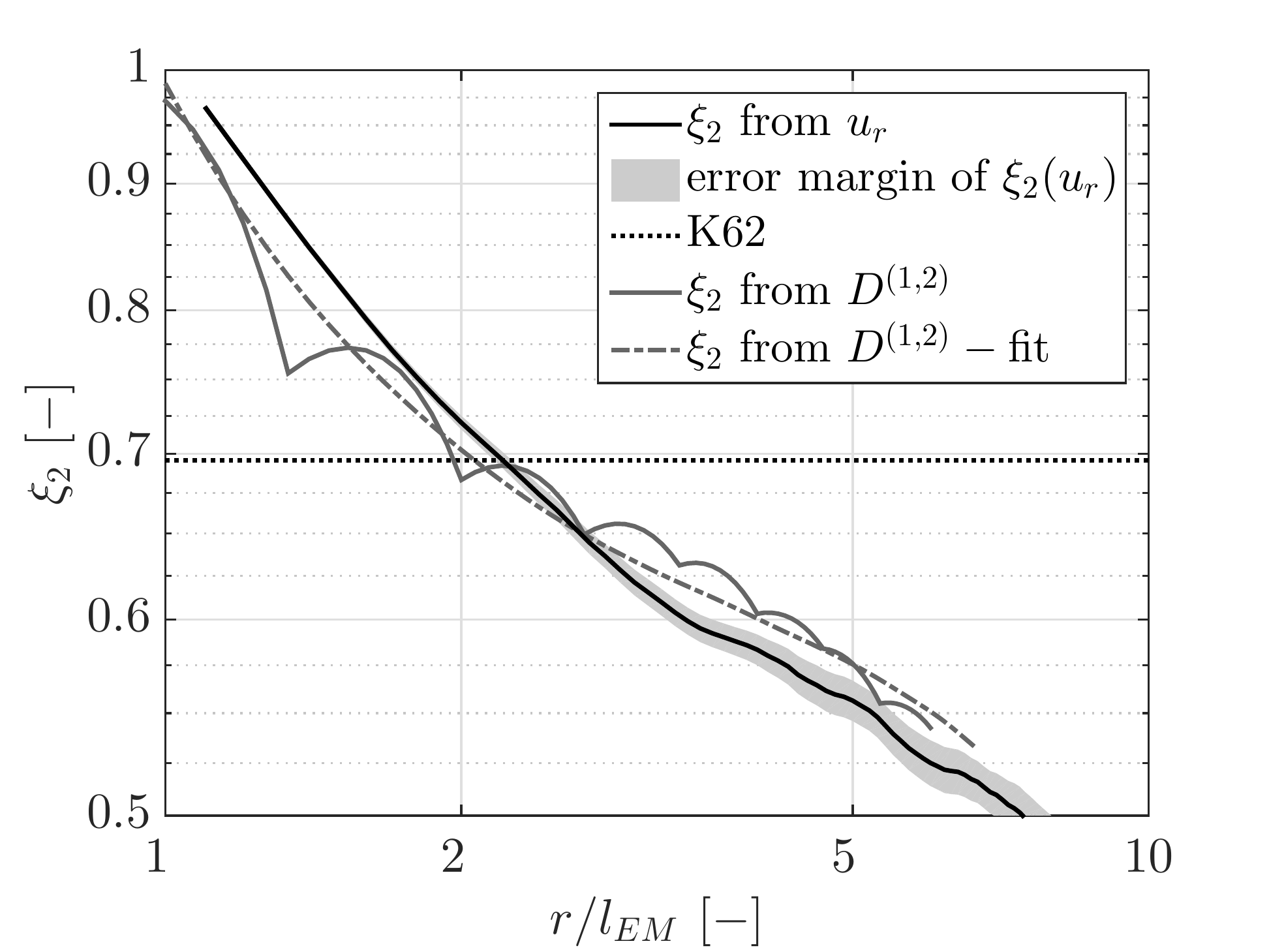}
b)\includegraphics[width=0.6\textwidth]{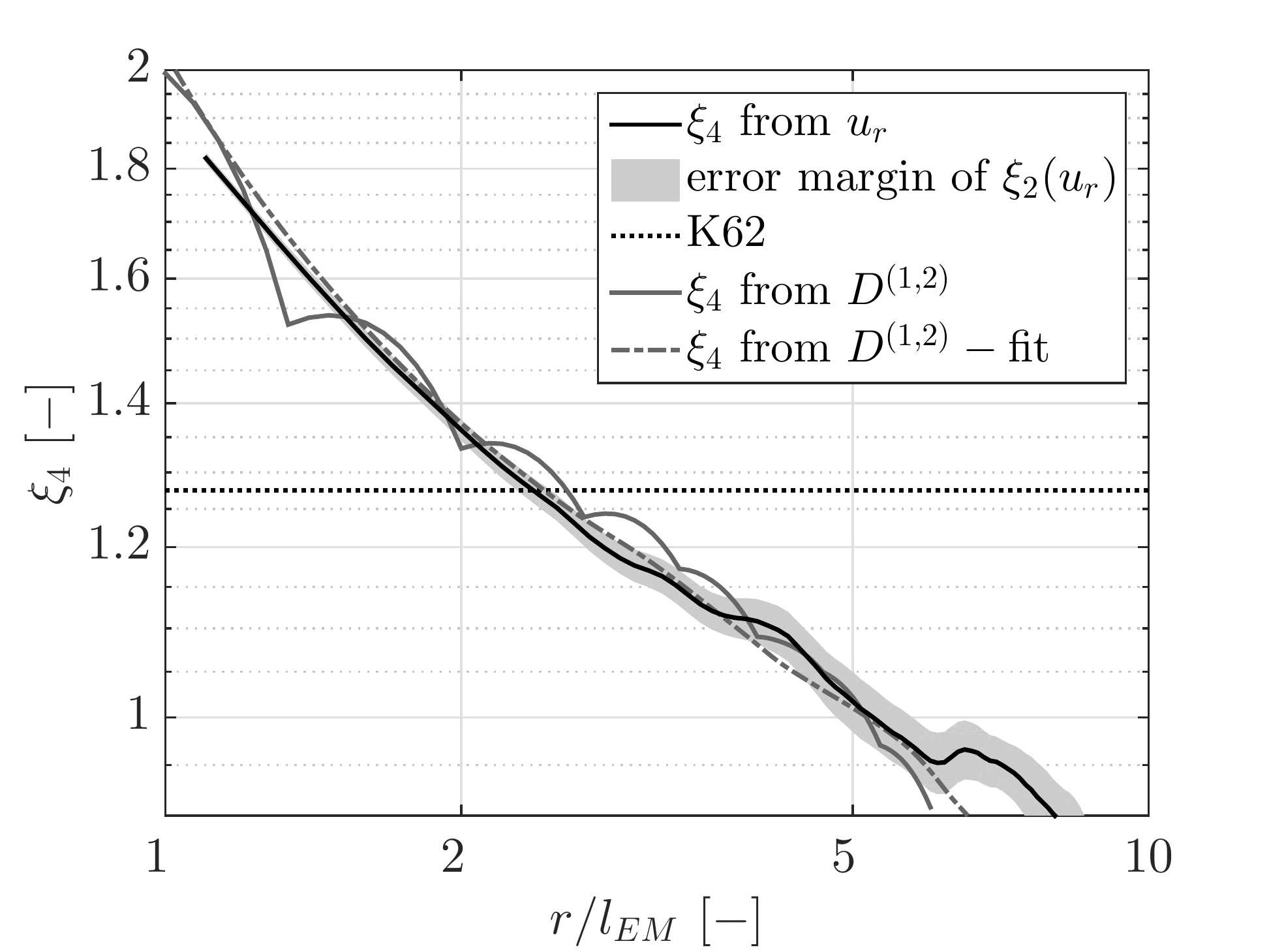}
c)\includegraphics[width=0.6\textwidth]{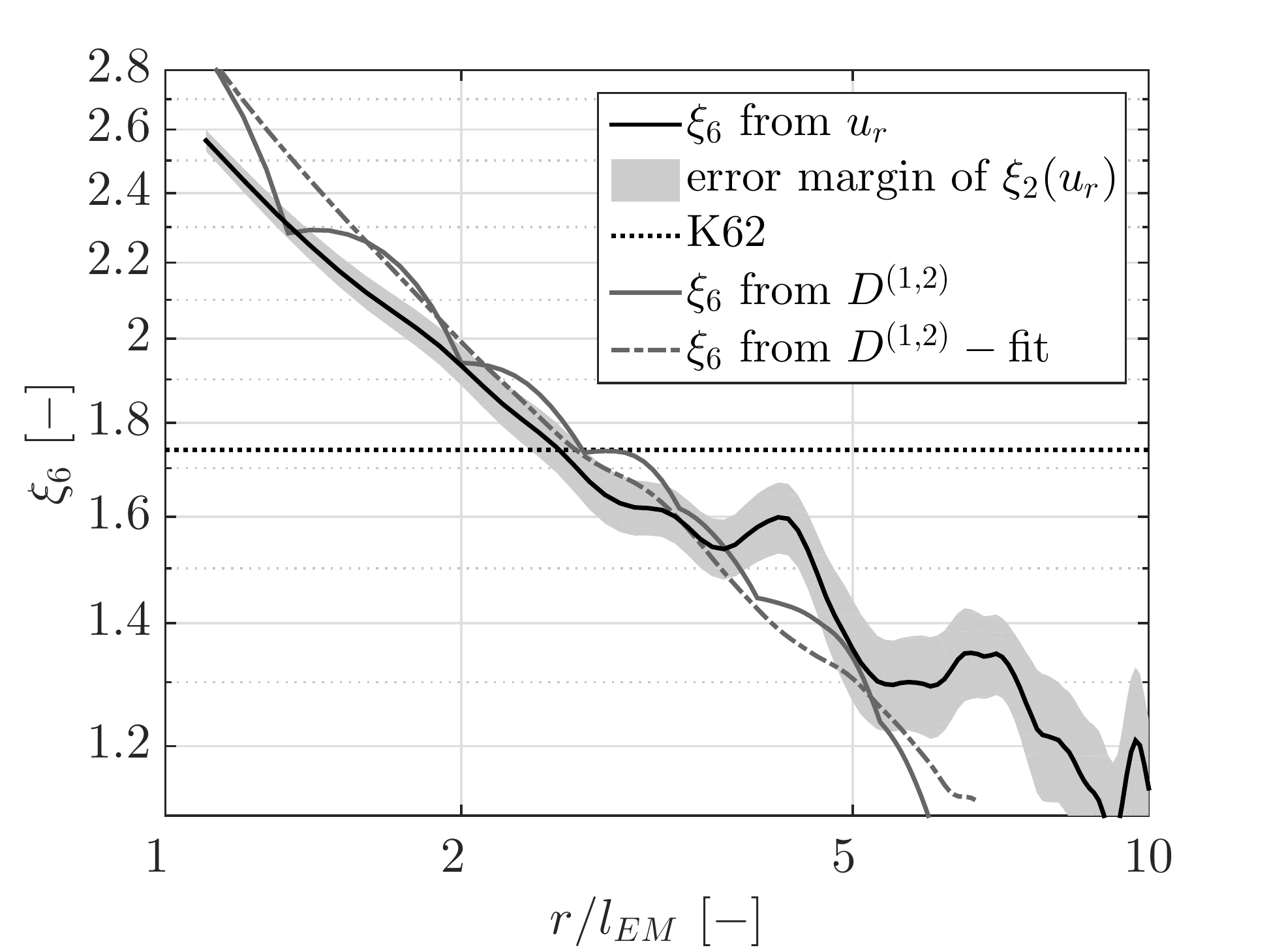}
\caption{Local scaling exponent $\xi_\kappa(r)$ deduced from experimental data, reconstructed by $D^{(1,2)}$ as well as their fits. Also Kolmogorov's prediction is shown (K62) and an error margin of the experimental data. Grid (i) is used as exemplary dataset, characterised in table \ref{table:data} and \ref{table:dij_parameter}.  }
\label{fig:structure_D12}
\end{figure}

Eq. (\ref{eq:momentsFP2}) allows to calculate the structure functions and $\xi_{\kappa}(r)$ from the parameters $d_{ij}$. \mbox{Figure \ref{fig:fits}} shows our approaches of proper parametrisations according to $d_{11}=(\alpha_{11}\frac{r}{l_{EM}}+\beta_{11})^{-\frac{1}{2}}$, $d_{22}=\alpha_{22}\ln(\frac{r}{l_{EM}})+\beta_{22}$ and $d_{20}=(\alpha_{20}\ln(\frac{r}{l_{EM}})+\beta_{20})^{\frac{1}{3}}$. For this illustration three different datasets are used. The  $Re_\lambda$-values and the
corresponding fit parameters $\alpha_{ij}$ and $\beta_{ij}$ as well as the flow type are summarised in table \ref{table:dij_parameter}. 
\begin{table}
\centering
  \begin{tabular}{l|c|cc|cc|cc} 
Type  &  $Re_\lambda$ &  $\alpha_{11}$ &  $\beta_{11}$ &  $100\alpha_{22}$ &  $100\beta_{22}$ &  $100\alpha_{20}$ &  $100\beta_{20}$  \\
  \hline
grid            & 28              & 3.85               & -0.29               & -0.47              &  0.93               & 2.19               & 1.15              \\
grid (i)        & 153            & 3.97               & -0.29              &  -0.56             & 1.68              & 2.27               & 1.34               \\
free jet (i) & 166           & 4.19               & -0.91               & -0.85               & 2.2              & 2.16              & 0.98               \\
\end{tabular}
\caption{$d_{ij}$ parametrisation in terms of $\alpha_{ij}$ and $\beta_{ij}$, see also table \ref{table:data}.}
\label{table:dij_parameter}
\end{table}
\\
The local scaling exponents are shown in figure \ref{fig:structure_D12}, which are deduced directly from the velocity increments and from the parameters using eq. (\ref{eq:momentsFP2}). The values of the optimised $D^{(1,2)}$ as well as the mentioned fit functions of parameters $d_{ij}$ are used.
Here - as an exception - with $l_{EM}=\lambda$ to highlight the crossing of K62.
Note, the \textit{jumpy development of $\xi_\kappa$} is caused by a linear interpolation of $D^{(1,2)}$.
An error margin $\Delta \xi_\kappa(r)$ of $\xi_\kappa(r)$ is derived from the experimental data. This is done by splitting the measured data in $n$-groups and determining $n$-time $\xi_\kappa(r)$, from which the standard deviation $\sigma( \xi_\kappa^n(r))$ is calculated. In this way ($n=50$)
\begin{eqnarray} 
\label{eq:Fehler_SF}
\Delta \xi_\kappa(r) = \frac{\sigma( \xi_\kappa^n(r))}{\sqrt{n}}
\end{eqnarray}
is determined.
As an orientation also Kolmogorov's prediction according to eq. (\ref{eq:se}) is shown, labelled K62, likewise to figure \mbox{\ref{fig:xi2} a)}. 

\subsection*{A5: Interpretation of $D^{(1)}$ and $D^{(2)}$ with the short time propagator}
\begin{figure}
\centering  
a)\includegraphics[width=0.47\textwidth]{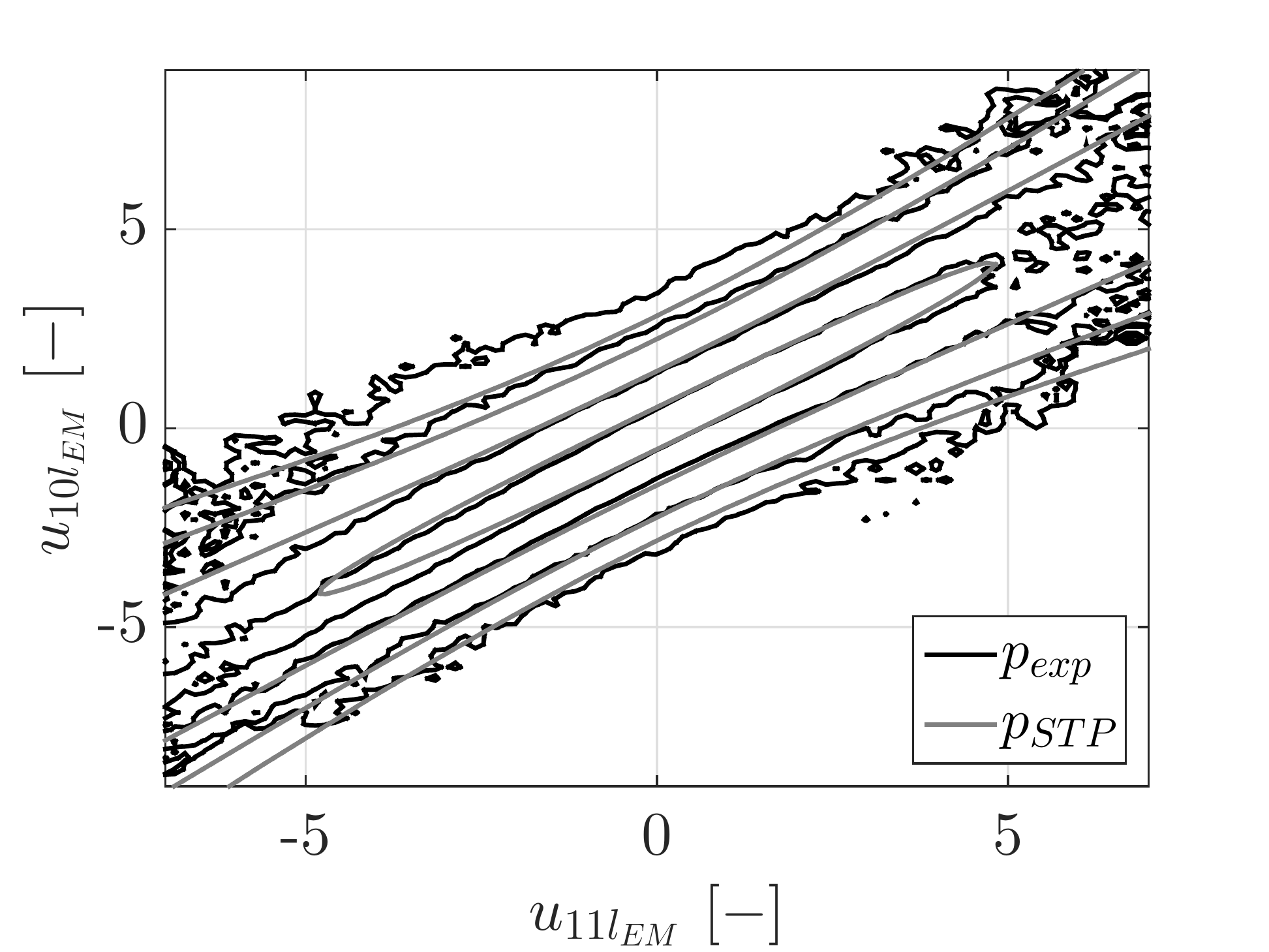}
b)\includegraphics[width=0.47\textwidth]{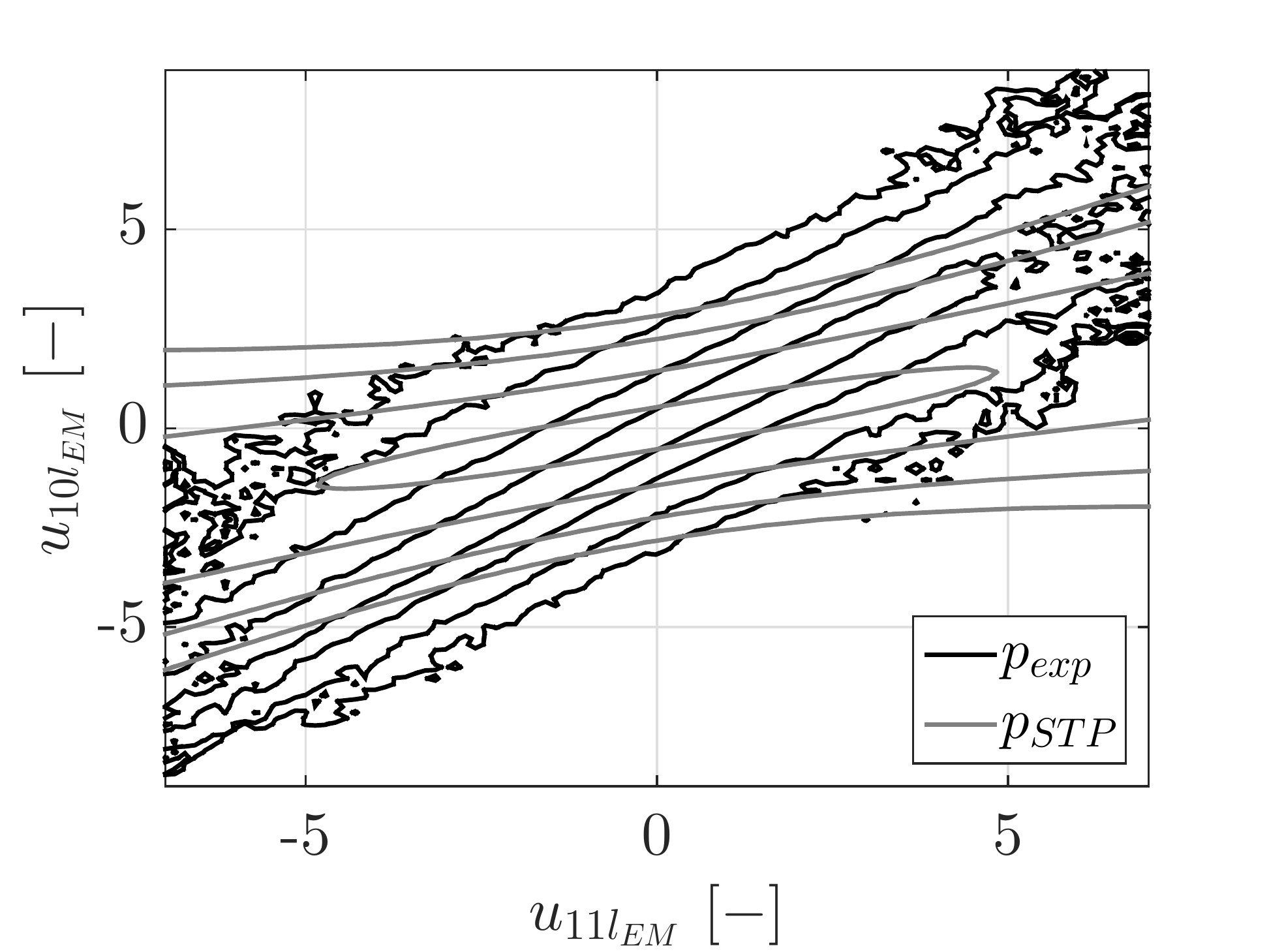}
c)\includegraphics[width=0.47\textwidth]{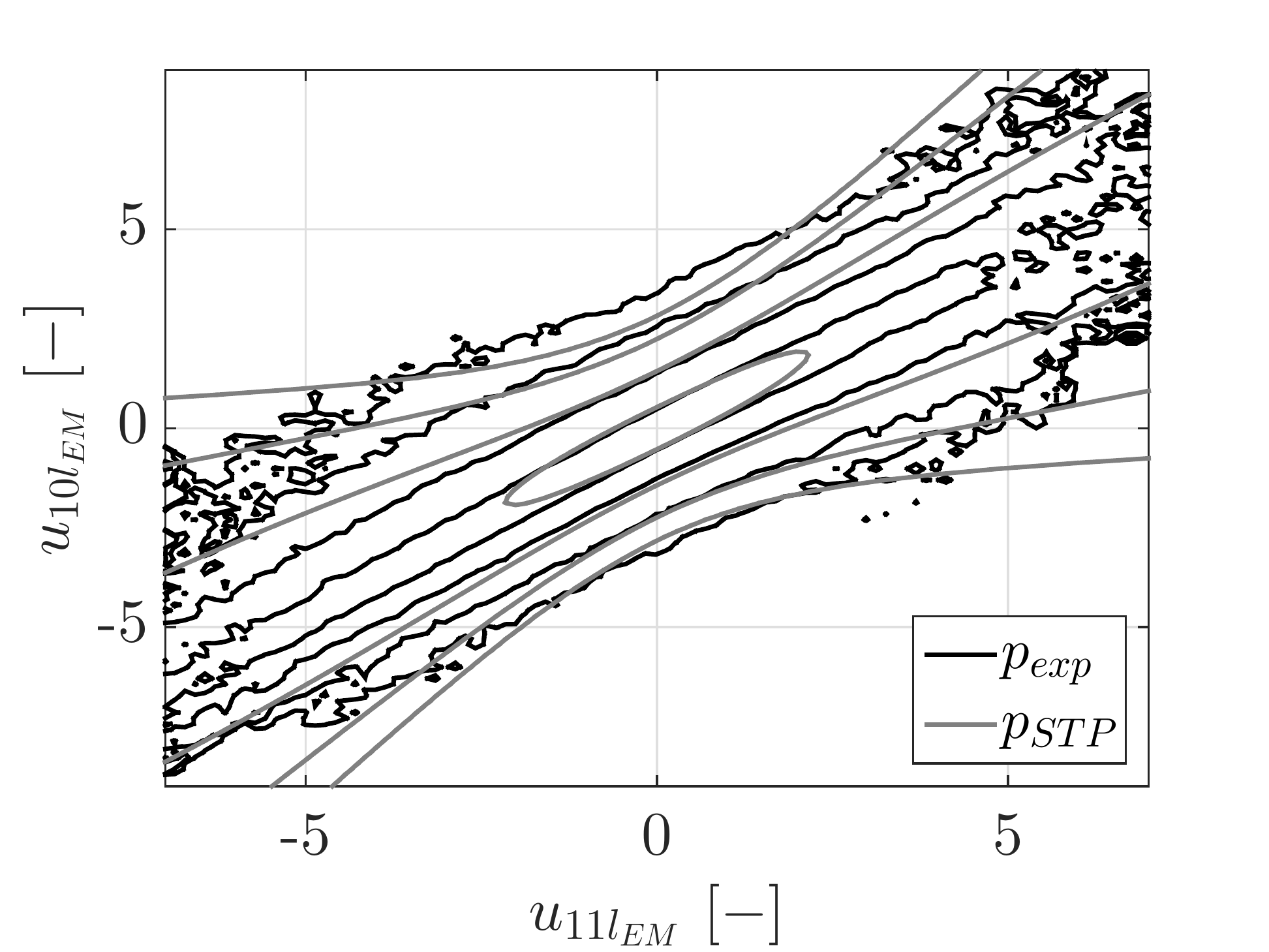}
d)\includegraphics[width=0.47\textwidth]{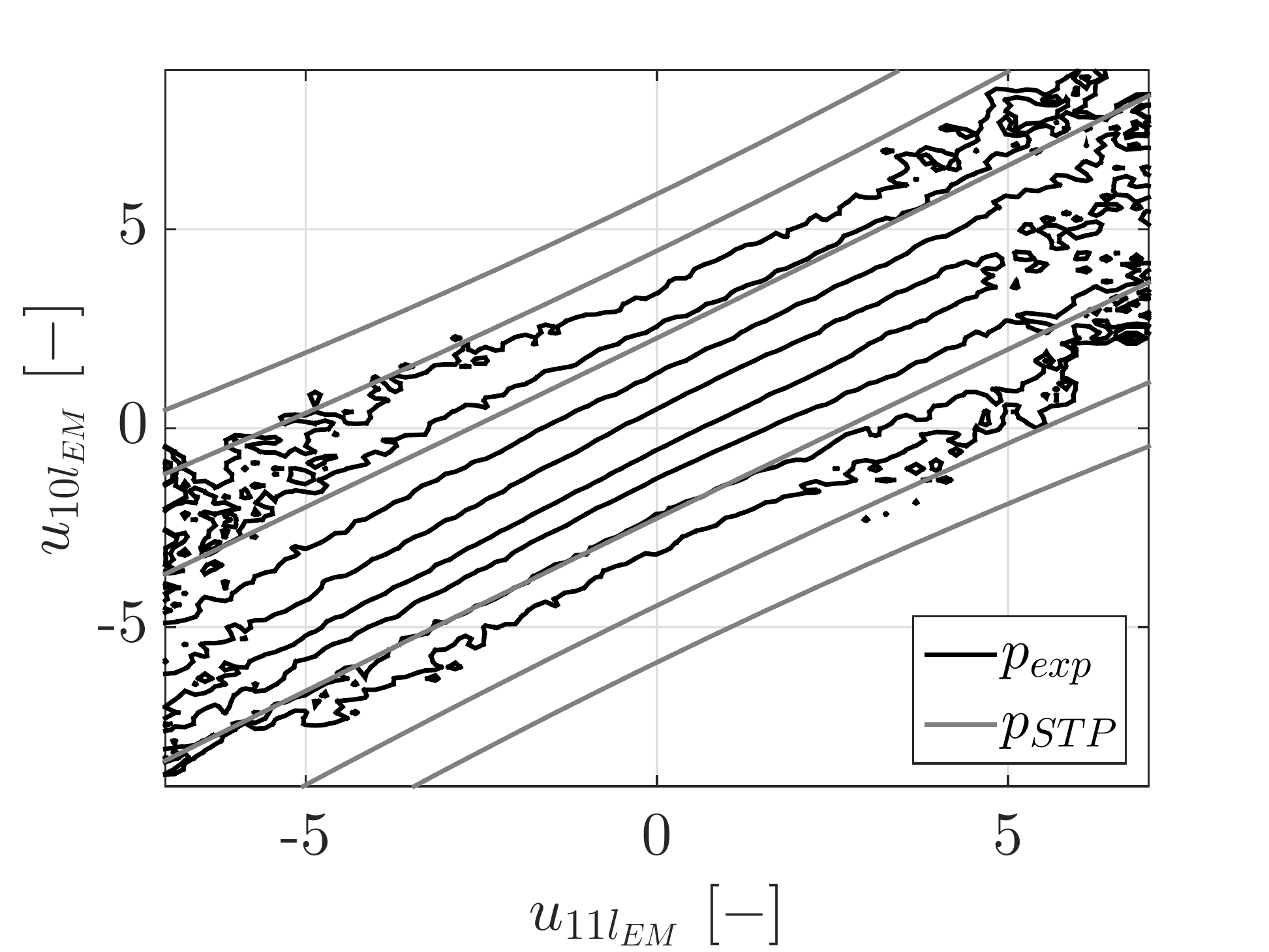}
\caption{One exemplary experimentally found conditional PDF, $p_{exp}$ (black) (subgroup (v), $\Rey_\lambda=475$) and four different results of the short time propagator $p_{STP}$ (grey). 
Figure a) shows the short time propagator regarding to optimised $D^{(1,2)}$. 
In figure b) the same is shown as in a) except that $d_{11}$ is multiplied by 5.
In figure c) only $d_{22}$ is multiplied by 5 and in \mbox{figure d)} only $d_{20}$ is multiplied by 5.}
\label{fig:STP_bsp2}
\end{figure}

The interpretation of $D^{(1,2)}$ and $d_{11}$, $d_{22}$, $d_{20}$ is done by modifying optimised $D^{(1,2)}$ and presenting the corresponding short time propagator.
The short time propagator deduced by unmodified $d_{11}$, $d_{22}$ and $d_{20}$ (regular optimised) is shown in figure \ref{fig:STP_bsp2} a). 
The agreement is good  between short time propagator and the experimental reference data.
In figure \ref{fig:STP_bsp2} b) $d_{11}$ is multiplied by 5, whereas $d_{22}$ and $d_{20}$ are unmodified.
This multiplication twists the diagonal of short time propagator in comparison to the experimental reference data. 
Thus, we see that the magnitude of $d_{11}$ controls the diagonal.
In figure \ref{fig:STP_bsp2} c) $d_{22}$ is multiplied by 5, whereas $d_{11}$ and $d_{20}$ are unmodified.
An enlarged $d_{22}$ increases the curvature of the isolines in the contour plot. Therefore, the probability for extreme events increases, which is equivalent to an increased intermittency and to more fat tailed unconditional PDFs.
In figure \ref{fig:STP_bsp2} d) $d_{20}$ is multiplied by 5, whereas $d_{11}$ and $d_{22}$ are unmodified.
Here the reconstructed distribution gets broader around the distribution diagonal.

\subsection*{A6: Single developments of $d_{11}$, $d_{22}$ and $d_{20}$ in scale}
\begin{figure}
  \centering    
a)\includegraphics[width=0.47\textwidth]{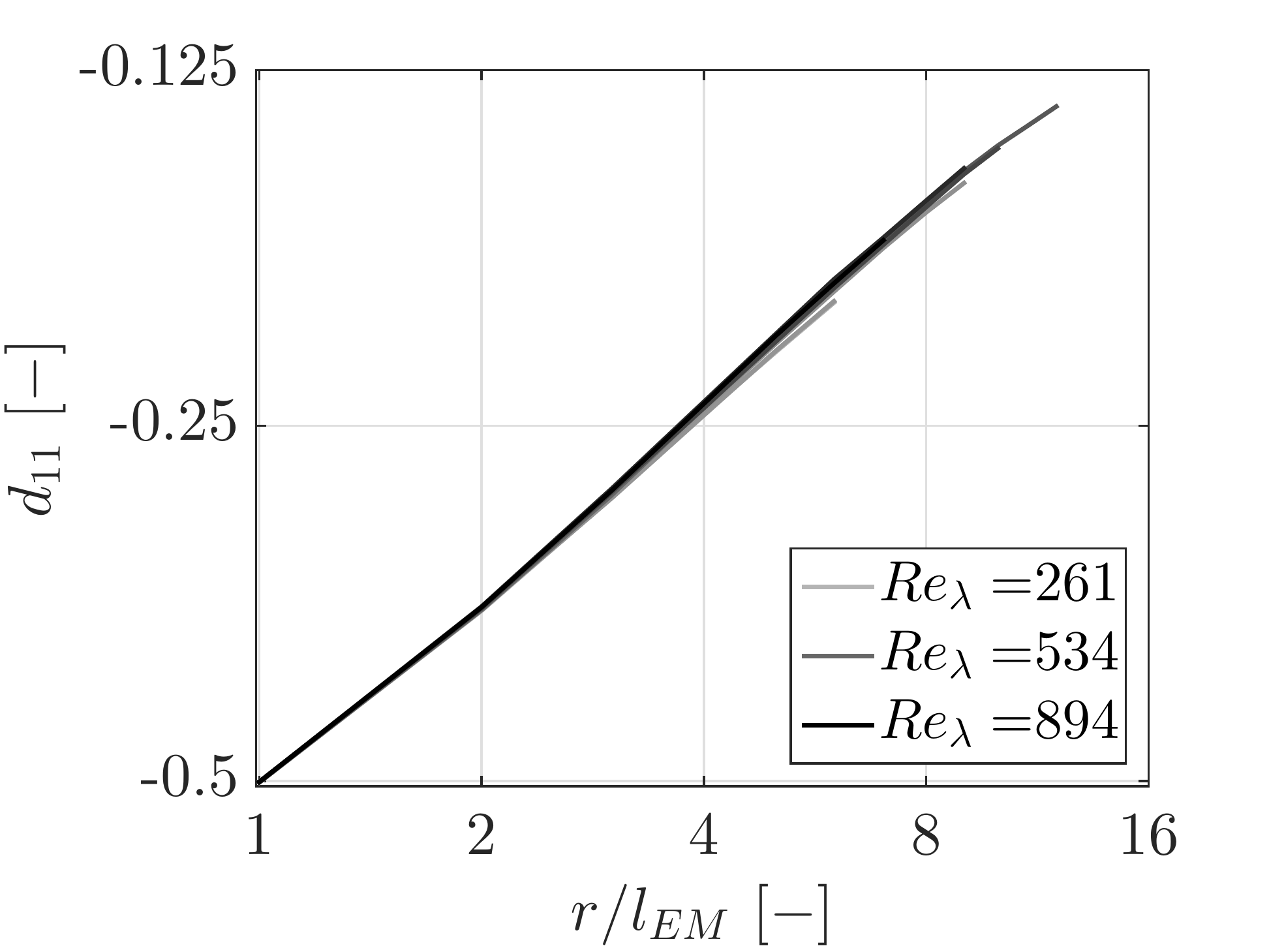}
b)\includegraphics[width=0.47\textwidth]{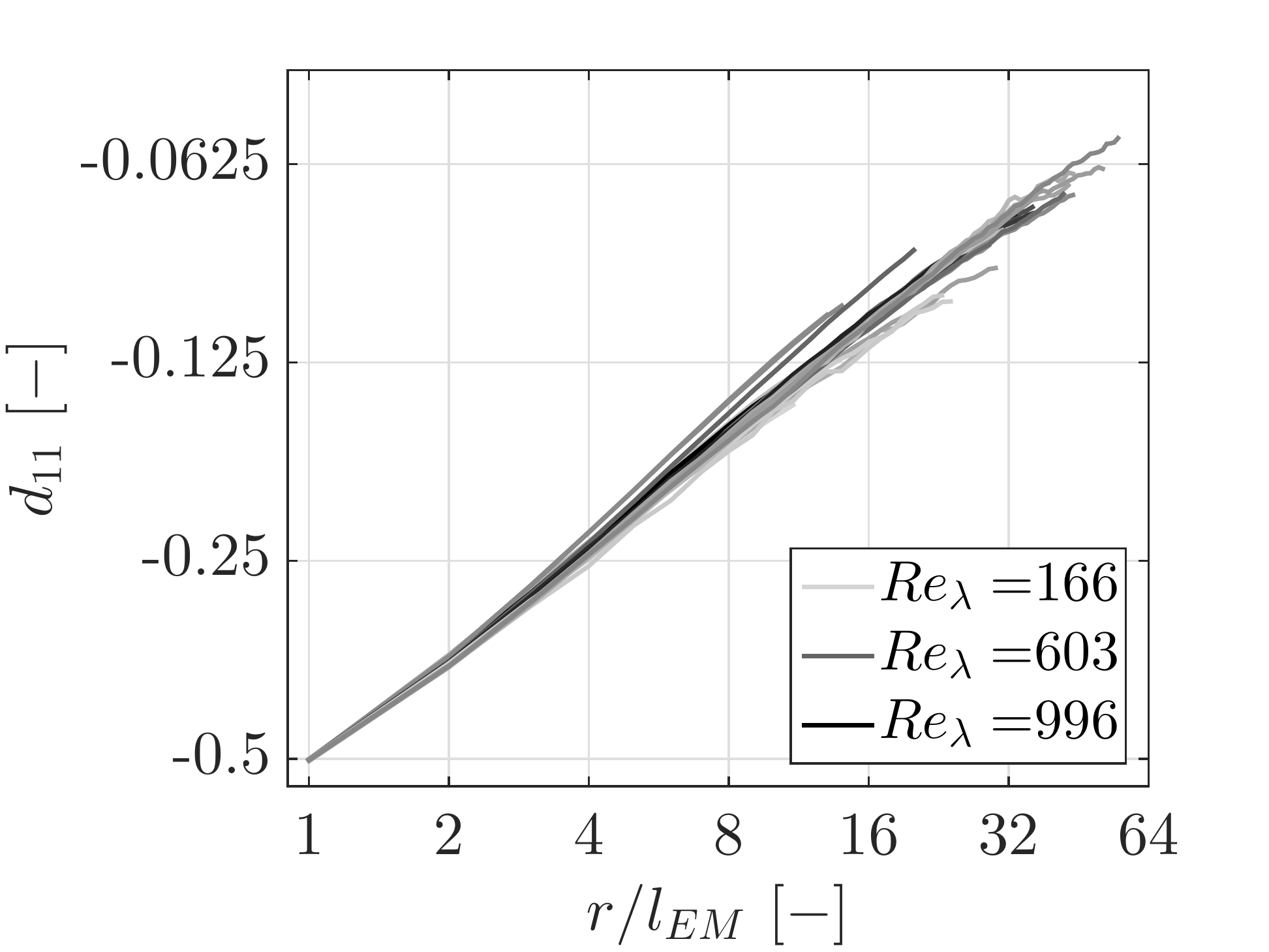}
  \caption{$d_{11}$ as function of scale, of cylinder flows (a) and free jet flows (b). Lines' greyscale indicates $\Rey_\lambda$.
  }
\label{fig:cylinder_jet_d11_all}
\end{figure}

\begin{figure}
  \centering    
a)\includegraphics[width=0.47\textwidth]{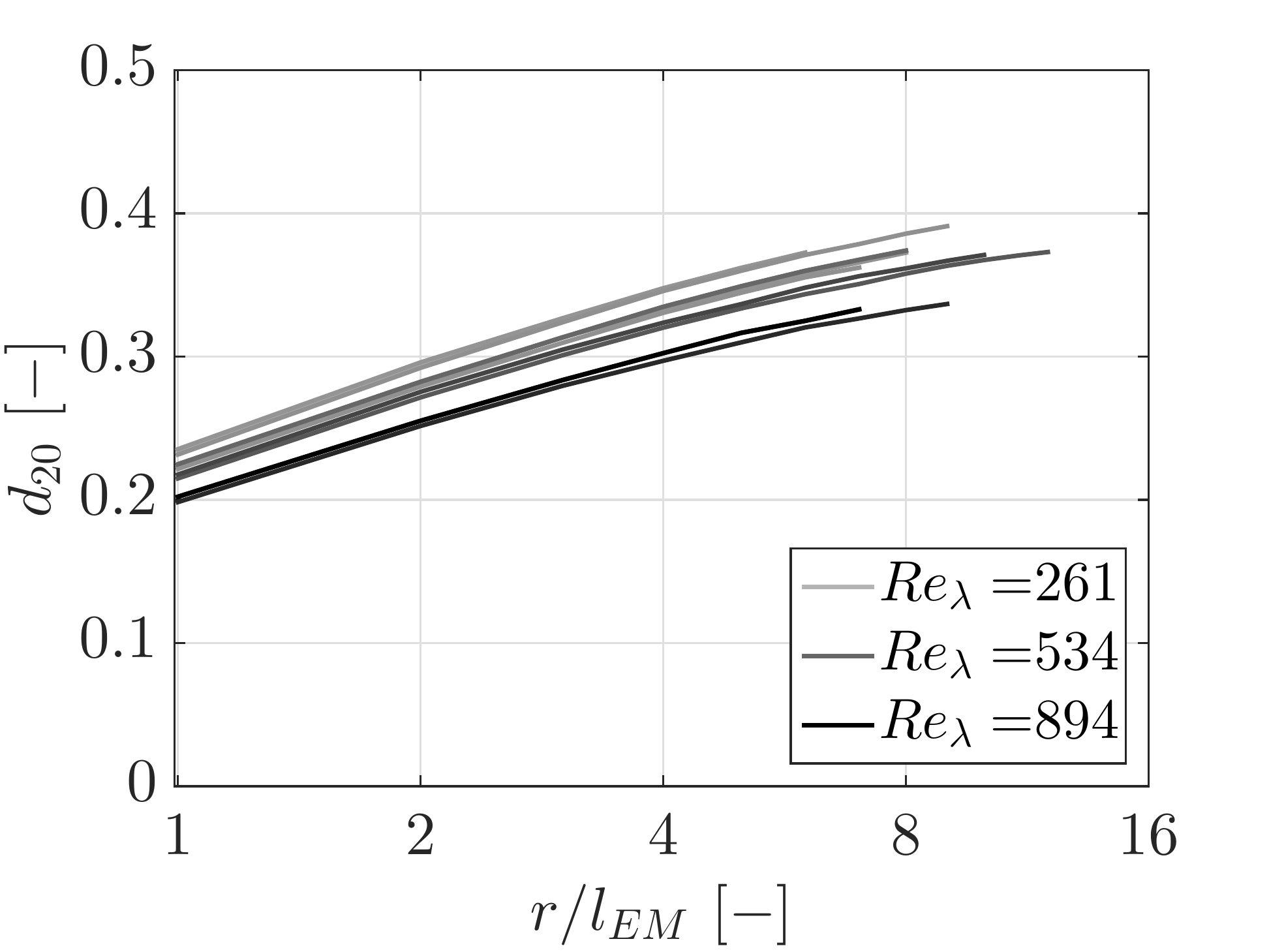}
b)\includegraphics[width=0.47\textwidth]{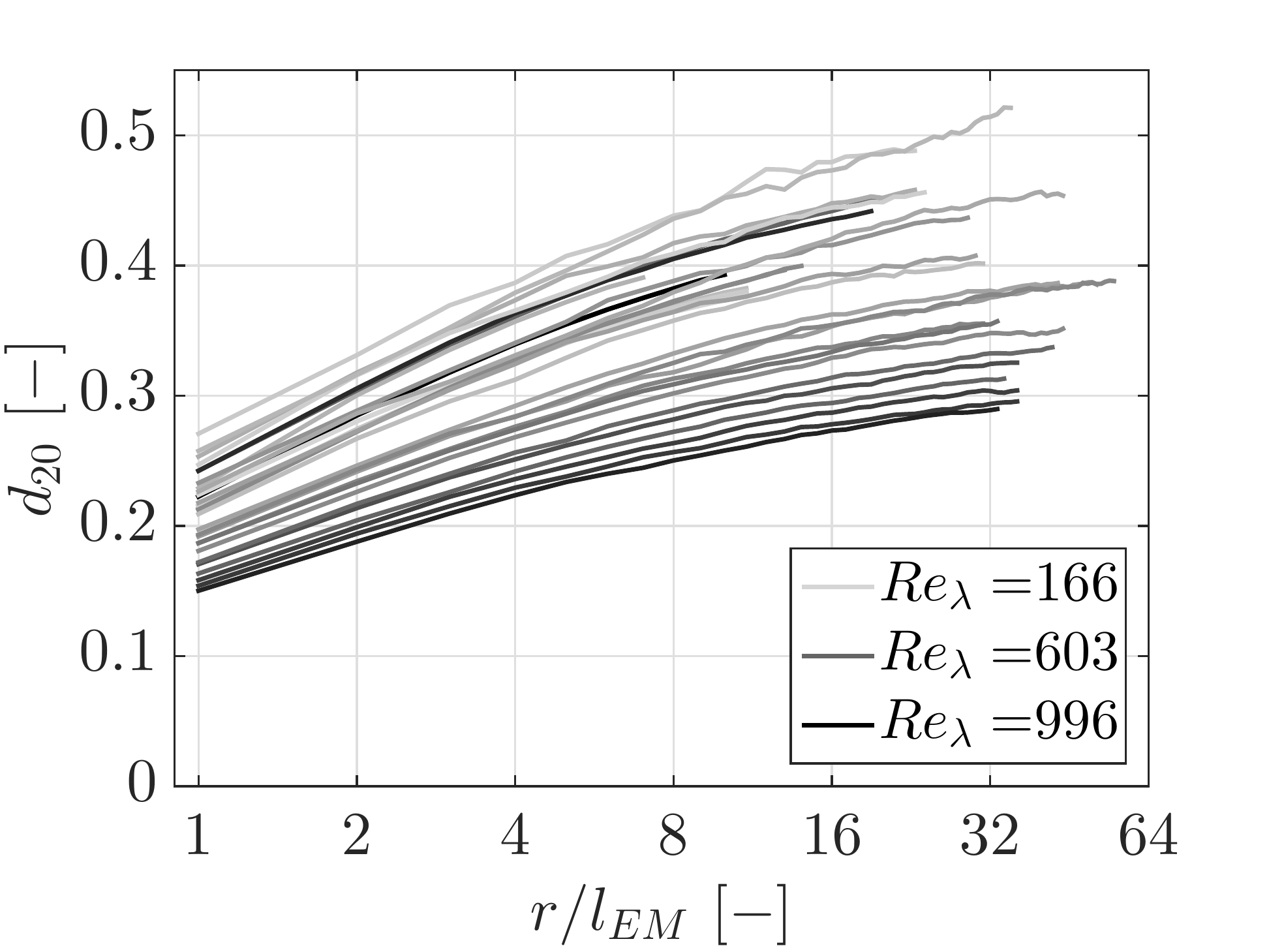}
  \caption{$d_{20}$ as function of scale, of cylinder flows (a) and free jet flows (b). Lines' greyscale indicates $\Rey_\lambda$.}
  \label{fig:cylinder_jet_d20_all}
\end{figure}

\begin{figure}
  \centering    
a)\includegraphics[width=0.47\textwidth]{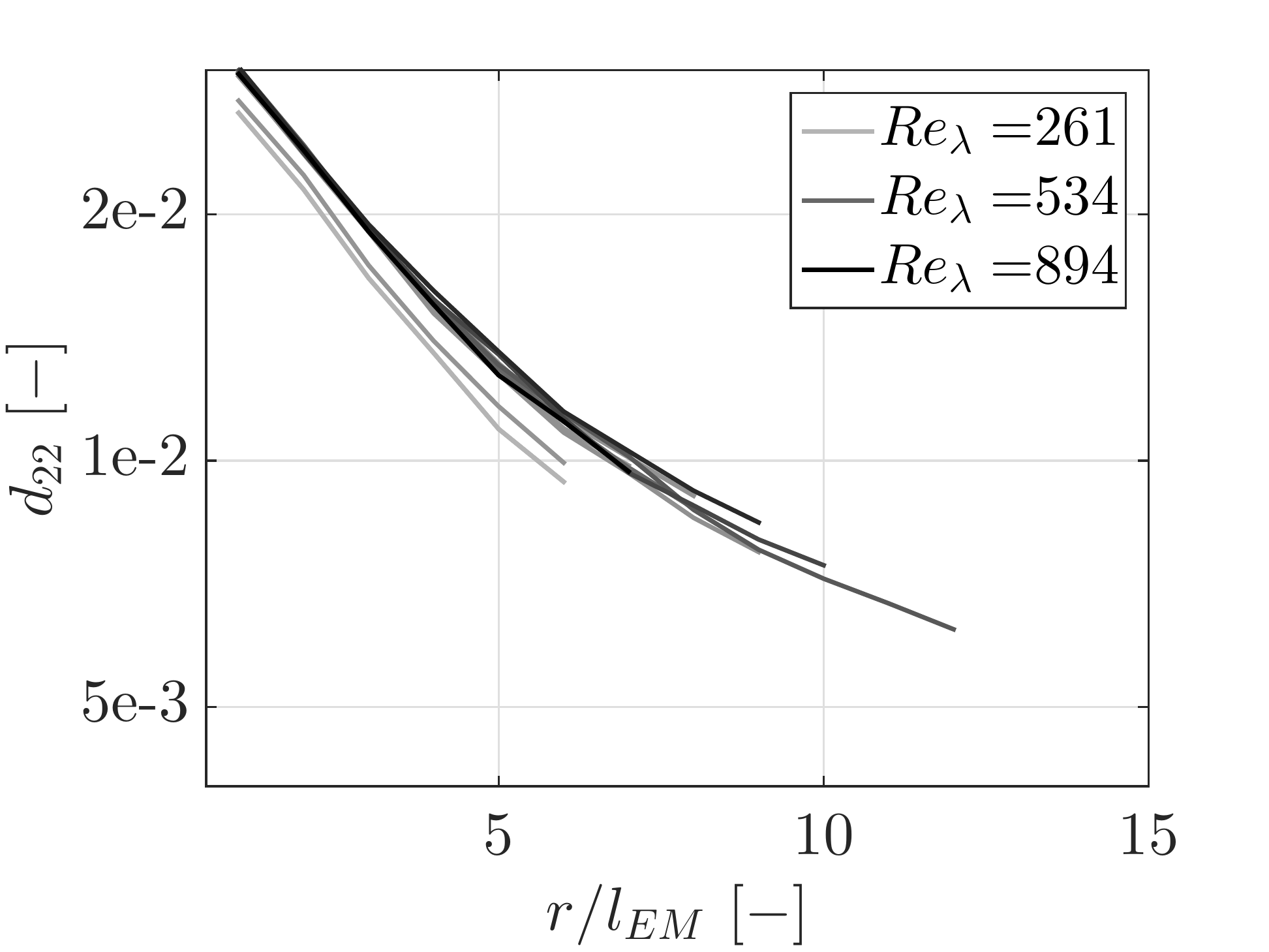}
b)\includegraphics[width=0.47\textwidth]{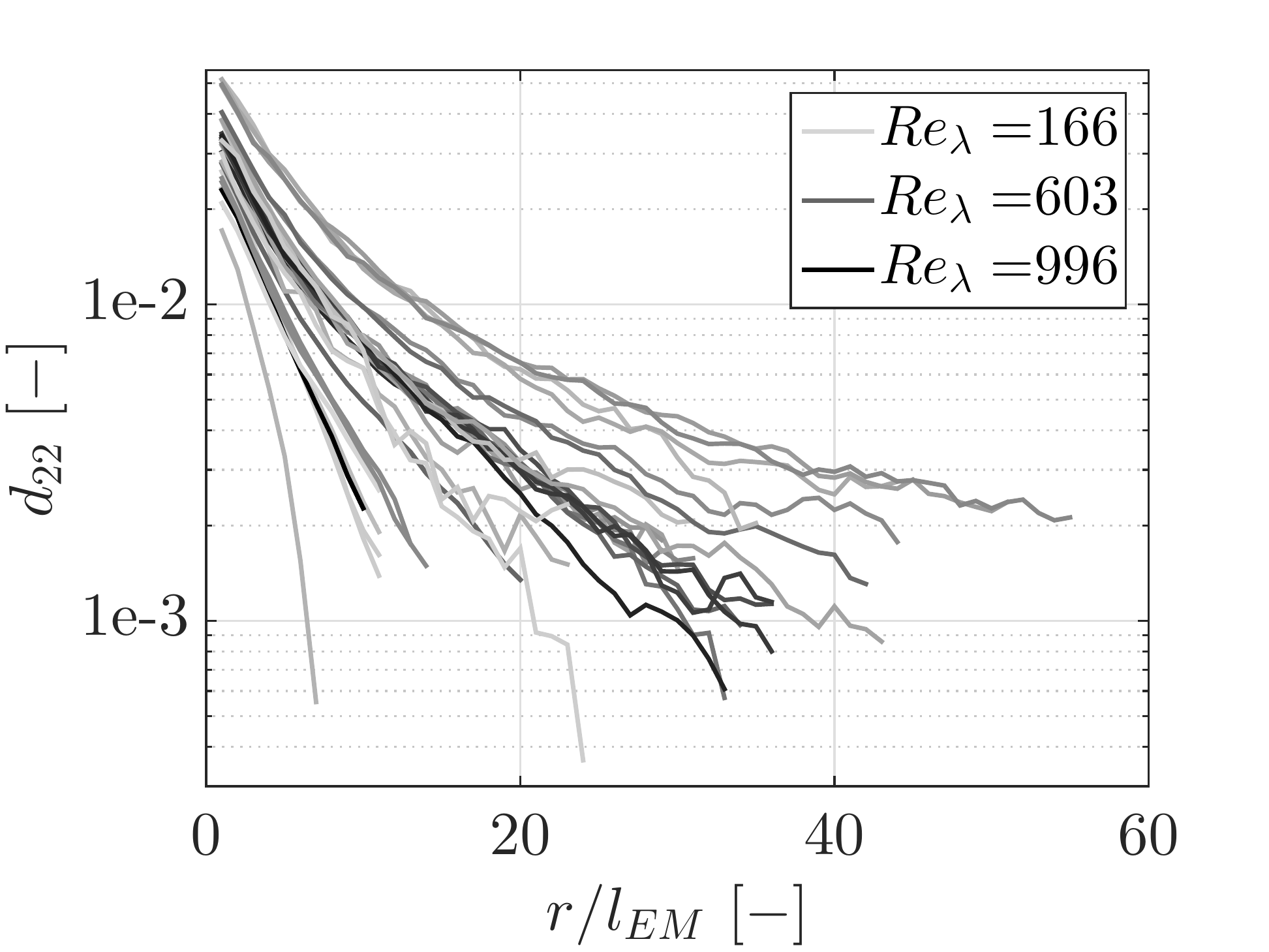}
  \caption{$d_{22}$ as function of scale, of cylinder flows (a) and free jet flows (b). Lines' greyscale indicates $\Rey_\lambda$.}
\label{fig:cylinder_jet_d22_all}
\end{figure}

For the sake of completeness, the coefficients $d_{11}$, $d_{22}$ and $d_{20}$ are
shown for cylinder and free jet flows as function of scale in figures \ref{fig:cylinder_jet_d11_all} - \ref{fig:cylinder_jet_d22_all}. These figures are just the same as figure \ref{fig:5_grid}.

\subsection*{A7: $d_{22}$ as function of $\frac{L}{\lambda}$}
\begin{figure}
\centering  
\vspace{0.2cm}
\includegraphics[width=0.7\textwidth]{figure/12.pdf}
\vspace{0.1cm}
\\
a)\includegraphics[width=0.45\textwidth]{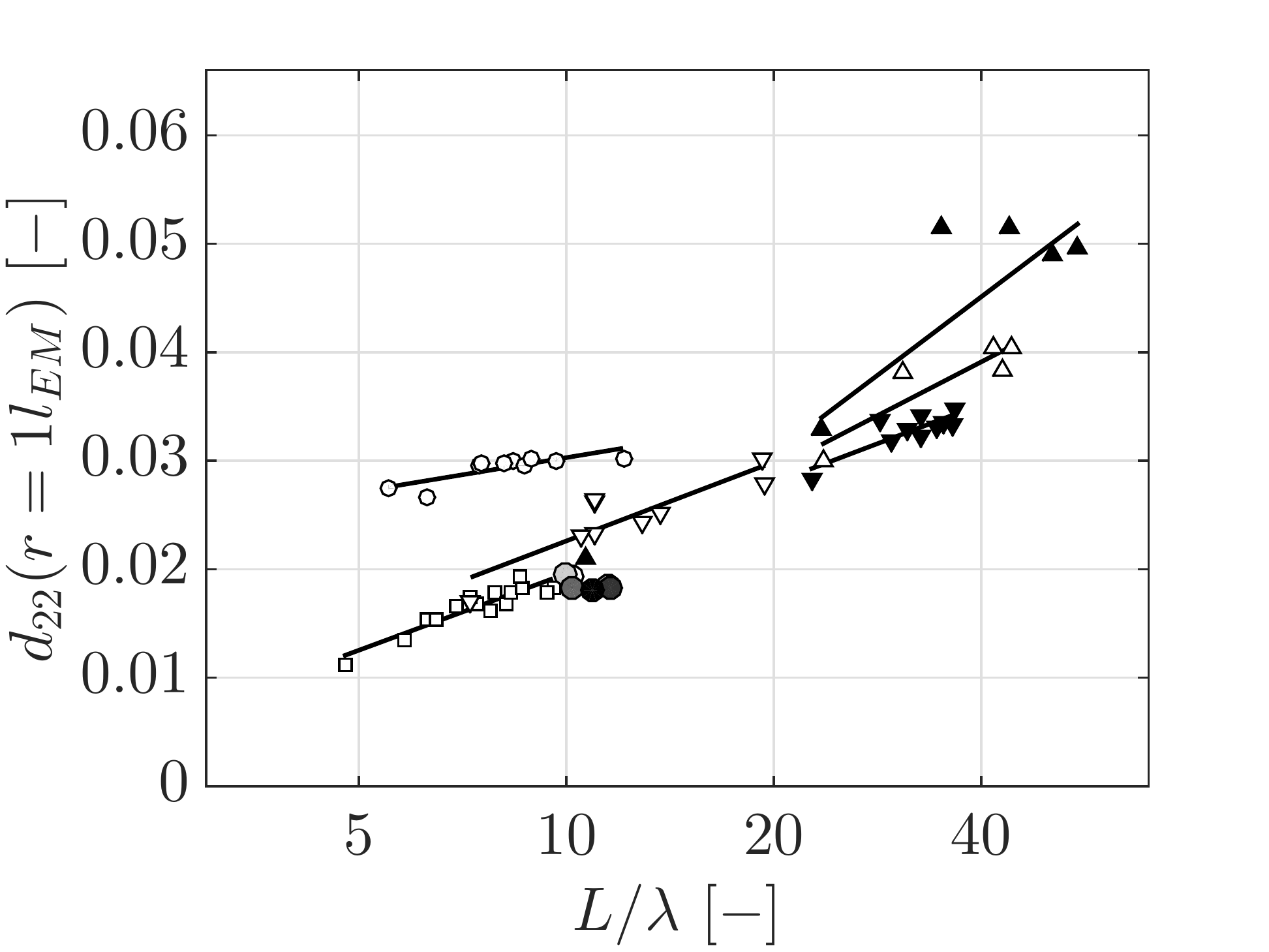}
b)\includegraphics[width=0.45\textwidth]{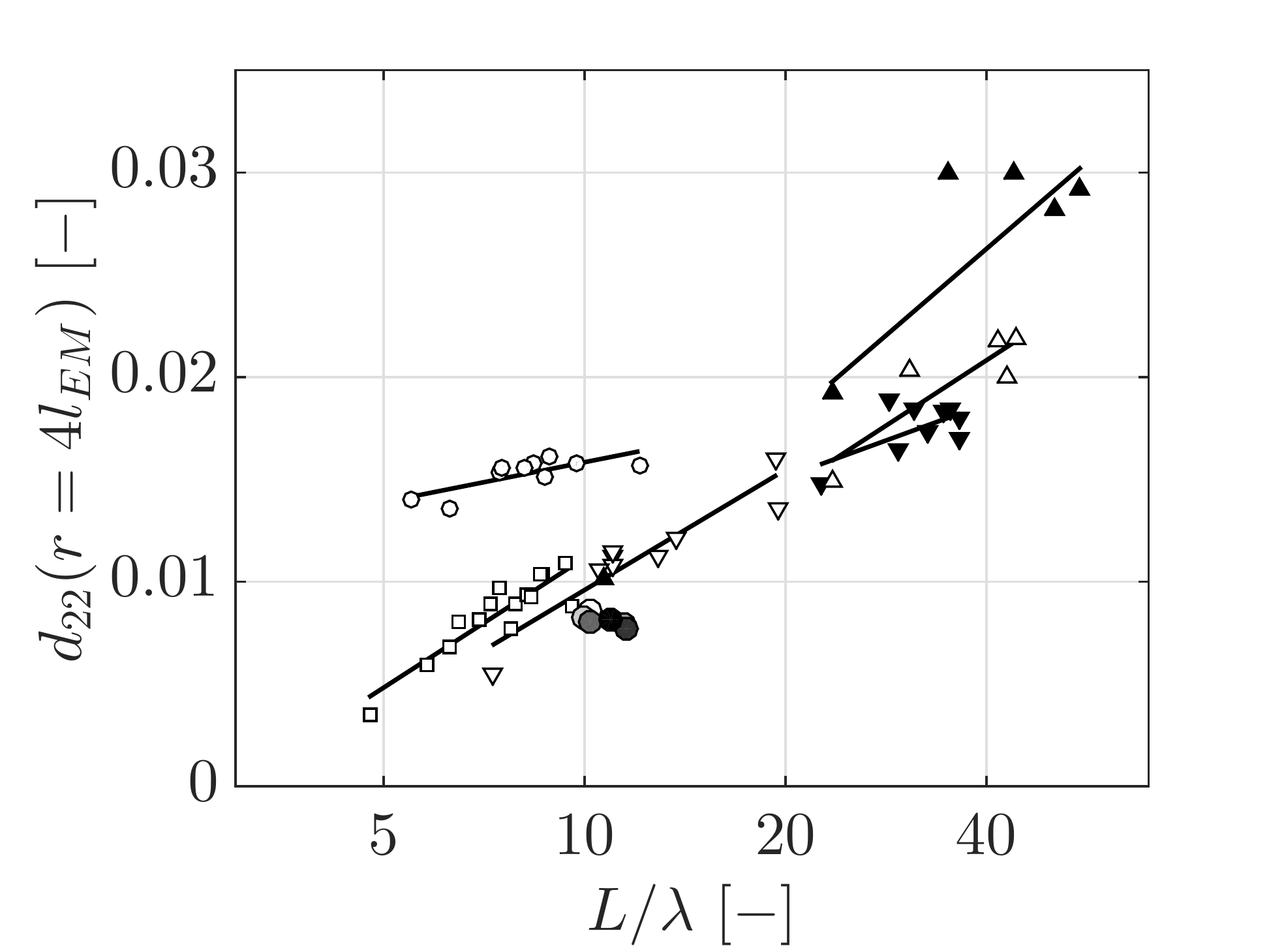}
\caption{Parameter $d_{22}$ as function of $L/\lambda$  at two fixed scales ($r= l_{EM}$ and $4l_{EM}$). Additionally, \textit{linear} fit functions indicate trends. 
Plots are semi-logarithmically plotted.
}
\label{fig:6_klein2}
\end{figure}

Findings of figures \ref{fig:6_klein} e) and f) can be further investigated in terms of universality by interpreting $d_{22}$ as function of a different quantity.
For instance, plotting $d_{22}$ as function of the dimensionless length $\frac{L}{\lambda}$, which is the relative length of the inertial range.
The results become newly sorted and give a much more uniform picture, see figures \ref{fig:6_klein2} a) and b).
Although, single free jet subgroups might be systematically distinguished in terms of the relative distance to the nozzle, the entire free jet datasets show a rather uniform group in contrast to figures \ref{fig:6_klein} e) and f). 
The relative location of the \textit{Cylinder} subgroup differs clearly from the other data in this presentation. 
Although $d_{22}$ is more sorted, we come to the same conclusion as above,
intermittency features in terms of $d_{22}$ are non-universal, and single flow types become apparent. 

\bibliographystyle{jfm}
\bibliography{FoT_49}

\end{document}